\documentclass[11pt]{article}
\usepackage{titlesec} 
\usepackage[font=small]{caption} 
\usepackage[margin=1in]{geometry}
\usepackage{authblk}
\usepackage{amsmath, amssymb}
\usepackage{graphicx}
\usepackage{hyperref}
\usepackage{url}
\usepackage{graphicx}
\DeclareUnicodeCharacter{223C}{\(\sim\)}
\usepackage{amsmath}
\usepackage{amssymb}
\usepackage{mathtools}
\usepackage{subcaption}
\usepackage{setspace}
\usepackage{booktabs}
\usepackage{footmisc}
\usepackage{graphicx}
\usepackage{indentfirst}
\usepackage{booktabs}
\usepackage{tabularx}
\usepackage{amsmath}  
\usepackage{siunitx}  %
\usepackage{algorithm}
\usepackage{algpseudocode}
\algrenewcommand\algorithmicrequire{\textbf{Input:}}   
\algrenewcommand\algorithmicensure{\textbf{Output:}}  

\usepackage[numbers,sort&compress]{natbib}
\usepackage{caption}
\usepackage{float}
\usepackage[table]{xcolor}
\definecolor{lightgreen}{rgb}{0.85,1.0,0.85}
\definecolor{lightred}{rgb}{1.0,0.85,0.85}
\usepackage[export]{adjustbox} % for trim & clip
\usepackage{algorithm}
\usepackage{algpseudocode}
\algrenewcommand\algorithmicdo{}
\usepackage{algpseudocode}
\usepackage{caption}
\usepackage{float}      
\usepackage{placeins}   
\usepackage{graphicx}   
\usepackage{booktabs} 

\renewcommand{\thesection}{\Roman{section}.}
\renewcommand{\thesubsection}{\thesection \Alph{subsection}}
\renewcommand{\thesubsubsection}{\thesubsection.\arabic{subsubsection}}

\titleformat{\section}[block]{\normalfont\Large\bfseries}
  {\thesection}{1em}{}

\titleformat{\subsection}[block]{\normalfont\large\bfseries}
  {\thesubsection}{1em}{}

\titleformat{\subsubsection}[block]{\normalfont\normalsize\bfseries}
  {\thesubsubsection}{1em}{}
\title{\textbf{Generative Quasi-Continuum Modeling of Confined Fluids at the Nanoscale}}

\author[1]{Bugra Yalcin}
\author[1,2]{Ishan Nadkarni}
\author[3]{Jinu Jeong}
\author[1,2]{Chenxing Liang}
\author[1,2*]{Narayana R. Aluru}

\affil[1]{Oden Institute for Computational Engineering and Sciences, The University of
Texas at Austin, Austin 78712, Texas, United States}
\affil[2]{Walker Department of Mechanical Engineering, The University of Texas at Austin,
Austin 78712, Texas, United States
}
\affil[3] {Department of Mechanical Science and Engineering, University of Illinois at
Urbana--Champaign, Urbana, Illinois, 61801, United States}
 
\date{}

\makeatletter
\let\@oldmaketitle\@maketitle
\def\@maketitle{%
  \@oldmaketitle%
  \vspace{-2.5em} 
  \begin{center}
    {\small\textbf{*Correspondence:} \texttt{aluru@utexas.edu (N.A.)}}
  \end{center}
  \vspace{0.5em}
}
\makeatother

\begin{document}
\maketitle

\begin{abstract}
We present a data-efficient, multiscale framework for predicting the density profiles of confined fluids at the nanoscale. While accurate density estimates require prohibitively long timescales that are inaccessible by ab initio molecular dynamics (AIMD) simulations, Machine Learned Molecular Dynamics (MLMD) offers a scalable alternative, enabling the generation of force predictions at ab initio accuracy with reduced computational cost. However, despite their efficiency, MLMD simulations remain constrained by femtosecond timesteps, which limit their practicality for computing long-time averages needed for accurate density estimation. To address this, we propose a conditional denoising diffusion probabilistic model (DDPM) based quasi-continuum approach that predicts the long-time behavior of force profiles along the confinement direction, conditioned on noisy forces extracted from limited AIMD dataset. The predicted smooth forces are then linked to continuum theory via the Nernst-Planck equation to reveal the underlying density behavior. We test the framework on water confined between two graphene nano slits and demonstrate that density profiles for channel widths outside of the training domain can be recovered with ab initio accuracy. Compared to AIMD and MLMD simulations, our method achieves orders-of-magnitude speed-up in runtime and requires significantly less training data than prior works.
\end{abstract}

\section{Introduction}

When fluids are confined at the nanoscale, they exhibit physical and chemical behavior that deviate significantly from their macroscopic (bulk) counterparts \cite{bocquetNanofluidicsBulkInterfaces2010c,aluruFluidsElectrolytesConfinement2023f,faucherCriticalKnowledgeGaps2019a}. The interplay between fluid molecules’ interaction with other molecules and the confining surface gives rise to inhomogeneous spatial arrangement of the fluid molecules. The spatial arrangement, known as interfacial layering, manifests as oscillatory behavior in structural properties of the fluid such as density and force profiles \cite{hansenTheorySimpleLiquids2007c,antmanMicroflowsNanoflows2005c}. These structural properties are altered drastically depending on the level of confinement \cite{ruiz-barraganNanoconfinedWaterGraphene2019c,fongInterplaySolvationPolarization2024} and they underpin many unique behavior of nanofluidic systems, as they directly influence properties such as transport \cite{josephWhyAreCarbon2008b,schochTransportPhenomenaNanofluidics2008c}, ion mobility \cite{wuUnderstandingElectricDoubleLayer2022c,ivanovNanoconfinementEffectsStructural2020c} and adsorption \cite{ilgenAdsorptionNanoconfinedSolid2023c,wordsworthInfluenceNanoconfinementElectrocatalysis2022c}—which are critical in understanding numerous nanoscale applications, such as nano power generation \cite{fengSinglelayerMoS2Nanopores2016c}, desalination \cite{heiranianWaterDesalinationSinglelayer2015c,qianNanoconfinementMediatedWaterTreatment2020c}, electrochemical energy storage \& conversion \cite{yangUltrafastLithiumIonTransport2025c,jiangOscillationCapacitanceNanopores2011c} and nanosensing \cite{zhangNanoconfinementEffectSignal2021c,chuNanofluidicSensingInspired2024c}. Therefore, it is of fundamental importance to be able to predict and control the interfacial layering observed in fluids under nanoscale confinement.

As experimental methods in probing molecular-scale layering and transport are challenging \cite{chantipmaneeNanofluidicManipulationSingle2024c}, and classical continuum theories often fail in capturing the effects of confinement \cite{bocquetNanofluidicsBulkInterfaces2010c,aluruFluidsElectrolytesConfinement2023g}, atomistic computational methods have been used for understanding confined fluid behavior. Monte Carlo simulations (MC) were among the first to gain insights into fluid behavior near interfaces. In MC simulations, equilibrium configurations are sampled from the Boltzmann distribution governed by the underlying potential energy surface, enabling access to equilibrium properties without resolving system dynamics. Monte Carlo methods have been widely used to study equilibrium properties of confined fluids, revealing phenomena such as layering near walls \cite{snookMonteCarloStudy1978c}, confinement-induced phase transitions in water \cite{meyerLiquidLiquidPhaseTransition1999c}, and squeezing behavior of liquid argon under nano confinement. \cite{lengComparativeStudyGrand2013c}. MC suffers from poor scalability and slow convergence in molecular systems where complex interactions such as hydrogen bonding and electrostatics are prominent, as trial moves frequently lead to energetically unfavorable configurations \cite{paquetMolecularDynamicsMonte2015,snookMonteCarloStudy1978c}. This necessitates the use of advanced move strategies and enhanced sampling techniques, increasing both algorithmic complexity and computational cost.

Molecular dynamics (MD) simulations provide atomistic insights into the structure and dynamics of molecular systems, capturing dynamics and time-dependent correlations absent in Monte Carlo methods \cite{frenkelUnderstandingMolecularSimulation2002c}. MD has been employed extensively to study the structural and dynamical behavior of fluids under confinement \cite{faucherCriticalKnowledgeGaps2019a,lynchWaterNanoporesBiological2020c}. A fundamental limitation of MD is the small time step (typically 1–2 fs) required to resolve fast atomic vibrations such as bond stretching. While this ensures numerical stability and accuracy, it puts restrictions on the accessible length and timescales by the simulation\cite{paquetMolecularDynamicsMonte2015}.

Multiscale methods aim to alleviate this problem by transferring information between finer (atomistic) and coarser (continuum or coarse-grained) representations of a system to combine the accuracy of the former with the computational efficiency of the latter. For instance, Coarse-Graining (CG) strategies reduce the computational complexity by grouping atoms and representing them as single interaction sites, then the interaction potential among these sites is optimized to retain the essential characteristics of the underlying all-atom (AA) system \cite{noidPerspectiveAdvancesChallenges2023,jinBottomupCoarseGrainingPrinciples2022}. CG have been used to investigate the structure and dynamics of ionic liquids and water under nano confinement \cite{sanghiCoarsegrainedPotentialModels2012a,motevaselianExtendedCoarsegrainedDipole2018a,mashayakCoarseGrainedPotentialModel2012a}. Quasi-continuum (QC) approaches investigated direct incorporation of atomistic information such as analytical form of pairwise interactions into continuum models for rapid simulation of nanoscale hydrodynamics. It has been shown to capture the structural and thermodynamical properties of simple Lennard-Jones (LJ) fluids \cite{raghunathanInteratomicPotentialbasedSemiclassical2007a}, and water \cite{mashayakThermodynamicStatedependentStructurebased2012a} under different levels of confinement and thermodynamic conditions.

Although the development in modern computational resources including High-Performance Computing (HPC), GPU acceleration, and parallel computing has drastically expanded the accessible length and time scales in MD and multiscale simulations \cite{stoneAcceleratingMolecularModeling2007c,shawAtomicLevelCharacterizationStructural2010c,jungScalingMolecularDynamics2019c} the accuracy of the underlying force field ultimately determines the realism of the system's predicted properties. In systems such as water confined in sub nanometer lengths, electronic properties and quantum effects become significant and therefore, predictions of aforementioned methods when they are equipped with classical force-fields may become unreliable \cite{wuInitioContinuumLinking2023a}. Ab-initio molecular dynamics (AIMD) addresses this limitation by computing interatomic forces on the fly using Density Functional Theory (DFT), inherently capturing effects such as charge transfer, polarization, and other quantum interactions that are neglected in classical models \cite{tuckermanInitioMolecularDynamics2002c}. As a result, it is well-suited for accurately simulating confined fluids at the nanoscale \cite{liangStructuralDynamicalProperties2023c}. For instance, it has been observed that the structural properties obtained by the AIMD simulations deviate significantly for those obtained by classical force-field (SPC/E water \cite{berendsenMissingTermEffective1987}) for water confined between graphene particularly for heavy confinement ($< 4$ nm) \cite{wuInitioContinuumLinking2023a}. However, because AIMD requires DFT calculations at every time step to compute forces, its computational cost is orders of magnitude higher than classical MD. As a result, AIMD is typically restricted to simulation of small systems (hundreds of atoms) and short trajectories (tens of picoseconds) \cite{heStatisticalVariancesDiffusional2018c}. This motivates the development of computational tools that can achieve AIMD-level accuracy at significantly lower cost.

Over the past decade, machine learning–driven molecular dynamics (MLMD) has emerged as an alternative tool for simulating molecular systems with quantum-level accuracy, orders of magnitude cheaper than ab initio methods such as DFT and AIMD \cite{choudharyUnifiedGraphNeural2023,zhangDeepPotentialMolecular2018c,
behlerGeneralizedNeuralNetworkRepresentation2007,schuttSchNetDeepLearning2018,unkePhysNetNeuralNetwork2019,parkAccurateScalableGraph2021,batznerE3equivariantGraphNeural2022,batatiaMACEHigherOrder2022,musaelianLearningLocalEquivariant2023c}. In MLMD, neural network potentials (NNPs) are trained  on quantum mechanical data to learn complex many-body interactions. Once trained, NNPs are integrated into MD simulations to predict interatomic forces based on high-dimensional representations of the atomic environments, thereby, by passing expensive quantum mechanical calculations at every time step . A major bottleneck of this method is the need for large, diverse, and high-quality training datasets, often requiring expensive ab initio simulations \cite{choudharyUnifiedGraphNeural2023,schuttSchNetDeepLearning2018,unkePhysNetNeuralNetwork2019,batznerE3equivariantGraphNeural2022,koRecentAdvancesOutstanding2023}. Despite replacing expensive force evaluations, MLMD inherits the small time step requirement of classical MD \cite{liGraphNeuralNetworks2022c,fuSimulateTimeintegratedCoarsegrained2022c}. Additionally, the training process itself could be computationally demanding \cite{liExploitingRedundancyLarge2023c,choudharyUnifiedGraphNeural2023}.

Several recent frameworks have combined the efficiency of multiscale methods with the predictive power of machine learning to study fluids under nanoconfinement. For example, \cite{nadkarniDataDrivenApproachCoarseGraining2023a} used neural networks to predict coarse-grained interaction parameters for simple liquids in confined environments. Building on earlier QC approaches, a deep learning–based framework \cite{wuDeepLearningbasedQuasicontinuum2022} was developed to predict structural properties of Lennard-Jones (LJ) fluids and water, where a convolutional encoder–decoder (CED) was trained to infer the potential profile along the nanochannel, from the knowledge of LJ parameters and thermodynamic conditions. More recently, \cite{wuInitioContinuumLinking2023a} extended this idea to predict AIMD-accurate density profiles of water under varying levels of confinement. The authors employed DeepMD \cite{zhangDeepPotentialMolecular2018c} to generate sufficient number of AIMD-accurate frames which were then used to train a CED that denoises noisy force profiles obtained from short AIMD trajectories.

In this work, a generative modeling based extension of the QC framework employed in \cite{wuInitioContinuumLinking2023a}, is proposed to obtain density profiles of confined liquids with AIMD accuracy under reduced data requirements. We train an Equivariant Graph Neural Network (EGNN) \cite{musaelianLearningLocalEquivariant2023c}  on a dataset which is randomly subsampled from the AIMD water confined in graphene nano slit data used in the previous work. The EGNN is trained under an active-learning (AL) loop, where the training dataset is iteratively expanded by short AIMD trajectories performed on frames that are selected based on deviations in atomic distance distributions relative to the original dataset. The AL iterations are performed until the MLMD simulation can be run until desired accuracy for the density of a given channel width is reached. An Adaptive Gaussian Denoising (AGD) \cite{dengAdaptiveGaussianFilter1993} strategy based on sampling uncertainty, which we refer to as Local Force Approximation (LFA) is proposed as an early stopping strategy to obtain smooth force profiles without resorting to long nano-second simulations required to obtain enough frames for averaging. The smooth forces are used to train a conditional denoising diffusion-model \cite{pmlr-v139-nichol21a,hoDenoisingDiffusionProbabilistic2020c,sohl-dicksteinDeepUnsupervisedLearning2015}, a generative model which we use to predict smooth force profiles for unseen confinement conditions during training, conditioned on the extremely noisy (due insufficient statistics) forces that are obtained from the QCT-inference algorithm adapted from \cite{wuInitioContinuumLinking2023a}. Finally, the predicted force profiles are used as input to the continuum theory given by the Nernst-Planck equation to obtain the density profiles. The key contributions of this paper are:
\clearpage
\paragraph{}
\begin{itemize}
    \item We present a data-efficient, generative, multiscale framework for predicting AIMD-accurate density profiles of nano confined fluids, which requires 70\% less training data than previous approaches.
    \item We propose a sampling-uncertainty-based adaptive denoising strategy that enables early termination of MLMD simulations and eliminates the need for long-time force sampling required to obtain smooth force profiles in nanoconfined fluids.
\end{itemize}

\section{Methods}

\subsection{Continuum Model (Nernst–Planck Equation)}

The system under study in this work (water sandwiched between graphene sheets) is modeled as a fluid confined between two infinite walls (in the $y$, $z$ directions). Therefore, the particle density only varies along the channel width ($x$-direction). Further, the system is at steady-state ($\frac{\partial \rho}{\partial t} = 0$). Hence, the equilibrium density profile along the channel, $\rho(x)$, can be obtained by solving the 1D steady-state Nernst--Planck equation \cite{teorellTransportProcessesElectrical1953}, which relates the equilibrium force profile $f(x)$ to $\rho(x)$.
\begin{equation}
\frac{d}{dx} \left( \frac{d\rho}{dx} - \frac{\rho}{RT} f(x) \right) = 0
\label{eq:np}
\end{equation}

subject to Dirichlet boundary conditions:

\begin{equation}
\rho(x=0) = \rho(x=L) = 0
\label{eq:bc}
\end{equation}

where $L$ is the channel width. For the solution to be physical, it must preserve the total number of particles in the channel, which can be mathematically imposed as:

\begin{equation}
\frac{1}{L} \int_0^L \rho(x) \, dx = \rho_{\text{avg}}
\label{eq:constraint}
\end{equation}

Here, $\rho_{\text{avg}}$ is the average density of the fluid inside the nanochannel. The solution of Eq.~\ref{eq:np} requires the input of $f(x)$, the force profile along the channel, which is related to the potential energy $U(x)$ via $f(x) = -\frac{dU(x)}{dx}$. In this framework, we seek to bridge AIMD accurate atomistic physics to the continuum model by making $f(x)$ AIMD-accurate. The details of the numerical solution technique used to solve Eq.~\ref{eq:np} can be found in Appendix~A.

\subsection{Computational Framework}

The proposed framework predicts density profiles of confined nanofluids with AIMD accuracy using a limited set of AIMD simulation data corresponding to different levels of confinement. Our approach consists of three stages:

\begin{enumerate}
    \item An equivariant graph neural network (EGNN) \cite{musaelianLearningLocalEquivariant2023c} is used to obtain long-time statistics of the force distribution along the confinement direction, which are then used to train a denoising diffusion probabilistic model (DDPM) \cite{hoDenoisingDiffusionProbabilistic2020c}.
    \item For a new channel width, the force distribution along the confinement dimension is approximated from the limited AIMD data using the methodology proposed in \cite{wuInitioContinuumLinking2023a}.
    \item The obtained noisy force profile is denoised via the reverse diffusion process, and the resulting smooth force is fed into the continuum model described by the Nernst–Planck equation (Eq.~\ref{eq:np}).
\end{enumerate}

First, we present the training workflow for the EGNN and DDPM. This is followed by a description of the inference pipeline, which uses the trained models to predict density profiles with ab initio accuracy under different confinement conditions.

\

\begin{figure}[htbp]
    \centering
    \includegraphics[width=0.95\textwidth, trim=50 100 0 50, clip]{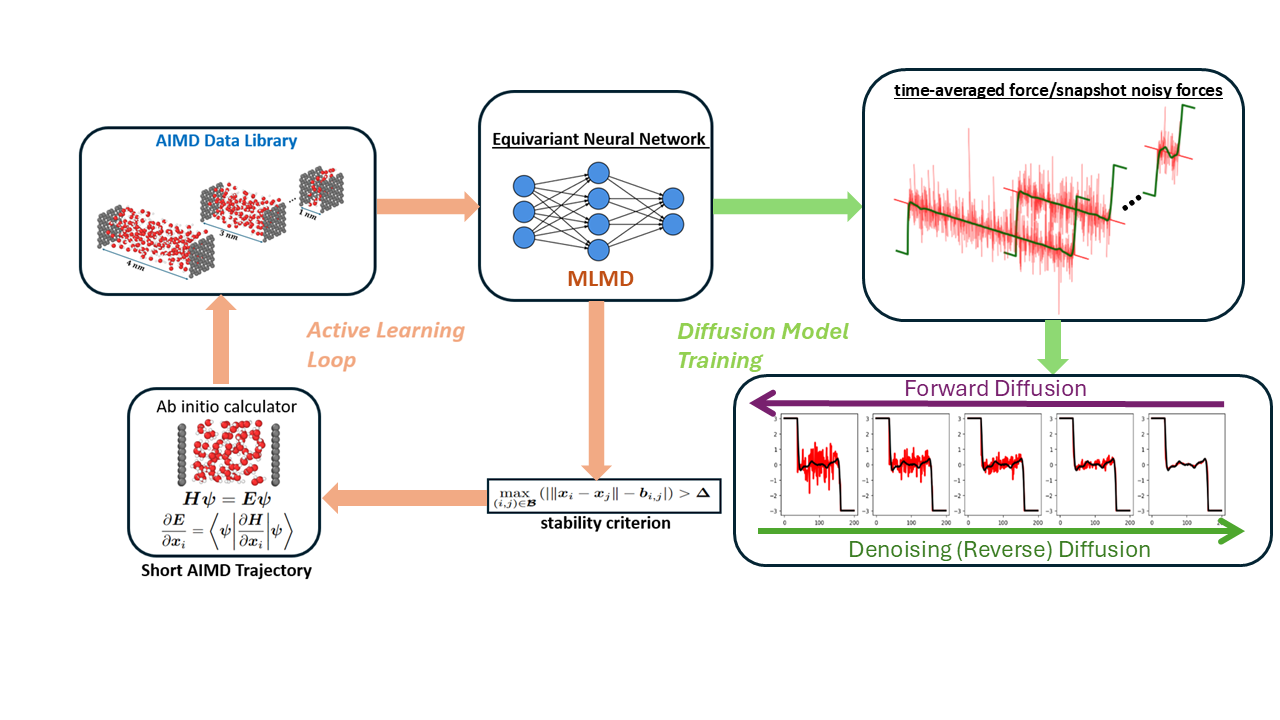}
    \caption{
    \textbf{Workflow for training the neural network force field and denoising diffusion model.} (left) An active learning loop is used to iteratively train an Equivariant Graph Neural Network (EGNN) on AIMD data of nanoconfined water. The EGNN serves as the force field for MD simulations. When a configuration violates the stability criterion, short AIMD simulations are performed on that frame, and the resulting trajectory is added to the training set. (right) Once sufficient number of frames have been generated to capture the desired force statistics, single-frame noisy forces and their corresponding time-averaged (smooth) counterparts are extracted from the MLMD simulations. These paired data are used to train the diffusion model, which learns a mapping from noisy to smooth force profiles.
    }
    \label{fig:qct_training}
\end{figure}

\subsubsection{Neural Network Potential}

In slit-like nanochannels, the equilibrium density profile is related to the underlying force profile via Eq~\ref{eq:np}. Obtaining ab initio--accurate density profiles through this equation therefore requires equally accurate force inputs. However, achieving statistical convergence in the force profile necessitates averaging over many AIMD frames (typically spanning nanosecond timescales), which is computationally infeasible due to the high cost of ab initio simulations.

In this work, we use Allegro \cite{musaelianLearningLocalEquivariant2023c} as the NNP and perform molecular dynamics simulations with the LAMMPS~\cite{thompsonLAMMPSFlexibleSimulation2022} package. In contrast to traditional Graph Neural Networks (GNNs), which rely on iterative message passing between neighboring atoms to build representations, Allegro constructs local, equivariant features directly using tensor products of spherical harmonics. This design eliminates the need for recursive updates through the graph, significantly improving computational efficiency and scalability (see \cite{musaelianLearningLocalEquivariant2023c} for details).
The Allegro model receives atomic positions $\mathbf{q}$ and species $\mathbf{z}$ as input to construct local neighborhood graphs around each atom within a prespecified cut-off radius, which we set as 6~\AA, and computes pairwise, rotation-equivariant representations of atomic environments. The network learns to map these representations to pairwise energy contributions, which are aggregated to yield the total potential energy of the system, $E_{\text{tot}}(\mathbf{q}, \mathbf{z})$. Forces on each atom are then obtained via the energy gradient: $\mathbf{F}_{\mathbf{q}} = -\nabla_{\mathbf{q}} E_{\text{tot}}(\mathbf{q}, \mathbf{z})$. Detailed architecture and training specifications are provided in Appendix~B.

\paragraph{Training Data.} The initial training dataset is constructed by randomly selecting 30\% of the AIMD-confined water configurations used in ~\cite{wuInitioContinuumLinking2023a}. This dataset comprises 14 systems spanning channel widths from 1~nm to 4~nm. A 90/10 train/test split is applied in both the initial and subsequent training rounds. Further dataset details are provided in Appendix~\ref{sec:AppendixB}.

\vspace{1em}
\paragraph{Active Learning.}

Test-set accuracy alone is insufficient to evaluate the robustness of a NNP during MLMD simulations \cite{fuForcesAreNot2022}. As the simulation progresses, atomic configurations that reside in regions of the potential energy surface (PES) lying outside the training distribution may be encountered. The network’s predictions on such configurations therefore become unreliable and could result in instabilities. Active Learning (AL) addresses this issue by identifying “critical” configurations encountered during MLMD simulations and iteratively augmenting the training set with their corresponding force and energy labels obtained from first principles calculations, allowing the network to learn previously unexplored regions of the PES~\cite{fuForcesAreNot2022,zhangDPGENConcurrentLearning2020}.

\vspace{1em}
\paragraph{AIMD Configuration Labeling.}

We adapt a bond-length–based selection criterion~\cite{liangStructuralDynamicalProperties2023c,fuForcesAreNot2022,rajaStabilityAwareTrainingMachine2024} to detect important frames. First, the mean ($\mu_\text{dist}$) and standard deviation ($\sigma_\text{dist}$) of  intramolecular (O--H) and intermolecular (O--O, C--O, C--H) distances are calculated from the initial dataset. During simulation, whenever a frame exhibits a distance deviation greater than $6\sigma_\text{dist}$ (denoted $\Delta$), the simulation is terminated. The frame is flagged as important and used as the initial condition for a short AIMD simulation (see Appendix~\ref{sec:AppendixD} for the simulation details). The resulting AIMD frames are appended to the training set, the network is retrained, and a new MLMD simulation is launched.

This procedure (Algorithm \ref{alg:train}) is carried out independently for each channel in the training set, and we define one AL iteration as a complete pass through all channels. After two AL iterations, we were able to run 1.5\,ns--long trajectories without encountering instabilities.

\begin{algorithm}[ht]
\caption{Active Learning for NNP Training}
\label{alg:train}
\begin{algorithmic}[1]
\State \textbf{Input:} Initial dataset $\mathcal{D}_0$ composed of channels $\{L_1, L_2, \dots\}$
\State Compute distance statistics the mean ($\mu_{\text{dist}}$) and standard deviation ($\sigma_{\text{dist}}$) from $\mathcal{D}_0$
\State Set deviation threshold $\Delta = 6\sigma_\text{dist}$
\State Train initial NNP $\mathcal{M}_0$ on $\mathcal{D}_0$
\For{each active learning iteration $k$}
    \For{each channel $L_i$ in the dataset}
        \State Run MLMD simulation using model $\mathcal{M}_{k-1}$
        \For{each frame in the trajectory}
            \State Determine the set $\mathcal{B}$ of relevant atomic pairs in the frame
            \State For each atomic pair compute distance $b_{i,j}\in \mathcal{B}$,
            \If{$\max_{(i,j) \in \mathcal{B}} \left| \|\mathbf{x}_i - \mathbf{x}_j\| - b_{i,j} \right| > \Delta$}
                \State Flag the frame; terminate the simulation
                \State Initialize short AIMD simulation from the flagged configuration
                \State Add resulting AIMD frames to dataset $\mathcal{D}_k$
            \EndIf
        \EndFor
    \EndFor
    \State Retrain model $\mathcal{M}_k$ on updated dataset $\mathcal{D}_k$
\EndFor
\end{algorithmic}
\end{algorithm}

\subsubsection{Local Force Approximation (LFA)}
\label{sec:lfa_theory}
In practice, the number of AIMD frames required to obtain the underlying smooth equilibrium force profile varies with the varying degree of confinement. We observed that for narrower channels, where molecular layering is more pronounced, density and force profiles obtained from spatial binning converge faster compared to larger channels. Moreover, convergence in density profiles is achieved faster than that of force profiles (see Appendix~\ref{sec:AppendixC}). Consequently, relying solely on the convergence of density profiles and using the corresponding force profile in Eq~\ref{eq:np} may lead to inaccurate density predictions and therefore cannot be used as a ground truth label for diffusion model training.

In this section, we propose a strategy based on Adaptive Gaussian Denoising (AGD)~\cite{dengAdaptiveGaussianFilter1993,silvermanDensityEstimationStatistics2018} that enables MLMD simulations to be terminated whenever convergence in density profiles is achieved, thereby bypassing the additional computational cost associated with force profile convergence.

\vspace{1em}
\paragraph{Statistical Convergence Criteria.}
To assess convergence, we compute the mean of the per-bin sampling uncertainties \(\sigma_\rho(x_i)\), estimated via block averaging, and normalize this quantity by the maximum of the binned density profile. This yields a relative uncertainty metric, and we assume that the density profile has converged when this normalized mean uncertainty falls below a threshold \(\tau_\rho\). Similarly, we assume that a force profile \(f(x)\) is sufficiently smooth if, upon solving Eq.~1, the resulting density profile \(\rho(x)\) yields a relative root mean squared error (RRMSE) below the same threshold \(\tau_\rho\) when compared to the binned density \(\rho_{\text{bin}}(x)\) obtained from the MLMD simulation. Throughout this work, we set \(\tau_\rho = 0.01\).

\vspace{1em}
\paragraph{Adaptive Gaussian Denoising (AGD).}

In MD simulations, inherent thermal fluctuations introduce aleatoric uncertainty in observables. Temporal averaging reduces this uncertainty, with the associated statistical error scaling approximately as \( 1/\sqrt{N_{\mathrm{eff}}} \), where \( N_{\mathrm{eff}} \) is the number of uncorrelated samples \cite{rizziUncertaintyQuantificationMD2012}. That is, to reduce the standard error by a factor of 2, one must increase the number of uncorrelated frames used for averaging by a factor of 4. One can express the time-averaged ground truth force profile at a thermodynamic state \( (T, \rho) \) and channel width \( L \) as
\begin{equation}
f_{\mathrm{GT}}(x; T, \rho, L) = \tilde{f}(x; T, \rho, L) + \epsilon_t(x; T, \rho, L)
\label{eq:additive_noise_model}
\end{equation}

where \( \tilde{f} \) denotes the noisy force profile obtained from the MLMD simulation, and \( \epsilon_t\) represents the time-dependent additive noise due to thermal fluctuations. 
 Since \( f_{\mathrm{GT}} \) is unknown during simulation, we accept \( \tilde{f} \) as a sufficiently accurate approximation under the condition that the resulting density profile \( \rho_{\tilde{f}} \) satisfies the statistical convergence criterion defined above.

One critical observation is that the dependence of \( \epsilon_t \) on position cannot be neglected, as thermal noise characteristics vary along the confinement direction due to local differences in sampling density and dynamics. Therefore, the problem of extracting the ground truth force from noisy MD data can be viewed as a signal recovery task under spatially varying (heteroscedastic) noise. Unlike traditional assumptions of independent noise, we account for spatial correlations by modeling 
\(\boldsymbol{\epsilon}_t \sim \mathcal{N}(\mathbf{0}, \boldsymbol{\Sigma})\), 
where the covariance matrix \(\boldsymbol{\Sigma}\) captures the correlation structure of noise between spatial bins. Assuming statistical independence between spatial bins may not only underestimate uncertainty in the denoised output but also result in suboptimal smoothing behavior~\cite{wandKernelSmoothing1994,opsomerNonparametricRegressinCorrelated2001}.  As a result, traditional denoising methods that assume noise characteristics are homoscedastic, such as fixed-kernel smoothing~\cite{rosenblattRemarksNonparametricEstimates1956,fixDiscriminatoryAnalysisNonparametric1989}, Wiener filtering~\cite{oppenheimSignalsSystemsInference2016,wienerExtrapolationInterpolationSmoothing1949}, and Savitzky–Golay filtering~\cite{savitzkySmoothingDifferentiationData1964,NumericalRecipesArt1992}, may result in blurring of important features such as force peaks. Adaptive kernel smoothing~\cite{breimanVariableKernelEstimates1977,wandKernelSmoothing1994} offers a straightforward and computationally efficient approach that adapts smoothing strength based on local signal characteristics. Inspired by prior work in image processing \cite{dengAdaptiveGaussianFilter1993}, where the Gaussian kernel width is modulated based on the local variance in the underlying data, we adopt a similar approach tailored to molecular simulation data.

Under the additive noise model, the observed signal \( y(t) \) can be expressed as:
\begin{equation}
y(t) = \tilde{y}(t) + \epsilon(t)
\end{equation}
where \( \tilde{y}(t) \) is the true underlying signal and \( \epsilon(t) \sim \mathcal{N}(0, \sigma^2) \) is additive Gaussian noise. In 1D, standard Gaussian denoising corresponds to convolving \( y(t) \) with a Gaussian kernel \( G(x) \), defined as:
\begin{equation}
G(x) = \frac{1}{\sqrt{2\pi\sigma^2}} \exp\left( -\frac{x^2}{2\sigma^2} \right)
\end{equation}
Here, \( \sigma \) controls the bandwidth of the filter: small \( \sigma \) preserves fine-scale structure but performs limited smoothing, while large \( \sigma \) increases smoothing at the expense of blurring finer features, such as force peaks or sharp gradients that may be characteristic of the true underlying signal. The Gaussian kernel assumes that the underlying signal is locally smooth, and that nearby points offer a more accurate approximation of the true value than distant ones. The denoised signal is given by:

\begin{equation}
\tilde{y}(t) = (y * G)(t) = \int_{-\infty}^{\infty} y(\tau) G(t - \tau) d\tau
\label{eq:gaussian_convolution}
\end{equation}

Adaptive Gaussian Denoising (AGD) extends standard Gaussian Denoising by allowing the filter bandwidth to vary spatially. Specifically, the fixed kernel width \( \sigma \) is replaced by a position-dependent function \( \sigma(x) \), enabling the degree of smoothing to adapt based on local signal characteristics (\( \sigma \rightarrow \sigma(x) \)).

We leverage the sampling uncertainty in the force profile, reported as the standard error of the mean obtained via block averaging \cite{flyvbjergErrorEstimatesAverages1989}—to:
\begin{enumerate}
    \item guide the denoising process, and
    \item quantify the propagated uncertainty in the resulting density profiles.
\end{enumerate}

To quantify the sampling uncertainty in the time-averaged force profile, we employ a standard block-averaging approach \cite{flyvbjergErrorEstimatesAverages1989}, where the trajectory is partitioned into independent segments. For each spatial bin \( x_i \), the standard error of the mean (SEM) is computed as:
\begin{equation}
\bar{\sigma}_{\tilde{f}}(x_i) = \frac{1}{\sqrt{n_B}} 
\sqrt{ \frac{1}{n_B - 1} \sum_{j=1}^{n_B} \left( \langle \tilde{f} \rangle_j^B(x_i) - \langle \tilde{f} \rangle^B(x_i) \right)^2 }
\label{eq:sem_force}
\end{equation}

Here, \( \langle \tilde{f} \rangle_j^B(x_i) \) is the average force in bin \( x_i \) within block \( j \), and \( \langle \tilde{f} \rangle^B(x_i) \) is the average across all \( n_B \) blocks. The resulting quantity \( \bar{\sigma}_{\tilde{f}}(x_i) \) provides a local estimate of the statistical uncertainty due to thermal fluctuations and finite sampling. By the Central Limit Theorem, the per-bin force distribution approximates a Gaussian distribution as the number of blocks increases. Therefore, we model the additive noise term \( \epsilon_t(x_i) \) in Eq.~\eqref{eq:additive_noise_model} as a zero-mean Gaussian, \( \epsilon_t(x_i) \sim \mathcal{N}\left(0, \bar{\sigma}_{\tilde{f}}^2(x_i)\right) \).

With the knowledge of the uncertainty associated with spatial bins, an uncertainty-based kernel width \( \sigma_s^i \) can be defined as:
\begin{equation}
\sigma_s^i = \sigma_{\text{base}} \cdot \left( \frac{ \bar{\sigma}_{\tilde{f}}(x_i) }{ \min\limits_{x} \bar{\sigma}_{\tilde{f}}(x) } \right)
\label{eq:adaptive_sigma}
\end{equation}

Here, \( \sigma_{\text{base}} \) defines the minimum kernel width. The normalization ensures that \( \sigma_s = \sigma_{\text{base}} \) at the bin location of minimum uncertainty.

\paragraph{Choosing the Base Kernel Width.}
The base kernel width $\sigma_{\text{base}}$ acts as a minimum smoothing scale. It controls the amount of smoothing in the most confident regions of the force profile. Intuitively,
\begin{itemize}
    \item If $\sigma_{\text{base}}$ is too large, even bins with low uncertainty will be excessively smoothed, potentially blurring important physical features such as force peaks.
    \item If $\sigma_{\text{base}}$ is too small, the overall smoothing may be insufficient in moderate- and high-uncertainty regions.
\end{itemize}

This trade-off represents the classical bias–variance dilemma: larger values reduce variance in the force estimate but may introduce bias by over-smoothing physically relevant structure. Conversely, smaller values preserve signal details but allow more noise to persist. To balance this trade-off, we adopt an L-curve-based strategy \cite{hansenAnalysisDiscreteIllPosed1992} that characterizes the trade-off between data fidelity and smoothness. For each candidate $\sigma_{\text{base}}$ value, we apply AGD and evaluate two competing quantities:

\begin{itemize}
    \item \textbf{Weighted chi-squared residual:} a statistical measure that quantifies how well the smoothed force profile $\tilde{f}_{\mathrm{smooth}}$ fits the original noisy data $\tilde{f}(x)$, while explicitly accounting for correlated uncertainty. Let $\mathbf{r} = \tilde{{f}} - \tilde{{f}}_{\mathrm{smooth}}$ denote the residual vector between the noisy and smoothed force profiles, and let $\Sigma$ be the estimated covariance matrix of the noise. The residual is defined as:
    \begin{equation}
        \chi^2 = {r}^\top \left( \Sigma + \epsilon I \right)^{-1} {r}
        \label{eq:weighted_residual}
    \end{equation}
    where $\epsilon = 1 \times 10^{-8}$ is a small constant added to the diagonal for numerical stability. A lower $\chi^2$ indicates closer fidelity to the original force data, implying less smoothing.

    \item \textbf{Roughness:} the integrated squared gradient of $\tilde{f}_{\mathrm{smooth}}$, i.e., $\sum_i \left( \frac{d\tilde{f}_{\mathrm{smooth}}}{dx} \right)^2$, which quantifies the total smoothness of the profile.
\end{itemize}

These competing objectives define a parametric curve in log--log space, known as the L-curve. The base kernel width is selected as the $\sigma_{\text{base}}$ value corresponding to the point of maximum curvature, which identifies a favorable balance between fidelity and smoothness.

 \paragraph{Curvature-aware reweighting for Bias Correction.}

It is desired to mitigate the bias introduced by kernel smoothing in high curvature regions, particularly near walls where molecular ordering induces sharp force oscillations. The sampling uncertainty in these regions is smaller compared to regions farther from the confining walls, where molecules exhibit more disordered motion. However, small discrepancies in these high-curvature regions can propagate into significant deviations in the resulting density predictions. A variety of bias mitigation strategies have been proposed, including two-step kernel estimators which aim to improve the estimation accuracy without significantly increasing the variance.

In the two-step kernel estimator proposed in~\cite{hengartnerAsymptoticUnbiasedDensity2009}, bias is reduced by first computing an initial (pilot) estimate via conventional kernel smoothing, followed by a multiplicative bias correction using a ratio-based reweighting scheme. Building on their approach, we construct the initial estimate using the uncertainty-based bandwidths (Eq.~\eqref{eq:adaptive_sigma}). In our discrete setting, the convolution operation becomes a weighted summation over noisy force values.
\begin{equation}
\tilde{f}_{\text{pilot}}(x_i) = \frac{1}{Z_i^{\text{pilot}}} \sum_{j=1}^{n} K_{\sigma_i^s}(x_j - x_i) \cdot \tilde{f}(x_j),
\end{equation}
where \( K_{\sigma_i^s} \) is a Gaussian kernel with adaptive bandwidth \( \sigma_i^s \) centered at bin \( x_i \),\( n \) is the total number of bins in the domain, and \( Z_i^{\text{pilot}} = \sum_{j=1}^n K_{\sigma_i^s}(x_j - x_i) \) ensures normalization. Following this, we compute the multiplicative correction factor \( \hat{\alpha}(x_i) \) using the same formulation as in~\cite{hengartnerAsymptoticUnbiasedDensity2009}:
\begin{equation}
\hat{\alpha}(x_i) = \frac{1}{Z_i^{\text{corr}}} \sum_{j=1}^{n} K_{\sigma_c^i}(x_j - x_i) \cdot \frac{1}{\tilde{f}_{\text{pilot}}(x_j)},
\end{equation}
where \( \sigma_c^i \) denotes the bandwidth used in the second kernel pass and \( Z_i^{\text{corr}} = \sum_{j=1}^n K_{\sigma_i^c}(x_j - x_i) \) is the normalization constant. The bias-corrected estimate $\tilde{f}_{\text{smooth}}$ is then obtained by multiplying the pilot by the correction factor:
\begin{equation}
\tilde{f}_{\text{smooth}}(x_i) = \frac{1}{Z_i^{\text{corr}}} \sum_{j=1}^{n} K_{\sigma_c^i}(x_j - x_i) \cdot \frac{\tilde{f}_{\text{pilot}}(x_i)}{\tilde{f}_{\text{pilot}}(x_j)}.
\end{equation}

Here, \( \sigma_c^i = \sigma_s^i \cdot \eta(x_i) \) is a curvature-reweighted bandwidth. The procedure for obtaining the curvature-induced weights is described below:

We define the curvature at bin \( x_i \) as \( \kappa(x_i) = \left. \frac{d^2 \tilde{f}^{\mathrm{pilot}}}{dx^2} \right|_{x = x_i} \), where \( \tilde{f}^{\mathrm{pilot}}(x_i) \) is the pilot estimate obtained via initial uncertainty-adaptive smoothing of the raw force profile. The reason for estimating the curvature from the pilot-smoothed force profile instead of the raw force data is that numerical differentiation behaves like a high-pass filter and amplifies high-frequency noise present in the original signal. To further suppress this amplified noise and reduce the effect of outliers, we apply a geometric mean filter over a window centered at each bin. Specifically, for each bin \( x_i \), we collect the curvature values within a local window \( \mathcal{W}_i \) of size equal to the correlation length \( \ell_c \) and compute:
\begin{equation}
    \bar{\kappa}(x_i) = \exp\left( \frac{1}{|\mathcal{W}_i|} \sum_{x_j \in \mathcal{W}_i} \log \left( \kappa(x_j) + \epsilon \right) \right),
\end{equation}
where \( \epsilon \) is a small constant to prevent numerical instability when \( \kappa(x_j) \) is close to zero.

Finally, the smoothed curvature \( \bar{\kappa}(x_i) \) is mapped to a bandwidth modulation factor using a nonlinear transformation \( \Phi \), defined to decrease with increasing curvature:
\begin{equation}
    \eta(x_i) = \Phi\left( \bar{\kappa}(x_i) \right),
\end{equation}
where \( \Phi : \mathbb{R}_{\geq 0} \rightarrow [\eta_{\min}, \eta_{\max}] \) is a monotonic function—such as an inverted sigmoid—designed to assign narrower bandwidths in high-curvature regions and broader ones in flatter regions.We define:
\begin{equation}
    \Phi(\kappa) = \eta_{\min} + (\eta_{\max} - \eta_{\min}) \cdot
    \left[ \frac{1}{1 + \exp\left( \alpha (\kappa - \kappa_{\text{flat}}) \right)} \right],
\end{equation}
where \( \alpha > 0 \) controls the steepness of the transition, and \( \kappa_{\text{flat}} \) defines the curvature threshold below which the signal is considered locally flat. We set \( \eta_{\min} = 0.1 \), equal to the bin spacing, and \( \eta_{\max} = 1 \), implying that the lowest-curvature regions are assigned full bandwidth (i.e., \( \sigma_c^i = \sigma_s^i \)). The steepness parameter is set to \( \alpha = 10 \), and the flatness threshold is chosen as \( \kappa_{\text{flat}} = 10^{-3} \).

\paragraph{Calculating Uncertainty in Denoised Forces.} The uncertainty in the smoothed force is computed via uncertainty propagation for weighted sums \cite{kleinUncertaintyPropagationGaussian2021}, explicitly accounting for spatial correlations across bins:

\begin{equation}
\bar{\sigma}_{\tilde{f}_{\mathrm{smooth}}}(x_i) = \sqrt{ \mathbf{w}_i^\top \, \Sigma \, \mathbf{w}_i }
\label{eq:uncertainty_with_covariance}
\end{equation}

Here, \( \mathbf{w}_i = [w_{i1}, w_{i2}, \dots, w_{iN}]^\top \) is the weight vector used to smooth at location \( x_i \),  
\( \Sigma \in \mathbb{R}^{N \times N} \) is the covariance matrix of the block-averaged force estimates \( \tilde{f}(x_j) \),  
and $\bar{\sigma}_{\tilde{f}_{\mathrm{smooth}}}(x_i)$ denotes the total propagated uncertainty in the smoothed force at location \( x_i \). The covariance matrix \( \Sigma \) is estimated from the block-averaged force values as:
\begin{equation}
\Sigma_{jk} = \frac{1}{n_B - 1} \sum_{\ell=1}^{n_B} 
\left( \langle \tilde{f} \rangle_\ell^B(x_j) - \langle \tilde{f} \rangle^B(x_j) \right)
\left( \langle \tilde{f} \rangle_\ell^B(x_k) - \langle \tilde{f} \rangle^B(x_k) \right)
\label{eq:covariance_blocks}
\end{equation}

To facilitate numerical stability and computational efficiency, we assume zero covariance between bins separated by a distance larger than the spatial correlation length \( \ell_c \), i.e., if \( \Delta x = |x_j - x_k| > \ell_c \). The correlation length \( \ell_c \) is estimated from the spatial autocorrelation function of the block-averaged force profile \( \tilde{f}(x) \), defined as:
\begin{equation}
C(\Delta x) = \frac{ \sum_{i} \left( \tilde{f}(x_i) - \bar{f} \right)\left( \tilde{f}(x_i + \Delta x) - \bar{f} \right) }{ \sum_{i} \left( \tilde{f}(x_i) - \bar{f} \right)^2 }
\label{eq:spatial_acf}
\end{equation}
where \( \bar{f} \) is the spatial mean of the force profile. We define the correlation length \( \ell_c \) as the first zero-crossing of the normalized autocorrelation function.

\paragraph{Propagating Force Uncertainty into Density Predictions.}

 To assess how uncertainty in the smoothed force propagates into the predicted density, we perform Monte Carlo sampling using the per-bin uncertainty $\sigma_{\hat{f}}(x_i)$ and the full spatial covariance matrix $\Sigma_{\hat{f}}$ of the smoothed force. Specifically, we draw $N$ samples from the multivariate normal distribution:
\[
\hat{f}^{(k)}(x) \sim \mathcal{N}(\tilde{f}_{\mathrm{smooth}}(x), \Sigma), \quad k = 1, \dots, N,
\]
and solve the Nernst–Planck equation for each sampled force profile to obtain a corresponding density profile $\rho^{(k)}(x)$. The empirical mean and standard deviation across these samples yield the predicted density $\bar{\rho}(x)$ and its uncertainty $\sigma_{\rho}(x)$:
\[
\bar{\rho}(x) = \frac{1}{N} \sum_{k=1}^{N} \rho^{(k)}(x), \quad
\sigma_{\rho}(x) = \sqrt{ \frac{1}{N - 1} \sum_{k=1}^{N} \left( \rho^{(k)}(x) - \bar{\rho}(x) \right)^2 }.
\]

It is important to note that this approach quantifies uncertainty due to thermal noise and finite sampling, as captured by the propagated variance in the force profile. However, it does not account for systematic bias introduced by the smoothing process itself, which may underestimate or distort physically relevant features in the force and, consequently, the density. As such, the reported uncertainty bounds reflect only the variance component of the total prediction error. Algorithm~\ref{alg:uq} describes this procedure.

\vspace{1em}

\begin{algorithm}[htbp]
\caption{Monte Carlo Uncertainty Quantification for Density Profile}
\label{alg:uq}
\begin{algorithmic}[0]
\Require Denoised force profile $\tilde{f}_{\mathrm{smooth}}(x)$, force uncertainties $\sigma_{\hat{f}}(x)$, number of samples $N$
\Ensure Mean density profile $\mu_\rho(x)$, uncertainty profile $\sigma_\rho(x)$
\For{$k = 1$ to $N$}
    \State Sample \( \hat{f}^{(k)}(x) \sim \mathcal{N}(\tilde{f}_{\mathrm{smooth}}(x), \Sigma) \)
    \State Obtain \( \rho^{(k)}(x) \) by solving Eq.~\eqref{eq:np} with \( \hat{f}^{(k)}(x) \)
\EndFor
\State \noindent\textbf{Compute:} 
\(\textstyle \mu_\rho(x) = \frac{1}{N} \sum_{k=1}^{N} \rho^{(k)}(x), \quad
\sigma_\rho(x) = \sqrt{ \frac{1}{N - 1} \sum_{k=1}^{N} \left( \rho^{(k)}(x) - \mu_\rho(x) \right)^2 }\)
\end{algorithmic}
\end{algorithm}
\subsubsection{Conditional Denoising Diffusion Probabilistic Model for Force Denoising}
\label{sec:ddpm_training}

Our goal is to predict average force inside a nano channel of a particular width given a noisy force profile corresponding to that channel width, obtained from limited AIMD data. This denoising task can be formulated as learning the underlying conditional probability distribution. Specifically, given a noisy force profile $\tilde{f}(x) \in \mathbb{R}^d$, we seek to learn the conditional probability distribution $q(f \mid \tilde{f})$, where $f(x) \in \mathbb{R}^d$ denotes the long-time averaged force profile and $d$ is the number of spatial bins across the channel. A sample, ${f}(x) \sim q(f \mid \tilde{f})$ from this distribution would then provide a denoised estimate of the smooth force profile using the noisy input as the conditioning. In this work we use DDPMs, that have shown remarkable performance on denoising tasks for a variety of applications. 

Denoising diffusion models \cite{pmlr-v139-nichol21a,hoDenoisingDiffusionProbabilistic2020c,sohl-dicksteinDeepUnsupervisedLearning2015} approximate complex data distributions by defining a two-stage generative process: a forward diffusion process that progressively perturbs the data by adding noise over a fixed number of steps, and a learned reverse process that reconstructs the data by iteratively denoising the corrupted inputs. This framework enables the model to sample from the target distribution by reversing the diffusion trajectory from pure noise to coherent data. More precisely, the forward process is defined as a fixed Markov Chain that iteratively adds Gaussian noise to clean data $f_0$ to approximate the conditional distribution $q(f_{1:T} \mid f_0)$, where $f_1, f_2, \dots, f_T$ are latent variables of the same dimension as $f_0$, according to a noise variance schedule $\beta_1, \beta_2, \dots, \beta_T$, such that the final state $f_T$ approaches an isotropic Gaussian:

\begin{equation}
q(f_{1:T} \mid f_0) \coloneqq \prod_{t=1}^T q(f_t \mid f_{t-1}) 
\end{equation}

The kernel $q(f_t \mid f_{t-1})$ can be written as:

\begin{equation}
q(f_t \mid f_{t-1}) \coloneqq \mathcal{N}(\sqrt{1 - \beta_t} \, f_{t-1}, \beta_t \, \mathbf{I}) 
\end{equation}

Here, the conditional distribution is Gaussian, with a mean given by a scaled version of the previous state. This formulation ensures that noise is gradually added to the data in a tractable manner over successive steps. Moreover, one can directly compute and sample $f_t$ at an arbitrary step $t$ as:

\begin{equation}
f_t \coloneqq \sqrt{\bar{\alpha}_t} \, f_0 + \sqrt{1 - \bar{\alpha}_t} \, \epsilon, \quad \epsilon \sim \mathcal{N}(0, \mathbf{I}) 
\end{equation}

where $\alpha_t = 1 - \beta_t$ and $\bar{\alpha}_t = \prod_{i=1}^t \alpha_i$.

The reverse process aims to recover $f_0$ starting from $f_T \sim \mathcal{N}(0, \mathbf{I})$ by a learnable Markov Chain such that:

\begin{equation}
\label{eq:ogreverse}
p_\theta(f_{0:T}) \coloneqq p(f_T) \prod_{t=1}^T p_\theta(f_{t-1} \mid f_t)
\end{equation}

Here the reverse kernel $p_\theta(f_{t-1} \mid f_t)$ can be written as:

\begin{equation}
p_\theta(f_{t-1} \mid f_t) \coloneqq \mathcal{N}(\mu_\theta(f_t, t), \sigma_t^2 \, \mathbf{I}) 
\end{equation}

The learnable parameter $\mu_\theta$ can be rewritten in terms of the model’s approximation of the noise $\epsilon_\theta$, which is added at each step, as:

\begin{equation}
\mu_\theta(f_t, t) \coloneqq \frac{1}{\sqrt{\alpha_t}} \left( f_t - \frac{\beta_t}{\sqrt{1 - \bar{\alpha}_t}} \, \epsilon_\theta(f_t, t) \right), \quad \sigma_t^2 = \beta_t 
\end{equation}

The training objective is thus to learn $\epsilon_\theta(f_t, t)$, which approximates the noise $\epsilon$ added during the forward process using a neural network. The network can then be trained by minimizing the loss function:

\begin{equation}
\mathcal{L}(\theta) \coloneqq \mathbb{E}_{f_0, \epsilon, t} \left[ \left\| \epsilon - \epsilon_\theta(f_t, t) \right\|^2 \right] 
\end{equation}

\begin{equation}
= \mathbb{E}_{f_0, \epsilon, t} \left[ \left\| \epsilon - \epsilon_\theta\left( \sqrt{\bar{\alpha}_t} f_0 + \sqrt{1 - \bar{\alpha}_t} \, \epsilon, t \right) \right\|^2 \right] 
\end{equation}

The expectation operator represents the average value of the squared noise error averaged over the three random variables $t$, $f_0$ and $\epsilon$, which are sampled during training.

\paragraph{Conditioning on forces}

Inspired by \cite{nadkarniMolecularDenoisingUsing2025,sahariaImageSuperResolutionIterative2022,liDeScoDECGDeepScoreBased2024}, we inform the denoising process by modifying the reverse process to take single-snapshot noisy forces $\hat{f}$ obtained from the MLMD simulation as conditioning information. This leads to an extended formulation of the reverse process that takes this conditioning procedure into account by modifying Eq.~\eqref{eq:ogreverse} as:

\begin{equation}
\label{eq:reverse}
p_\theta(f_{t-1} \mid f_t, \hat{f}) \coloneqq \mathcal{N}(\mu_\theta(f_t, t, \hat{f}), \sigma_t^2 \, \mathbf{I}) 
\end{equation}
\clearpage
Conditioning on force $p_\theta(f_{t-1} \mid f_t, \hat{f})$ aims to iteratively reconstruct the smooth force $f_0$ from a random Gaussian noise distribution, conditioned on the noisy observed force data. Thus, the denoising function $\epsilon_\theta$ takes an additional conditioning input and the loss function takes the form:

\begin{equation}
\mathcal{L}(\theta) \coloneqq \mathbb{E}_{f_0, \epsilon, t} \left[ \left\| \epsilon - \epsilon_\theta\left(\sqrt{\bar{\alpha}_t} f_0 + \sqrt{1 - \bar{\alpha}_t} \, \epsilon, \hat{f}, t \right) \right\|^2 \right]
\end{equation}

This framework enables conditional generation of smooth force profiles $f_0$, consistent with noisy single-frame observations $\hat{f}$ obtained from MLMD simulation data.
\begin{figure}[htbp]
    \centering
    \includegraphics[width=0.95\textwidth, trim=0 50 0 50, clip]{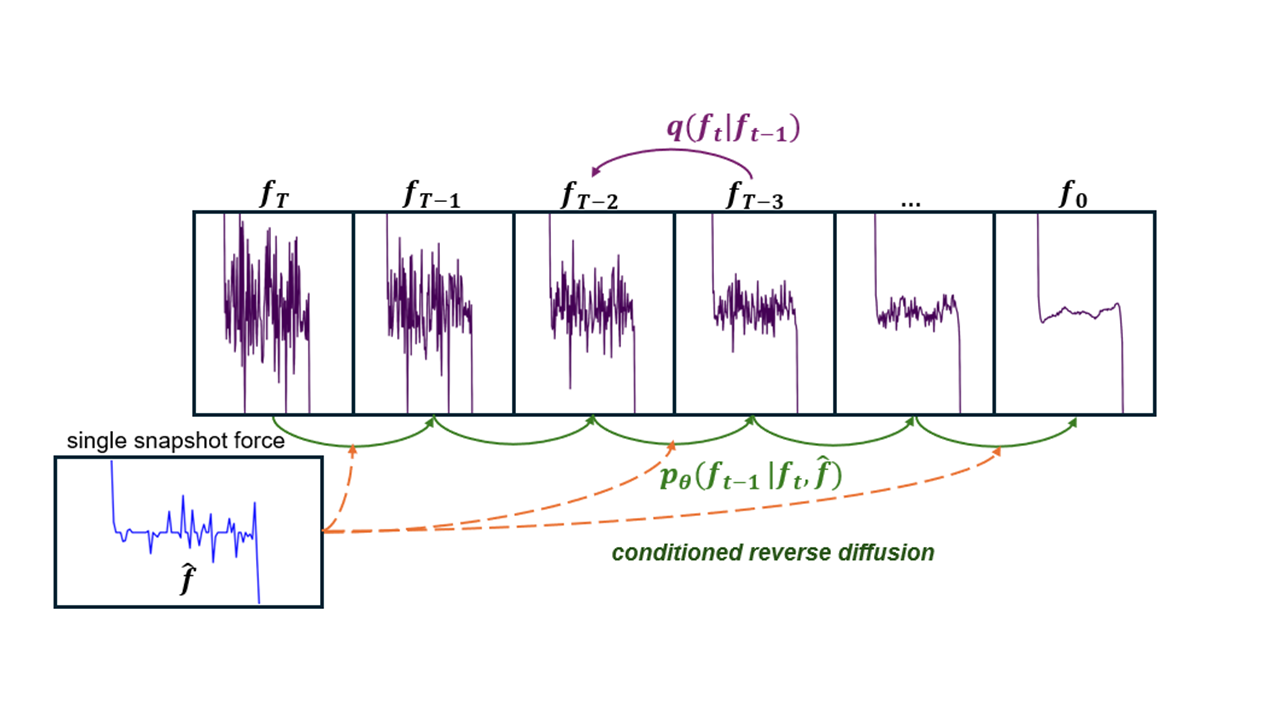}
    \caption{The forward diffusion ($q$) and the conditioned reverse diffusion ($p_\theta$) processes. The forward process corrupts smooth force profiles, and the reverse process aims to estimate and remove the noise added at each step. The reverse process is conditioned on single-snapshot noisy forces observed during the MLMD simulation. Once the model is trained, it is used to generate smooth force profiles from Gaussian noise, conditioned on noisy force profiles obtained from short AIMD trajectories.}
    \label{fig:reverse_diffusion}
\end{figure}
\subsubsection{Generative Quasi-Continuum Inference Algorithm}
\label{sec:GQCT}
Let the available dataset $\mathcal{D}$ consist of AIMD trajectories corresponding to $n$ different channel widths, denoted $\{L_1, L_2, \dots, L_n\}$. We seek to construct a mapping:

\begin{equation}
\phi: \mathbb{R} \times \mathcal{D} \rightarrow \mathbb{R}^d, \quad \text{such that} \quad \phi(L, \mathcal{D}) = \rho_L
\label{eq:phi_mapping}
\end{equation}

Here, $\phi$ maps a given channel width $L \in \mathbb{R}$—not necessarily in the set $\{L_1, L_2, \dots, L_n\}$—along with the trajectory dataset $\mathcal{D}$, to a corresponding density profile $\rho_L \in \mathbb{R}^d$, where $d$ denotes the number of spatial bins along the channel. We use a spatial binning resolution of 0.1~\AA.
\clearpage
In our implementation, $\phi$ is realized as a three-stage algorithm:

\begin{enumerate}
    \item \textbf{Identifying representative atomic neighborhoods} corresponding to continuum bins.
    \item \textbf{Inferring forces} on  the central water molecule of these neighborhoods and averaging to identify the average force in the corresponding bins.
    \item \textbf{Denoising the averaged force and solving} the Nernst–Planck equation to obtain the final density profile.
\end{enumerate}

\begin{figure}[htbp]
    \centering
    \includegraphics[width=0.95\textwidth, trim=0 0 0 0, clip]{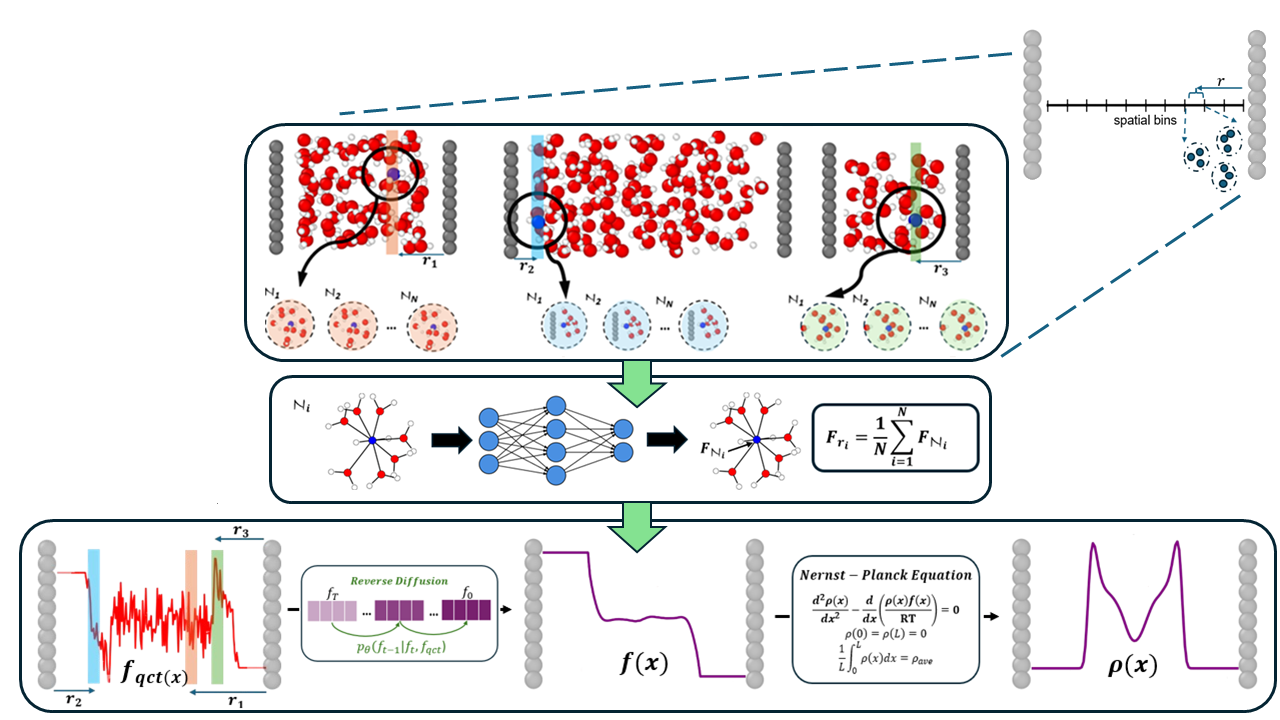}
    \caption{Quasi-continuum inference. The algorithm predicts the density profile of a channel for a confinement width specified by the user, using an AIMD dataset that contains trajectories for several channel widths. 
    \textbf{(top)} First, the channel dimension is spatially divided into bins (continuum points), and atomistic neighborhoods corresponding to these bins are identified. 
    \textbf{(middle)} These atomistic neighborhoods are fed into the neural network, and force on each continuum point is obtained by averaging the network's prediction of the central water molecule's force. 
    \textbf{(bottom)} The resulting noisy force profile is denoised by a reverse diffusion process; the smoothed force is then used in the Nernst–Planck equation to yield the density profile.}
    \label{fig:qct_inference}
\end{figure}

\textbf{Identifying Representative Neighborhoods.}  
Following \cite{wuInitioContinuumLinking2023a}, we divide the nanochannel of width $L$ into $N$ spatial bins of width $\Delta x = L/N$, where $x_{\text{min}} = 0$ and $x_{\text{max}} = L$. Each bin is defined as $\text{Bin}_i = [x_{\text{min}} + i \Delta x,\; x_{\text{min}} + (i + 1) \Delta x)$ for $i = 0, 1, \dots, N-1$, and the bin center is given by $x_i^{\text{center}} = x_{\text{min}} + (i + \tfrac{1}{2}) \Delta x$. Each bin defines a continuum point, and to approximate the force at a position $r$ from the wall, we identify the atomic neighborhoods $\{\mathcal{N}_i\}_{i=1}^{N_r}$ around water molecules that satisfy the relation $|x_{\mathrm{COM}} - r| < \Delta x$. Here, \( x_{\mathrm{COM}} \) denotes the x-coordinate of the center of mass of a water molecule. Each neighborhood includes all atoms within a 6~\AA~cutoff radius, consistent with the cutoff used during NNP training.
\clearpage

\textbf{Obtaining Force Values on the Continuum Points.}  
The force at a continuum point that is located $r$ away from the nearest graphene wall is computed by averaging the predicted molecular forces across the representative neighborhoods, i.e., $F_r = \frac{1}{\mathcal{N}_r} \sum_{i=1}^{\mathcal{N}_r} F(\mathcal{N}_i)
$. The force contribution from each neighborhood \( \mathcal{N}_i \) is defined as the center-of-mass force of the central water molecule, obtained by summing the NNP-predicted forces on its atoms: \( F(\mathcal{N}_i) = F_{H_1} + F_{H_2} + F_O = \text{NNP}(R_i, Z_i)\big|_{H_\text{center}, O_\text{center}} \), where \( R_i \) and \( Z_i \) are the atomic positions and atomic numbers within the neighborhood \( \mathcal{N}_i \), respectively.
To ensure density symmetry around the center of the channel, we enforce antisymmetry on the force $\bar{F}_r = \tfrac{1}{2} (F_r - F_{L - r})$, which satisfies $\bar{F}_{L - r} = -\bar{F}_r$. We refer to the resulting force profile corresponding to a channel width $L$ along the confinement direction (–$x$ direction) as the quasi-continuum force profile, denoted $f_{\text{qct}}^L(x) \in \mathbb{R}^d .$

\textbf{Density Prediction.}  
In practice, due to the limited number of AIMD configurations available for a given $L$, the estimated profile $f_{\text{qct}}^L(x)$ is inherently noisy. We denote this noisy estimate as \(\tilde{f}_{\text{qct}}^L(x)\).
Before it can be used to predict the density, $\tilde{f}_{\text{qct}}^L$ must be denoised. To do this, we invoke the conditional Denoising Diffusion Probabilistic Model (DDPM) described in Section~\ref{sec:ddpm_training}, which generates the smoothed force profile $f_{\text{qct}}^L(x)$ via the reverse generation process (Eq.~\ref{eq:reverse}), conditioned on $\tilde{f}_{\text{qct}}^L(x)$. Once the smooth force profile $f_{\text{qct}}^L(x)$ is obtained, it is fed into the Nernst–Planck equation (Eq.~\ref{eq:np}) to obtain the final density profile $\rho_L(x)$ across the channel. The full procedure for quasi-continuum inference is summarized in Algorithm~\ref{alg:qct}.
\begin{algorithm}[H]
\caption{Generative Quasi-Continuum Inference Algorithm}
\label{alg:qct}
\begin{algorithmic}[0]
\Require Channel width $L$, dataset $\mathcal{D}$ of AIMD trajectories, trained NNP, trained DDPM
\Ensure Density profile $\rho_L(x)$

\State Divide the channel domain $[0, L]$ into $d$ spatial bins of width $\Delta x$
\For{each bin center $x_i$}
    \State Identify atomistic neighborhoods $\mathcal{N}_i$ centered at distance $x_i$ from confining wall in $\mathcal{D}$
    \State Compute center-of-mass force on the central water molecule $F(\mathcal{N}_i)$ using the NNP
\EndFor
\State Construct noisy force profile: $\tilde{f}_{\text{qct}}^L(x)$ 

\State Generative Denoising: $\tilde{f}_{\text{qct}}^L(x) \xrightarrow{\text{DDPM}} f_{\text{qct}}^L(x)$
\State Solve NP equation (Eq.~\ref{eq:np}) using $f_{\text{qct}}^L(x)$ as input
\State \Return $\rho_L(x)$
\end{algorithmic}
\end{algorithm}

\section{Results}

\subsection{Assessment of the Local Force Approximation for Density Prediction}
\label{sec:LFA}
In this section, we assess the performance of the LFA strategy described in Section~2. This approach serves as a tool to reduce the computational cost associated with running MLMD simulations. The LFA strategy exploits the observation that density profiles, our primary quantity of interest, converge significantly faster than force profiles. While density reflects the time-averaged positional distribution of molecules, force measurements are much more sensitive to instantaneous atomic configurations and short-range fluctuations (see Figures Appendix~\ref{fig:lfa_convergence} and~\ref{fig:lfa_convergence2}). To quantitatively assess the performance of the method, 1.5~ns-long MLMD simulations were performed, and the resulting density and force profiles are used as ground truth. By “ground truth,” we imply that the uncertainty in the density is assumed to be negligible due to long-time averaging.
\clearpage

For each MLMD simulation performed across channels of different confinement widths, we first determine the minimum trajectory length required to meet a predefined convergence criterion for the density profile (see Section~\ref{sec:lfa_theory}). Once this condition is met, the corresponding force profile is extracted, and a density profile, along with its associated uncertainty, is computed by solving the Nernst–Planck equation (Eq.~\ref{eq:np}). Subsequently, the LFA denoising procedure is applied to the force profile. Using the denoised force as input, the density profile and its uncertainty are recomputed. 

\paragraph*{Evaluation metrics.} To evaluate the accuracy of predicted density profiles, we report two complementary metrics: the mean integrated squared error (MISE) and the relative root mean squared error ($\epsilon_\rho$). The MISE is a standard measure of global accuracy in nonparametric estimation. In our setting, it is defined as the expected integrated squared difference between the predicted and ground truth density profiles, i.e., $\text{MISE}_\rho = \mathbb{E} \left[ \int \left( \rho^{(i)}(x) - \rho_{\text{GT}}(x) \right)^2 \, dx \right]$, where $\rho^{(i)}(x)$ denotes the density profile obtained from the $i$-th Monte Carlo sample obtained from a denoised force realization, and $\rho_{\text{GT}}(x)$ is the ground truth density profile. We also report the normalized root mean squared error ($\epsilon_\rho$), which provides a pointwise measure of deviation, defined as $\epsilon_\rho = \text{RMSE}(\bar{\rho}, \rho_{\text{GT}})/\max_x \rho_{\text{GT}}(x)$.

\begin{figure}[H]

  \centering

  \begin{subfigure}[t]{0.3\textwidth}
    \centering
    \includegraphics[width=\linewidth]{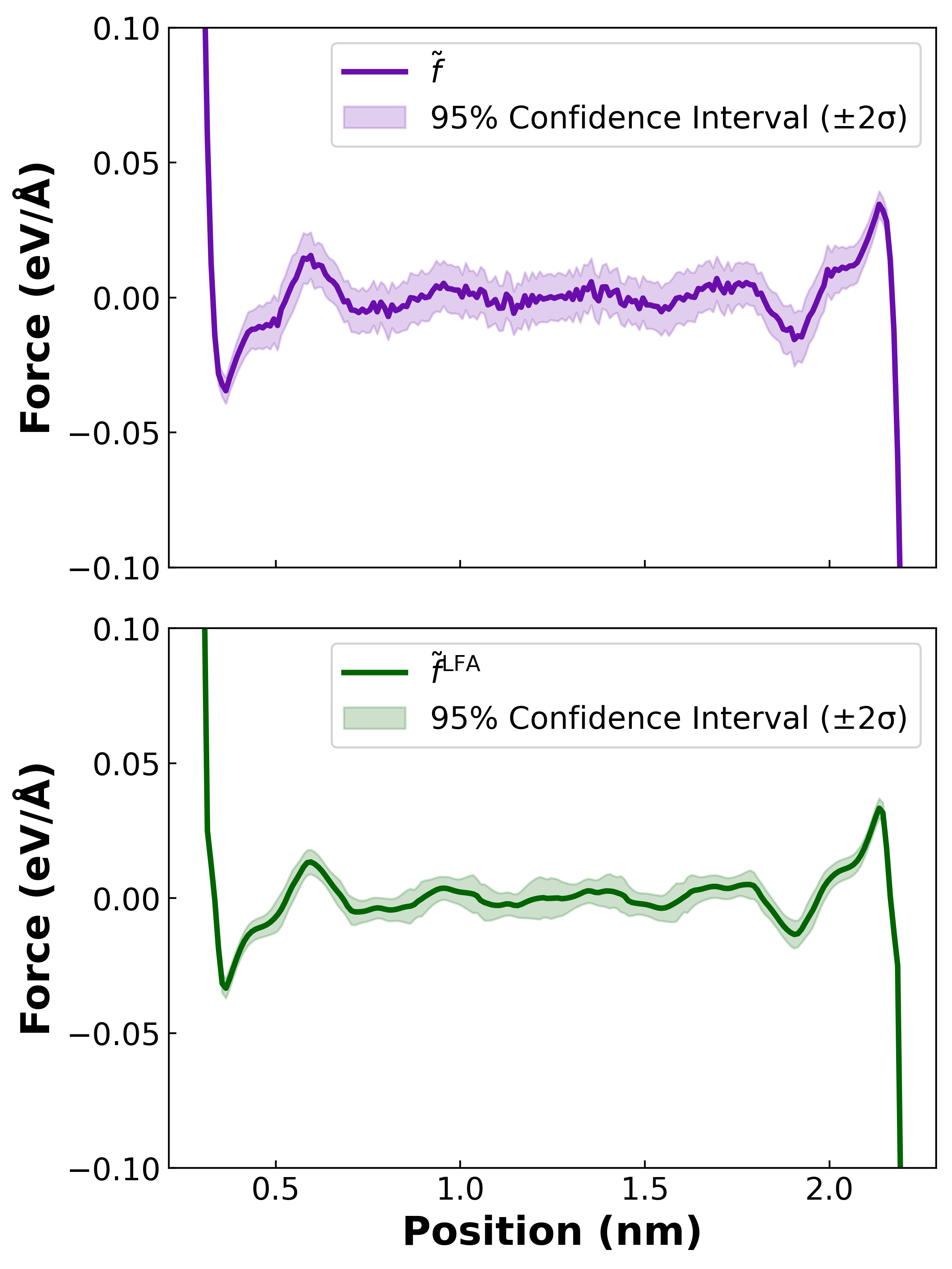}\\[-0.5ex]
    \includegraphics[width=\linewidth]{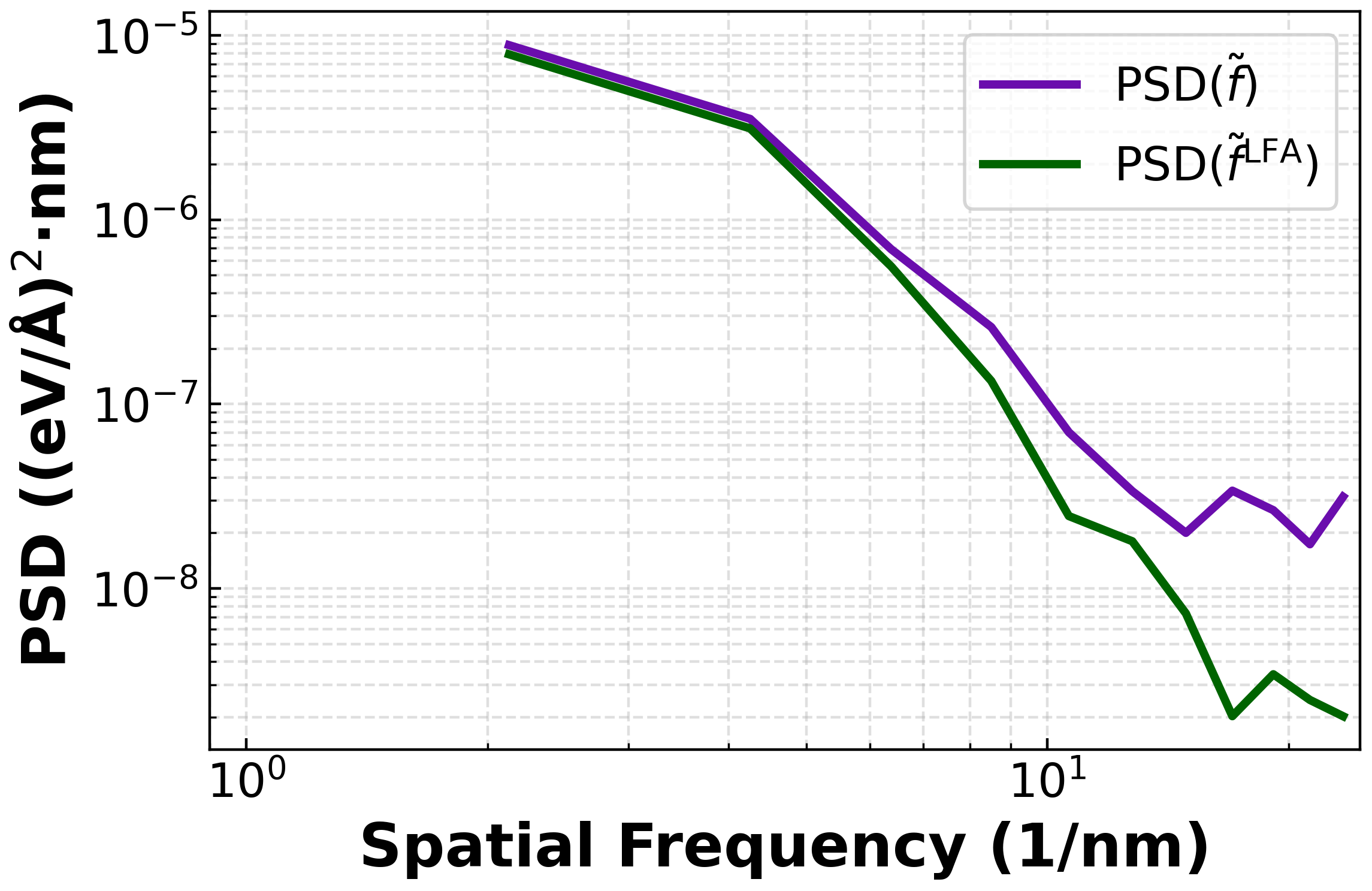}
    \caption*{(a)}
  \end{subfigure}
  \hfill
  \begin{subfigure}[t]{0.3\textwidth}
    \centering
    \includegraphics[width=\linewidth]{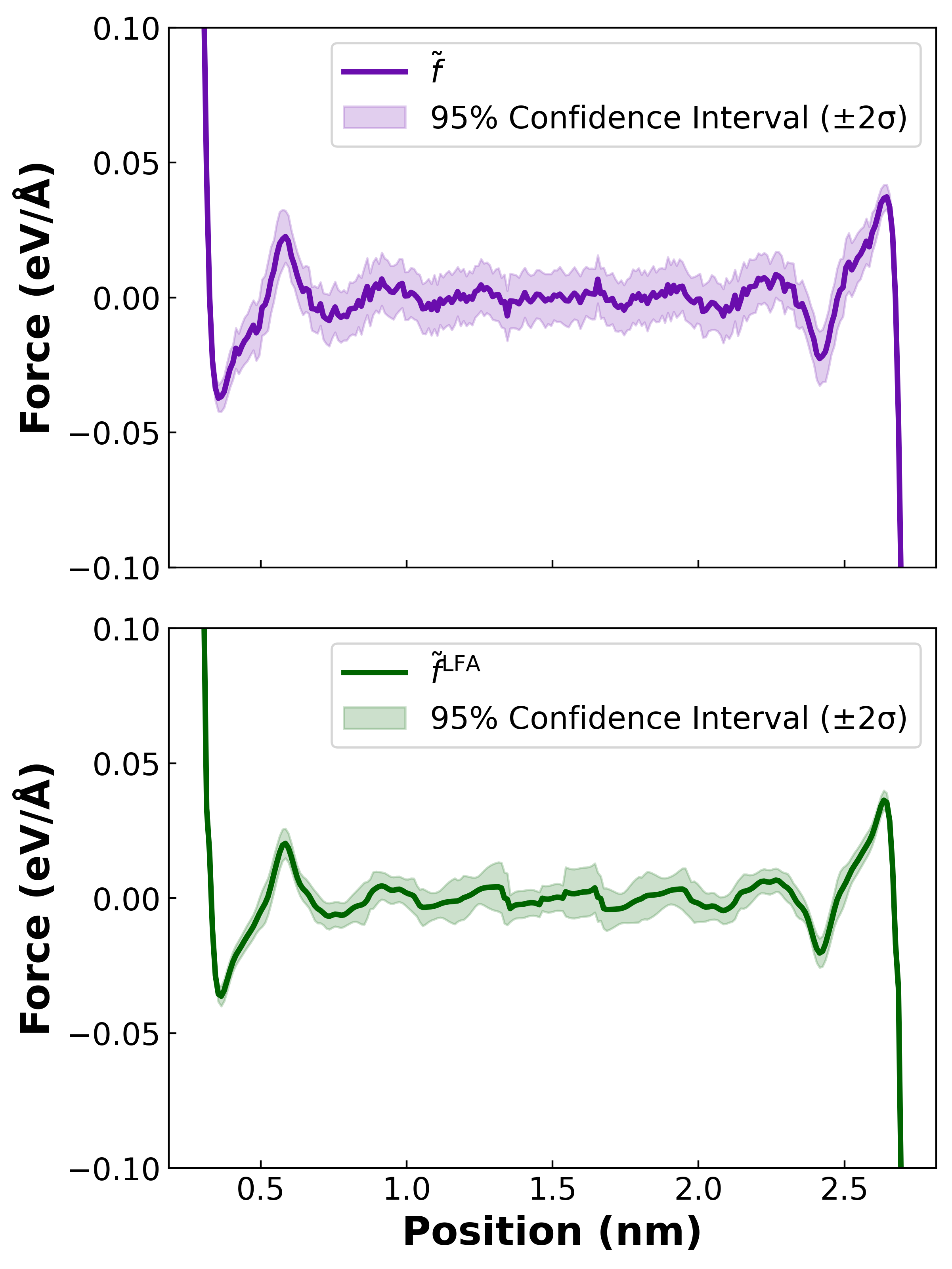}\\[-0.5ex]
    \includegraphics[width=\linewidth]{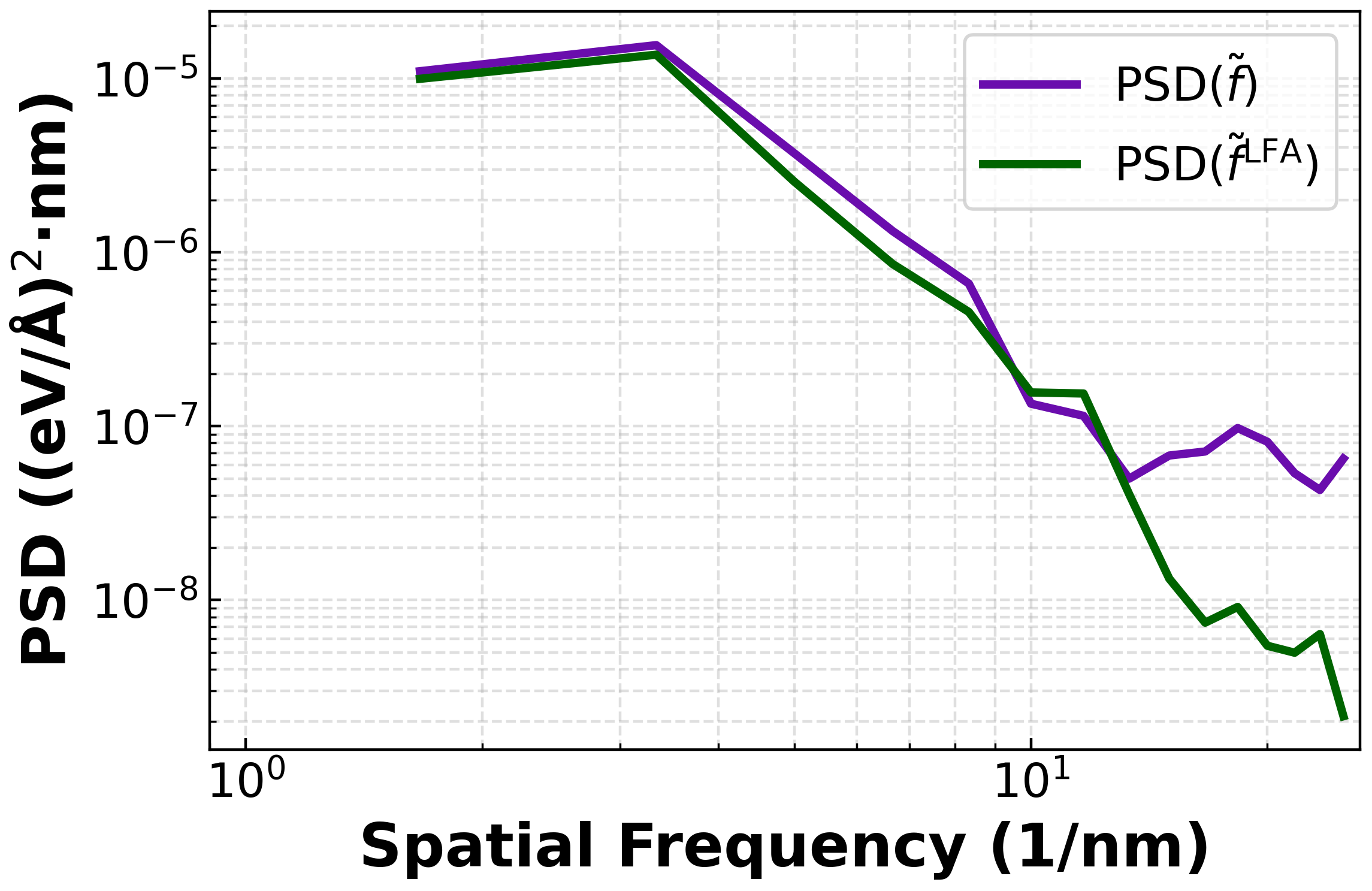}
    \caption*{(b)}
  \end{subfigure}
  \hfill
  \begin{subfigure}[t]{0.3\textwidth}
    \centering
    \includegraphics[width=\linewidth]{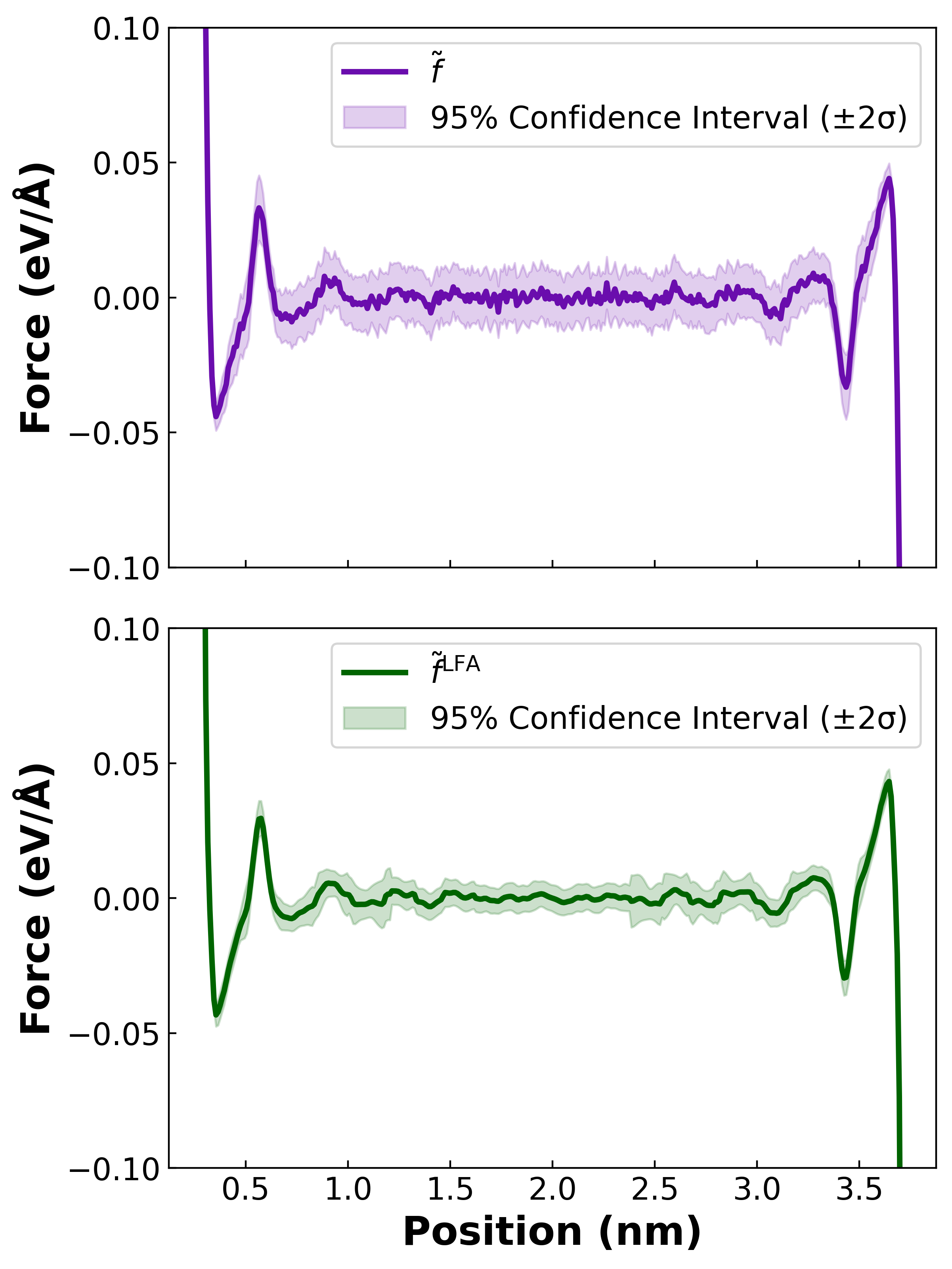}\\[-0.5ex]
    \includegraphics[width=\linewidth]{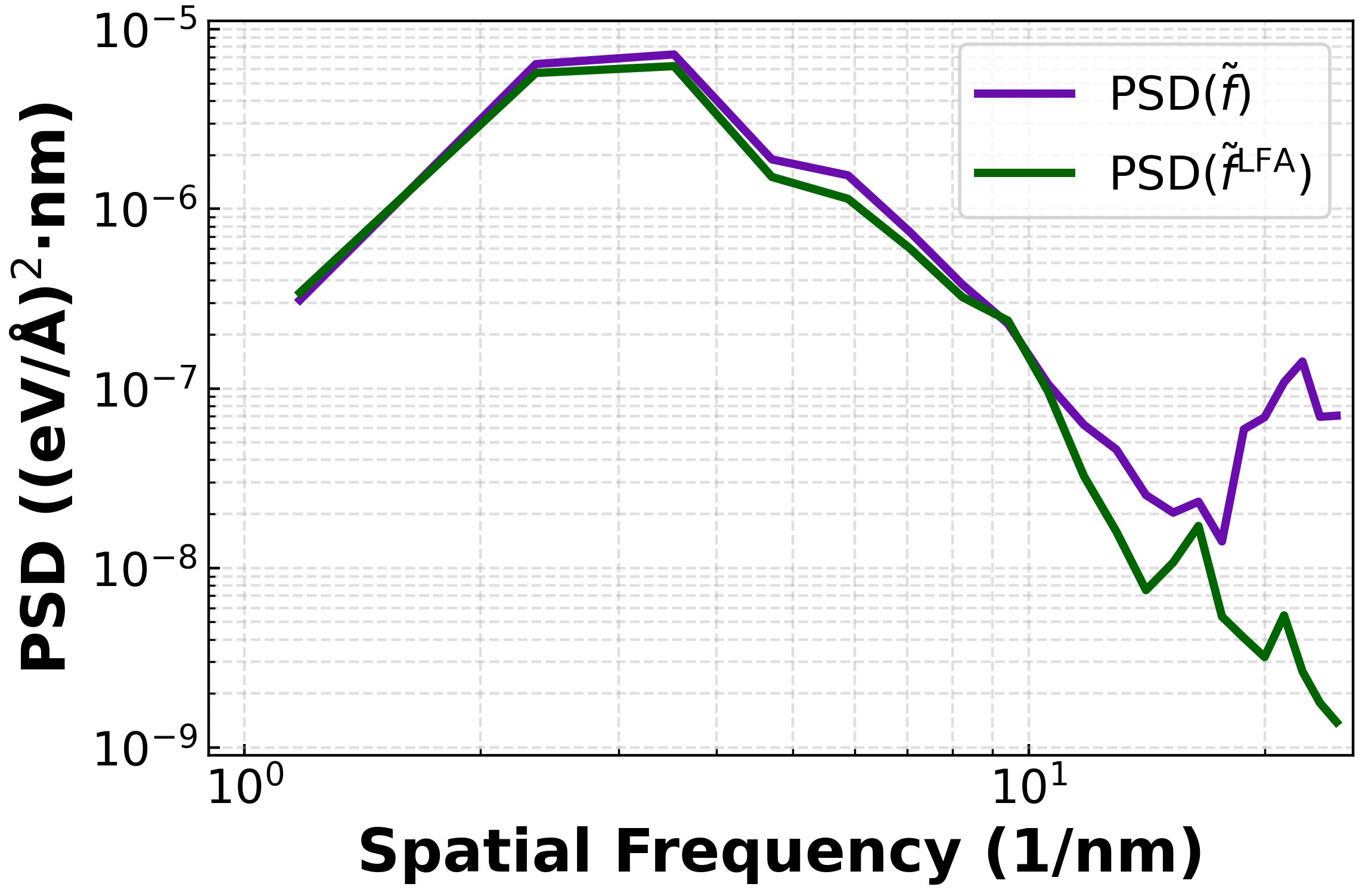}
    \caption*{(c)}
  \end{subfigure}

  \caption{
    Comparison of denoised force profiles and corresponding power spectral densities (PSDs) for water confined in graphene channels of varying widths. Each column represents a different channel width: 2.5~nm (left), 3.0~nm (center), and 4.0~nm (right). The top row shows MLMD-predicted force profiles \( \tilde{f} \) obtained by averaging simulation frames up to the point where the density profiles have converged. The middle row displays the same force profiles after denoising via the LFA, denoted \( \tilde{f}^{\mathrm{LFA}} \). The bottom row compares the spatial power spectral densities of \( \tilde{f} \) and \( \tilde{f}^{\mathrm{LFA}} \).
  }
  \label{fig:force_psd_columns}
\end{figure}
\begin{figure}[H] % Consider using [htbp] for better float placement
  \centering

  \begin{subfigure}[t]{0.325\textwidth} % Increased from 0.3
    \centering
    \includegraphics[width=\linewidth]{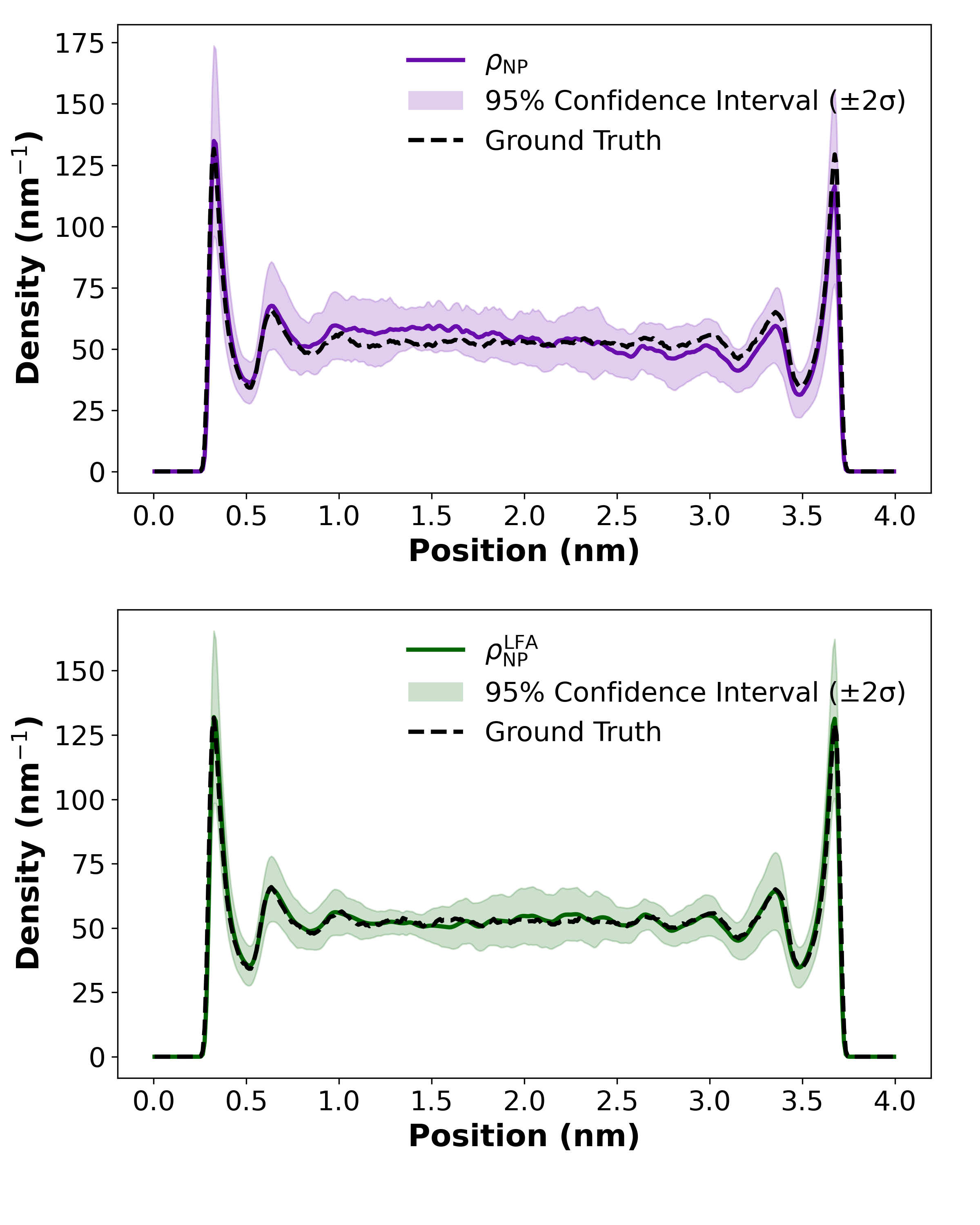}
    \caption*{(a)}
  \end{subfigure}
  \hfill
  \begin{subfigure}[t]{0.325\textwidth} % Increased from 0.3
    \centering
    \includegraphics[width=\linewidth]{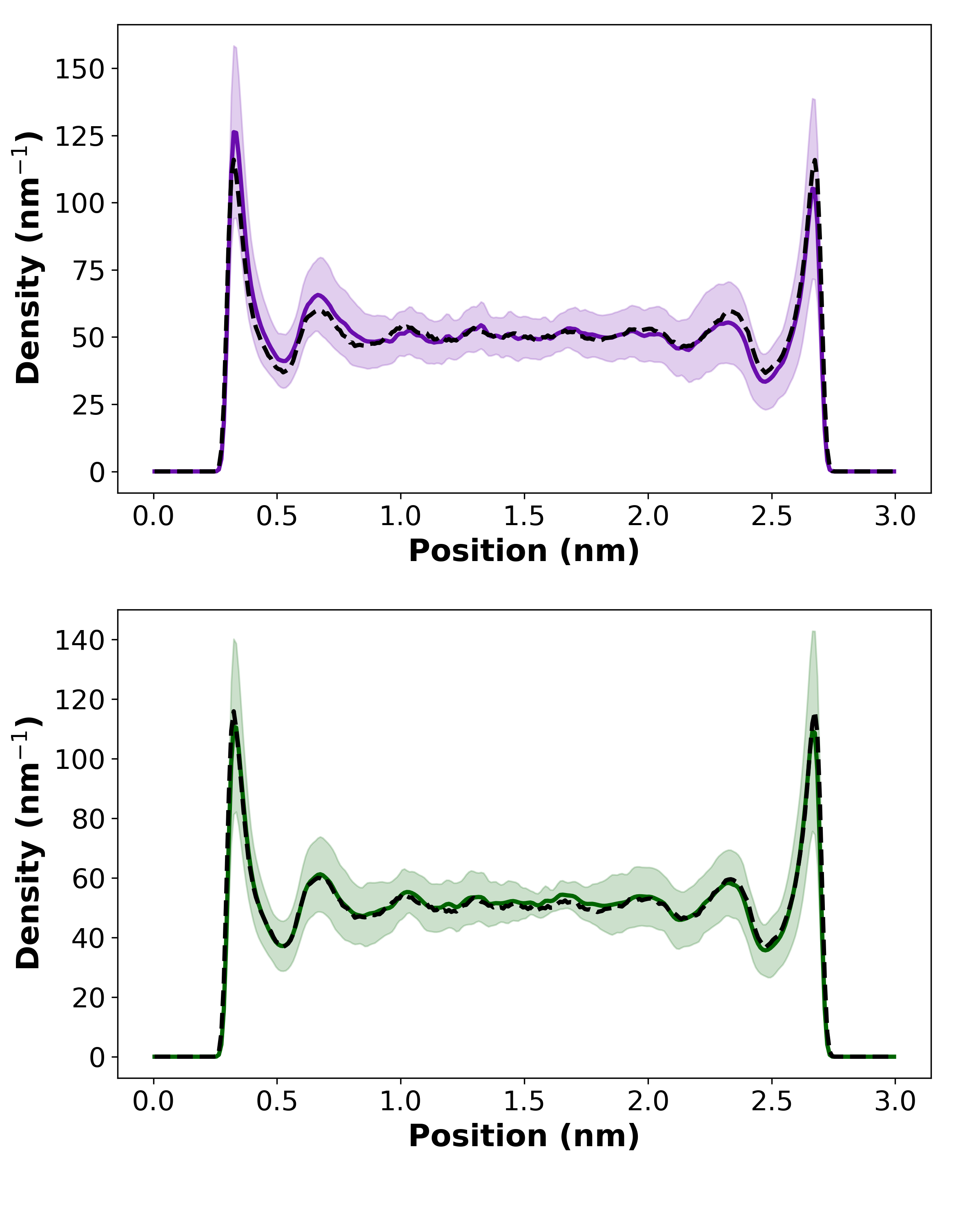}
    \caption*{(b)}
  \end{subfigure}
  \hfill
  \begin{subfigure}[t]{0.325\textwidth} % Increased from 0.3
    \centering
    \includegraphics[width=\linewidth]{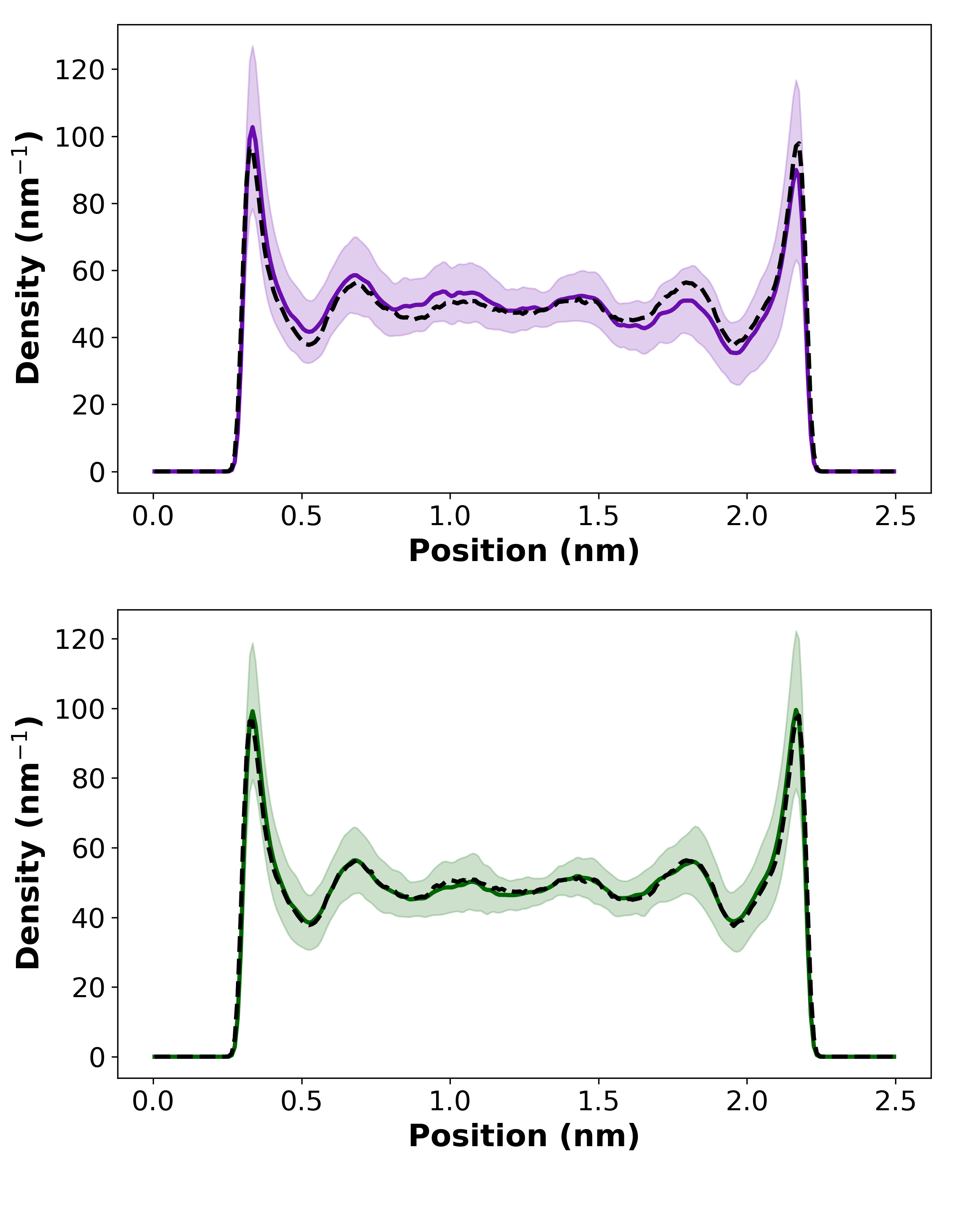}
    \caption*{(c)}
  \end{subfigure}

  \caption{
    Comparison of MLMD-predicted and smoothed density profiles for water confined in graphene channels of varying widths. Each column corresponds to a different channel width: 4~nm (left), 3.0~nm (center), and 2.5~nm (right).
  }
  \label{fig:density_comparison} 
\end{figure}

\clearpage
We observed that, at the point of apparent density convergence, the force profiles corresponding to the three widest channels, 2.5, 3 and 4 nm, produced density predictions that exceeded the desired accuracy threshold $\tau_\rho$. In Figure~\ref{fig:force_psd_columns}, we present the force profiles for the three widest confinement conditions, before applying the Local Force Approximation (LFA), denoted as $\tilde{f}$, and after denoising, denoted as $\tilde{f}^{\mathrm{LFA}}$. The corresponding power spectral densities (PSDs) are also shown. At low spatial frequencies, the PSD of $\tilde{f}^{\mathrm{LFA}}$ closely follows that of the original $\tilde{f}$, indicating that LFA preserves the long-wavelength features of the force profile. In contrast, at higher frequencies, the PSD is significantly attenuated, demonstrating that LFA effectively suppresses high-frequency noise while retaining physically relevant structure. Figure~\ref{fig:density_comparison} illustrates the predicted density profiles resulting from the original force fields $\tilde{f}$ and their denoised counterparts $\tilde{f}^{\mathrm{LFA}}$. In each subplot, the predicted density is shown alongside the ground truth profile with associated confidence intervals. The errors associated with these predictions are $\epsilon_\rho = 4.26$, $1.63$, and $3.70$ for the 2.5, 3.0, and 4.0~nm channels, respectively. After applying LFA, the errors are reduced to $\epsilon_\rho^{\mathrm{LFA}} = 0.95$, $0.76$, and $0.70$, all of which satisfy the accuracy criterion for density predictions. It is evident that directly integrating $\tilde{f}$ can lead to both under- and over-prediction of density, particularly in the density layering peaks (Figure~\ref{fig:density_comparison}a--c) and in bulk-like regions (notably in Figure~\ref{fig:density_comparison}a). The LFA procedure effectively corrects these deviations, producing profiles that closely match the ground truth.

\vspace{1em}  % or try \vspace{2em} for more space

 We further evaluated the performance of LFA across all channels in the training dataset, even though the force profiles at the point of density convergence were already smooth enough to yield accurate density predictions. Table~\ref{tab:lfa_metrics} reports the evaluation metrics, along with the ratio \( N_{\text{conv}} / N_{\text{GT}} \), which quantifies the number of frames required to reach a 1\% mean relative standard error in density (\( \bar{\sigma}_\rho \)) and force (\( \bar{\sigma}_f \)). The quantity \( \bar{\sigma}_f \) is estimated by fitting a power-law to the force uncertainty convergence curve (see \ref{sec:AppendixC}). The reduction in the number of decorrelated MLMD frames required for accurate density prediction due to LFA can be computed as \( (1 - N_{\text{conv}} / N_{\text{GT}}) \cdot N_{\text{GT}} \), which corresponds to approximately 350{,}000 and 109{,}000 frames for the widest (4.0~nm) and narrowest (1.0~nm) channels, respectively. 

\begin{table}[H]
  \centering
  \small
  \begin{adjustbox}{max width=\textwidth, center}
    \begin{tabular}{c c c c c c}
\toprule
\(\mathbf{L}\) (nm) & \(\frac{N_{\text{conv}}}{N_{\text{GT}}}\) &
\(\text{MISE}_\rho\) & \(\text{MISE}_\rho^{\text{LFA}}\) &
\(\epsilon_\rho\) & \(\epsilon_\rho^{\text{LFA}}\) \\
\midrule
\textbf{1.0} & 0.128 & 5.300 & \cellcolor{lightgreen}2.589 & 0.37 & \cellcolor{lightgreen}0.36 \\
\textbf{1.2} & 0.278 & 1.449 & \cellcolor{lightgreen}0.708 & 0.68 & \cellcolor{lightred}0.98 \\
\textbf{1.3} & 0.253 & 2.501 & \cellcolor{lightgreen}1.656 & 0.84 & \cellcolor{lightgreen}0.70 \\
\textbf{1.4} & 0.435 & 1.458 & \cellcolor{lightgreen}0.817 & 0.42 & \cellcolor{lightred}0.54 \\
\textbf{1.6} & 0.403 & 2.321 & \cellcolor{lightgreen}1.574 & 0.60 & \cellcolor{lightgreen}0.55 \\
\textbf{1.7} & 0.383 & 4.976 & \cellcolor{lightgreen}3.814 & 0.91 & \cellcolor{lightgreen}0.58 \\
\textbf{1.8} & 0.444 & 3.602 & \cellcolor{lightgreen}2.258 & 0.49 & \cellcolor{lightgreen}0.41 \\
\textbf{1.9} & 0.455 & 4.018 & \cellcolor{lightgreen}2.561 & 0.97 & \cellcolor{lightgreen}0.77 \\
\textbf{2.0} & 0.475 & 3.449 & \cellcolor{lightgreen}2.628 & 0.89 & \cellcolor{lightgreen}0.52 \\
\textbf{2.5*} & 0.817 & 2.100 & \cellcolor{lightgreen}1.054 & 4.26 & \cellcolor{lightgreen}0.95 \\
\textbf{3.0*} & 0.665 & 6.718 & \cellcolor{lightgreen}1.773 & 1.63 & \cellcolor{lightgreen}0.76 \\
\textbf{4.0*} & 0.608 & 1.545 & \cellcolor{lightgreen}0.857 & 3.70 & \cellcolor{lightgreen}0.70 \\
\bottomrule
\end{tabular}
\end{adjustbox}
\vspace{1mm}
\footnotesize{\textit{* Channel widths for which the convergence criterion was not met.}}
\caption{Performance metrics of the Local Force Approximation (LFA) strategy across varying channel widths \(L\) (in nm). The first column reports the ratio of the number of frames required to reach the convergence criterion to the total number of frames used for generating the ground truth. The remaining columns report the mean integrated squared error (MISE) and normalized root mean squared error (\(\epsilon\)) for density predictions only. Superscripts ‘LFA’ denote metrics evaluated after applying the Local Force Approximation. All MISE values for density are reported in units of \(10^{-2}\), and \(\epsilon\) values represent percent relative errors. Green-shaded cells indicate improvement relative to the baseline, while red-shaded cells denote degradation.}
\label{tab:lfa_metrics}
\end{table}

Across all confinement levels, the application of \textit{LFA} reduced the $\mathrm{MISE}$. However, the improvement in pointwise accuracy, measured by $\epsilon_\rho$, was less pronounced in narrower channels. For the $1.2\,\mathrm{nm}$ and $1.4\,\mathrm{nm}$ channels, a slight degradation in $\epsilon_\rho$ was observed, likely due to systematic bias introduced by \textit{LFA}. These results suggest that the effectiveness of \textit{LFA} depends on the spectral characteristics of the underlying force signal: it performs better in wider channels, where high-frequency fluctuations are more prominent. Conversely, when such fluctuations are suppressed, as in narrower channels, the reduction in $\mathrm{MISE}$ may come at the expense of local accuracy.

\noindent\subsection{Performance of Generative Quasi-Continuum Inference}

In this section, we assess the performance of the quasi-continuum inference algorithm (Algorithm~\ref{alg:qct}). We evaluate the predicted density profiles for channel widths $L = 1.1$ and $1.5$~nm (Figure~\ref{fig:qct_results}). These channel widths are not represented in the DDPM training data, and neither the corresponding ground truth force profiles nor the noisy single-frame forces were included during model training (see Section~\ref{sec:ddpm_training}). For evaluation, we use the same ground truth forces and densities that were used in Section~\ref{sec:LFA}.
\FloatBarrier
\noindent\begin{minipage}{\linewidth}
    \centering
    \includegraphics[width=1.05\linewidth, trim=0 50 20 0, clip]{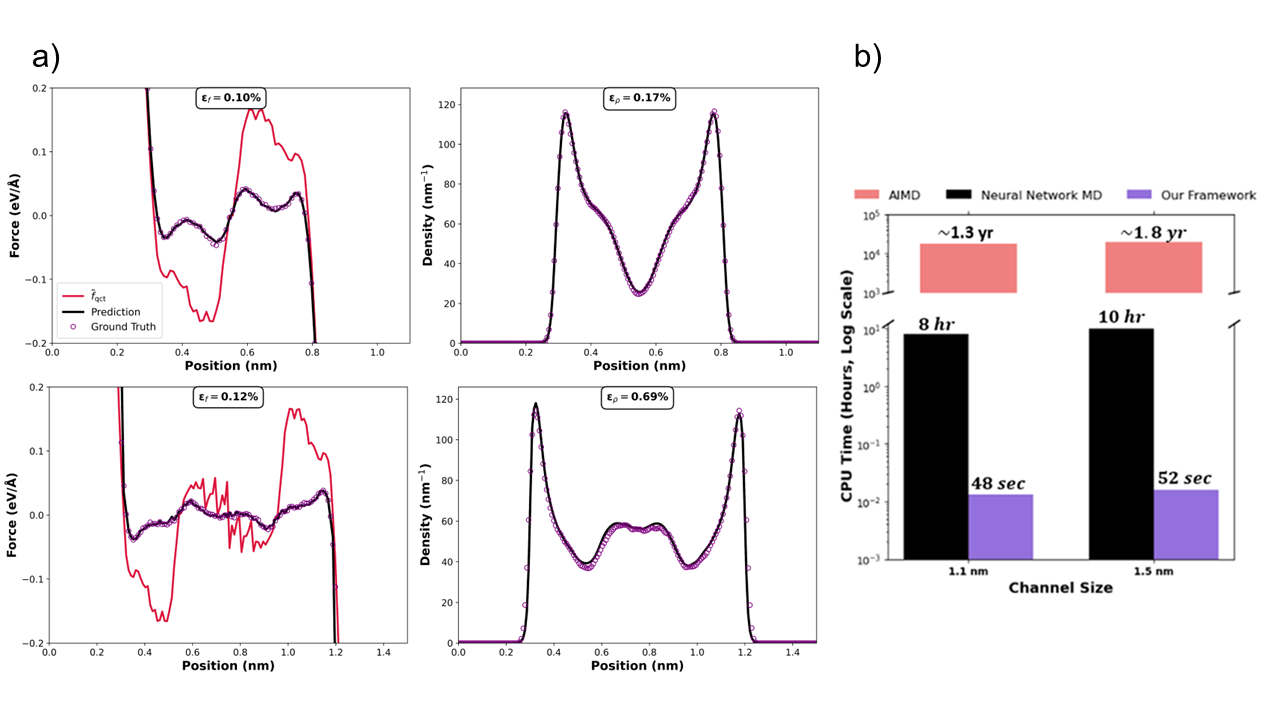}
    \vspace{0.5em}
    \captionsetup{format=plain, labelfont=bf, labelsep=colon, justification=justified}
    \captionof{figure}{ 
    (a) Comparison of forces and densities derived from these forces for unseen channel widths during training, 1.5 $nm$ (left) and 1.1 $nm$ (right), against ground truth data. Crimson curves are the noisy quasi-continuum force profiles. Black curves are the force profiles generated via reverse diffusion conditioned on noisy quasi-continuum forces. Density profiles (purple markers) are obtained by solving the Nernst-Planck equation \ref{eq:np} with the reverse diffusion output. 
    (b) Wall-clock time comparison for three methods (AIMD, MLMD, and our framework) in generating these density profiles.   AIMD values are extrapolated from the known CPU time to generate a single frame, scaled to match the ∼3 million steps required for a 1.5~ns simulation. Training time is not included in the reported runtimes.}
    \label{fig:qct_results}
\end{minipage}
\vspace{1em}  % ← add this to create natural spacing

\FloatBarrier
Given a target channel width $L$ that is not present in the AIMD dataset $\mathcal{D}$—which consists of trajectories for discrete channel widths $\{L_1, L_2, \dots, L_N\}$ (see Appendix~\ref{sec:AppendixB})—a noisy quasi-continuum force profile $f_{\text{qct}}^L(x)$ is constructed using the methodology described in Section~\ref{sec:GQCT}. This initial force profile is derived from matching atomistic neighborhoods in $\mathcal{D}$ that spatially align with bins in the new channel geometry. Next, the noisy quasi-continuum force is denoised using the trained DDPM model (Section~\ref{sec:ddpm_training}) via the conditional reverse diffusion process (Eq.~\ref{eq:reverse}). Finally, the smoothed force profile is inserted into the Nernst–Planck equation (Eq.~\ref{eq:np}) to yield the predicted density.

Figure~\ref{fig:qct_results} shows the resulting density predictions for $L = 1.1$ and $1.5$~nm, compared against ground truth reference densities.
As shown in Figure~\ref{fig:qct_results}a, the predicted denoised force profiles (black curves) closely match the ground truth forces (magenta markers) obtained from AIMD, yielding low relative RMSE values of 0.10\% and 0.12\% for the two channel sizes, respectively. When these forces are used in the solution of the Nernst--Planck equation, the resulting density profiles yield relative RMSEs of 0.17\% and 0.69\% compared to the AIMD reference, reflecting accurate reproduction of spatial density variations. In addition to accuracy, our framework’s computational efficiency is highlighted in Figure~\ref{fig:qct_results}b. AIMD simulations require $\sim$1.3–1.8 years of CPU time per system to converge to smooth density profiles, whereas MLMD simulations reduce this cost to 8 and 10 hours, respectively. The quasi-continuum inference algorithm, by contrast, predicts the density profiles in under a minute for both channels.

\begin{figure}[htbp]
    \centering

    % === Row a) ===
    \begin{minipage}{\textwidth}
        \begin{minipage}{0.05\textwidth}
            \raggedright \vtop{\vskip0pt\textbf{a)}}
        \end{minipage}
        \begin{minipage}{0.93\textwidth}
            \centering
            \includegraphics[width=0.9\textwidth]{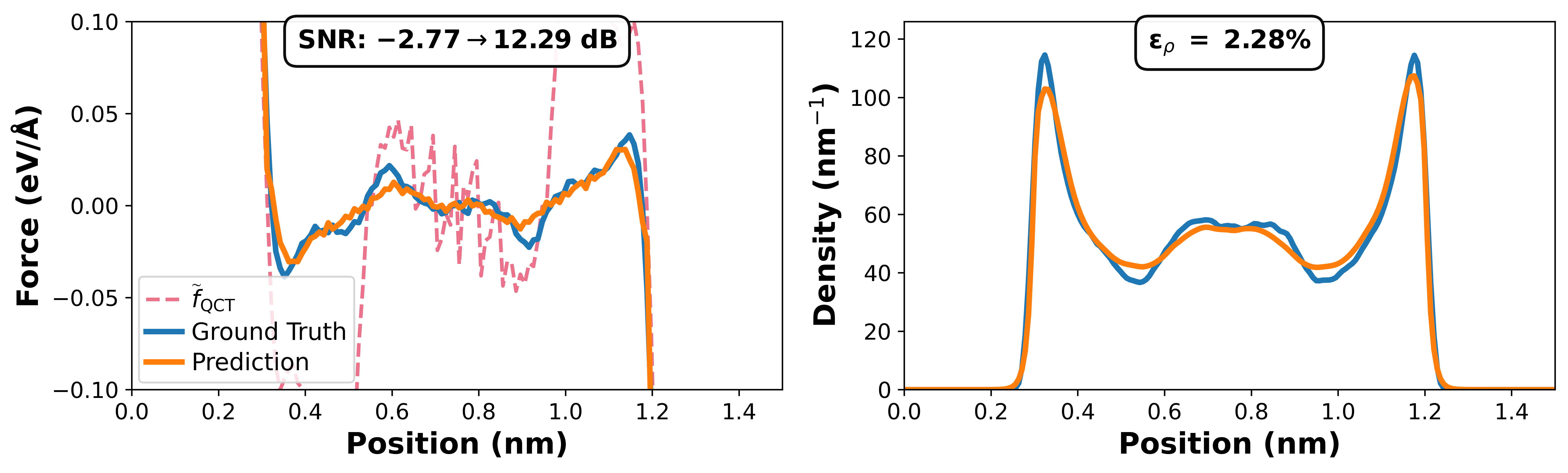}
        \end{minipage}
    \end{minipage}

    \vspace{1em}

    % === Row b) ===
    \begin{minipage}{\textwidth}
        \begin{minipage}{0.05\textwidth}
            \raggedright \vtop{\vskip0pt\textbf{b)}}
        \end{minipage}
        \begin{minipage}{0.93\textwidth}
            \centering
            \includegraphics[width=0.9\textwidth]{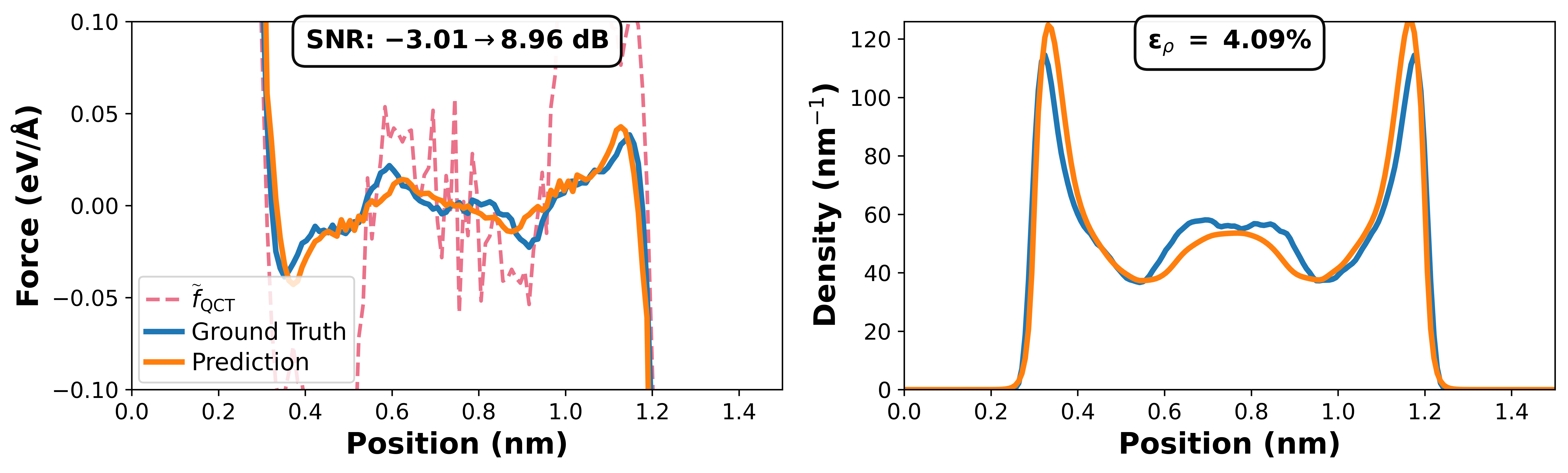}
        \end{minipage}
    \end{minipage}

    \vspace{1em}

    % === Row c) ===
    \begin{minipage}{\textwidth}
        \begin{minipage}{0.05\textwidth}
            \raggedright \vtop{\vskip0pt\textbf{c)}}
        \end{minipage}
        \begin{minipage}{0.93\textwidth}
            \centering
            \includegraphics[width=0.9\textwidth]{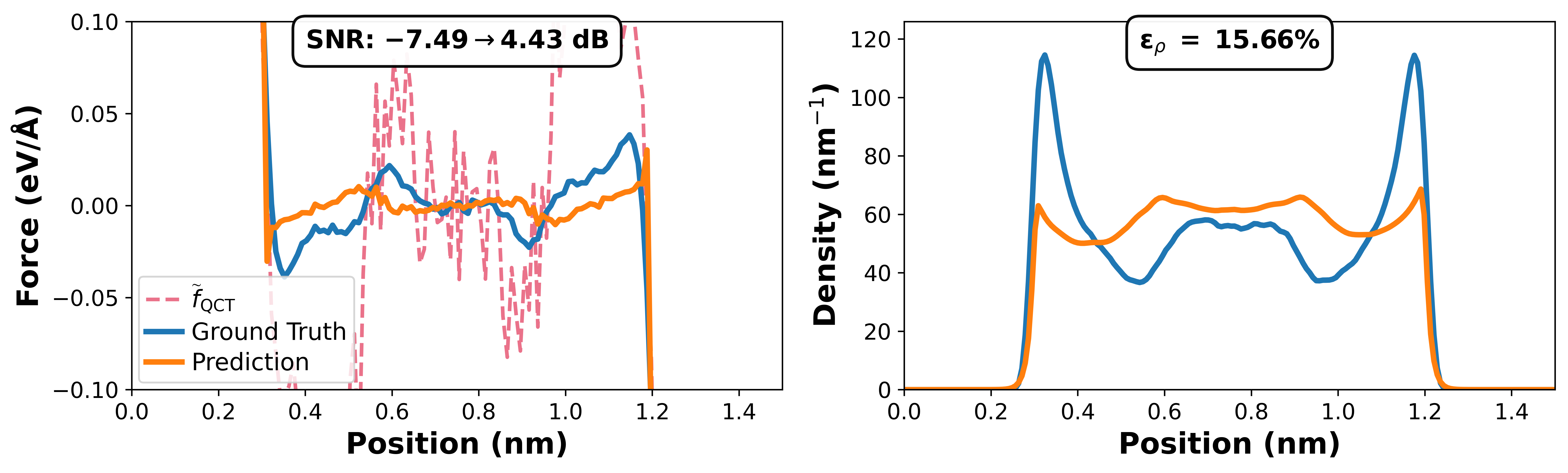}
        \end{minipage}
    \end{minipage}

    \caption{Quasi-continuum inference for force profiles obtained from varying fractions of the original dataset. 
(a) Force profile predicted using 70\% of the data compared with the ground truth, and the corresponding density profile obtained by solving the Nernst–Planck equation.
(b) using 40\% of the data. 
(c) using 10\% of the data.}

    \label{fig:force_density_snr_rows}
\end{figure}

\textbf{Data Requirement.} A natural question arises: \emph{What is the minimal AIMD dataset required to accurately recover the density profile?} To address this, we evaluate Algorithm~2 on randomly subsampled versions of the full AIMD dataset \( \mathcal{D} \). For  previously unseen channel widhts \( L = 1.5\,\text{nm} \), we generate noisy quasi-continuum force profiles \( \tilde{f}_{\text{qct}}^{\,L} \) (as defined in Section\ref{sec:GQCT}) using three randomly selected subsets comprising 10\%, 40\%, and 70\% of the original data. Figure~\ref{fig:force_density_snr_rows} illustrates that denoised force profiles degrade progressively with decreasing data availability, as reflected by declining SNR values in the force predictions. Taking the full dataset prediction as the baseline (15.96\,dB), subsampling to 70\% of atomistic neighborhoods reduces the SNR to 12.29\,dB, implying a 23.0\% reduction in force accuracy and a corresponding 3.1-fold increase in density error. Further reducing the dataset to 40\% lowers the SNR to 8.96\,dB, resulting in a 44.0\% reduction in force accuracy and a 5.5-fold increase in density error. When only 10\% of the data is used, the SNR drops to 4.43\,dB, decreasing the force accuracy by 72\% and amplifying the density error by 21×. These results suggest a nonlinear relationship between dataset size and predictive accuracy.

Interestingly, despite its overall lower accuracy, the 10\% case (Figure~7c) captures the correct number of local maxima in the density profile—four peaks for the 1.5\,nm channel—while both the 40\% and 70\% cases (Figures~7a and 7b, respectively) capture only three, failing to resolve the central minimum. Moreover, the 70\% case overpredicts the density in the first hydration layer, whereas the 40\% case underpredicts it. These qualitative differences further emphasize a  non-trivial dependence of diffusion model performance on the content of atomistic neighborhoods in the dataset. Our findings demonstrate that there is a limit to the diffusion models performance, and we leave the problem of finding the minimal dataset required to train the NNP while preserving downstream accuracy in force and density predictions for future work.

\clearpage

\section{Conclusion}

Density layering is a structural phenomenon that emerges when fluids are confined to nanometer-scale geometries and it is of fundamental importance for understanding the physics and chemistry of nanoscale systems. However, accurately capturing this behavior often requires ab initio methods that account for quantum and electronic effects, which are prohibitively expensive. In this work, we build upon prior developments in quasi-continuum theory and introduce a generative modeling framework that drastically reduces reliance on ab initio data. Our approach achieves ab initio accuracy in density predictions while maintaining the computational efficiency of continuum methods. Notably, we report a speed-up of more than two orders of magnitude compared to MLMD.

\section*{V. Data Availability}

The training and testing datasets generated and used in this work are available upon request to the corresponding authors.

\section{Code Availability}

The code developed for this work will be made publicly available through GitHub at:  
\url{https://github.com/multiscalenano/generativeQCT}

\section{Acknowledgements}

This work is supported by the Center for Enhanced Nanofluidic Transport (CENT), an Energy Frontier Research Center funded by the U.S. Department of Energy, Office of Science, Basic Energy Sciences under Award No. DE-SC0019112. We acknowledge the use of parallel computing resource Lonestar6  provided by the Texas Advanced Computing Center (TACC) at The University of Texas at Austin. 
\clearpage
\bibliography{Quasi-Continuum_references5_27_stripped}

\providecommand{\latin}[1]{#1}
\makeatletter
\providecommand{\doi}
  {\begingroup\let\do\@makeother\dospecials
  \catcode`\{=1 \catcode`\}=2 \doi@aux}
\providecommand{\doi@aux}[1]{\endgroup\texttt{#1}}
\makeatother
\providecommand*\mcitethebibliography{\thebibliography}
\csname @ifundefined\endcsname{endmcitethebibliography}  {\let\endmcitethebibliography\endthebibliography}{}
\begin{mcitethebibliography}{90}
\providecommand*\natexlab[1]{#1}
\providecommand*\mciteSetBstSublistMode[1]{}
\providecommand*\mciteSetBstMaxWidthForm[2]{}
\providecommand*\mciteBstWouldAddEndPuncttrue
  {\def\EndOfBibitem{\unskip.}}
\providecommand*\mciteBstWouldAddEndPunctfalse
  {\let\EndOfBibitem\relax}
\providecommand*\mciteSetBstMidEndSepPunct[3]{}
\providecommand*\mciteSetBstSublistLabelBeginEnd[3]{}
\providecommand*\EndOfBibitem{}
\mciteSetBstSublistMode{f}
\mciteSetBstMaxWidthForm{subitem}{(\alph{mcitesubitemcount})}
\mciteSetBstSublistLabelBeginEnd
  {\mcitemaxwidthsubitemform\space}
  {\relax}
  {\relax}

\bibitem[Bocquet and Charlaix(2010)Bocquet, and Charlaix]{bocquetNanofluidicsBulkInterfaces2010c}
Bocquet,~L.; Charlaix,~E. Nanofluidics, from Bulk to Interfaces. \emph{Chem. Soc. Rev.} \textbf{2010}, \emph{39}, 1073--1095\relax
\mciteBstWouldAddEndPuncttrue
\mciteSetBstMidEndSepPunct{\mcitedefaultmidpunct}
{\mcitedefaultendpunct}{\mcitedefaultseppunct}\relax
\EndOfBibitem
\bibitem[Aluru \latin{et~al.}(2023)Aluru, Aydin, Bazant, Blankschtein, Brozena, De~Souza, Elimelech, Faucher, Fourkas, Koman, Kuehne, Kulik, Li, Li, Li, Majumdar, Martis, Misra, Noy, Pham, Qu, Rayabharam, Reed, Ritt, Schwegler, Siwy, Strano, Wang, Yao, Zhan, and Zhang]{aluruFluidsElectrolytesConfinement2023f}
Aluru,~N.~R. \latin{et~al.}  Fluids and {{Electrolytes}} under {{Confinement}} in {{Single-Digit Nanopores}}. \emph{Chemical Reviews} \textbf{2023}, \emph{123}, 2737--2831\relax
\mciteBstWouldAddEndPuncttrue
\mciteSetBstMidEndSepPunct{\mcitedefaultmidpunct}
{\mcitedefaultendpunct}{\mcitedefaultseppunct}\relax
\EndOfBibitem
\bibitem[Faucher \latin{et~al.}(2019)Faucher, Aluru, Bazant, Blankschtein, Brozena, Cumings, Pedro De~Souza, Elimelech, Epsztein, Fourkas, Rajan, Kulik, Levy, Majumdar, Martin, McEldrew, Misra, Noy, Pham, Reed, Schwegler, Siwy, Wang, and Strano]{faucherCriticalKnowledgeGaps2019a}
Faucher,~S. \latin{et~al.}  Critical {{Knowledge Gaps}} in {{Mass Transport}} through {{Single-Digit Nanopores}}: {{A Review}} and {{Perspective}}. \emph{The Journal of Physical Chemistry C} \textbf{2019}, \emph{123}, 21309--21326\relax
\mciteBstWouldAddEndPuncttrue
\mciteSetBstMidEndSepPunct{\mcitedefaultmidpunct}
{\mcitedefaultendpunct}{\mcitedefaultseppunct}\relax
\EndOfBibitem
\bibitem[Hansen and McDonald(2007)Hansen, and McDonald]{hansenTheorySimpleLiquids2007c}
Hansen,~J.-P.; McDonald,~I.~R. \emph{Theory of Simple Liquids}, 3rd ed.; Elsevier / Academic Press: Amsterdam ; Boston, 2007\relax
\mciteBstWouldAddEndPuncttrue
\mciteSetBstMidEndSepPunct{\mcitedefaultmidpunct}
{\mcitedefaultendpunct}{\mcitedefaultseppunct}\relax
\EndOfBibitem
\bibitem[Antman \latin{et~al.}(2005)Antman, Marsden, and Sirovich]{antmanMicroflowsNanoflows2005c}
Antman,~S., Marsden,~J., Sirovich,~L., Eds. \emph{Microflows and {{Nanoflows}}}; Interdisciplinary {{Applied Mathematics}}; Springer-Verlag: New York, 2005; Vol.~29\relax
\mciteBstWouldAddEndPuncttrue
\mciteSetBstMidEndSepPunct{\mcitedefaultmidpunct}
{\mcitedefaultendpunct}{\mcitedefaultseppunct}\relax
\EndOfBibitem
\bibitem[{Ruiz-Barragan} \latin{et~al.}(2019){Ruiz-Barragan}, {Mu{\~n}oz-Santiburcio}, and Marx]{ruiz-barraganNanoconfinedWaterGraphene2019c}
{Ruiz-Barragan},~S.; {Mu{\~n}oz-Santiburcio},~D.; Marx,~D. Nanoconfined {{Water}} within {{Graphene Slit Pores Adopts Distinct Confinement-Dependent Regimes}}. \emph{The Journal of Physical Chemistry Letters} \textbf{2019}, \emph{10}, 329--334\relax
\mciteBstWouldAddEndPuncttrue
\mciteSetBstMidEndSepPunct{\mcitedefaultmidpunct}
{\mcitedefaultendpunct}{\mcitedefaultseppunct}\relax
\EndOfBibitem
\bibitem[Fong \latin{et~al.}(2024)Fong, Sumi{\'c}, O'Neill, Schran, Grey, and Michaelides]{fongInterplaySolvationPolarization2024}
Fong,~K.~D.; Sumi{\'c},~B.; O'Neill,~N.; Schran,~C.; Grey,~C.~P.; Michaelides,~A. The {{Interplay}} of {{Solvation}} and {{Polarization Effects}} on {{Ion Pairing}} in {{Nanoconfined Electrolytes}}. \emph{Nano Lett.} \textbf{2024}, \relax
\mciteBstWouldAddEndPunctfalse
\mciteSetBstMidEndSepPunct{\mcitedefaultmidpunct}
{}{\mcitedefaultseppunct}\relax
\EndOfBibitem
\bibitem[Joseph and Aluru(2008)Joseph, and Aluru]{josephWhyAreCarbon2008b}
Joseph,~S.; Aluru,~N.~R. Why {{Are Carbon Nanotubes Fast Transporters}} of {{Water}}? \emph{Nano Letters} \textbf{2008}, \emph{8}, 452--458\relax
\mciteBstWouldAddEndPuncttrue
\mciteSetBstMidEndSepPunct{\mcitedefaultmidpunct}
{\mcitedefaultendpunct}{\mcitedefaultseppunct}\relax
\EndOfBibitem
\bibitem[Schoch \latin{et~al.}(2008)Schoch, Han, and Renaud]{schochTransportPhenomenaNanofluidics2008c}
Schoch,~R.~B.; Han,~J.; Renaud,~P. Transport Phenomena in Nanofluidics. \emph{Reviews of Modern Physics} \textbf{2008}, \emph{80}, 839--883\relax
\mciteBstWouldAddEndPuncttrue
\mciteSetBstMidEndSepPunct{\mcitedefaultmidpunct}
{\mcitedefaultendpunct}{\mcitedefaultseppunct}\relax
\EndOfBibitem
\bibitem[Wu(2022)]{wuUnderstandingElectricDoubleLayer2022c}
Wu,~J. Understanding the {{Electric Double-Layer Structure}}, {{Capacitance}}, and {{Charging Dynamics}}. \emph{Chemical Reviews} \textbf{2022}, \emph{122}, 10821--10859\relax
\mciteBstWouldAddEndPuncttrue
\mciteSetBstMidEndSepPunct{\mcitedefaultmidpunct}
{\mcitedefaultendpunct}{\mcitedefaultseppunct}\relax
\EndOfBibitem
\bibitem[Ivanov \latin{et~al.}(2020)Ivanov, Poryvaev, Polyukhov, Prikhod'ko, Adonin, and Fedin]{ivanovNanoconfinementEffectsStructural2020c}
Ivanov,~M.~Y.; Poryvaev,~A.~S.; Polyukhov,~D.~M.; Prikhod'ko,~S.~A.; Adonin,~N.~Y.; Fedin,~M.~V. Nanoconfinement Effects on Structural Anomalies in Imidazolium Ionic Liquids. \emph{Nanoscale} \textbf{2020}, \emph{12}, 23480--23487\relax
\mciteBstWouldAddEndPuncttrue
\mciteSetBstMidEndSepPunct{\mcitedefaultmidpunct}
{\mcitedefaultendpunct}{\mcitedefaultseppunct}\relax
\EndOfBibitem
\bibitem[Ilgen \latin{et~al.}(2023)Ilgen, Leung, Criscenti, and Greathouse]{ilgenAdsorptionNanoconfinedSolid2023c}
Ilgen,~A.~G.; Leung,~K.; Criscenti,~L.~J.; Greathouse,~J.~A. Adsorption at {{Nanoconfined Solid}}--{{Water Interfaces}}. \emph{Annual Review of Physical Chemistry} \textbf{2023}, \emph{74}, 169--191\relax
\mciteBstWouldAddEndPuncttrue
\mciteSetBstMidEndSepPunct{\mcitedefaultmidpunct}
{\mcitedefaultendpunct}{\mcitedefaultseppunct}\relax
\EndOfBibitem
\bibitem[Wordsworth \latin{et~al.}(2022)Wordsworth, Benedetti, Somerville, Schuhmann, Tilley, and Gooding]{wordsworthInfluenceNanoconfinementElectrocatalysis2022c}
Wordsworth,~J.; Benedetti,~T.~M.; Somerville,~S.~V.; Schuhmann,~W.; Tilley,~R.~D.; Gooding,~J.~J. The {{Influence}} of {{Nanoconfinement}} on {{Electrocatalysis}}. \emph{Angewandte Chemie International Edition} \textbf{2022}, \emph{61}, e202200755\relax
\mciteBstWouldAddEndPuncttrue
\mciteSetBstMidEndSepPunct{\mcitedefaultmidpunct}
{\mcitedefaultendpunct}{\mcitedefaultseppunct}\relax
\EndOfBibitem
\bibitem[Feng \latin{et~al.}(2016)Feng, Graf, Liu, Ovchinnikov, Dumcenco, Heiranian, Nandigana, Aluru, Kis, and Radenovic]{fengSinglelayerMoS2Nanopores2016c}
Feng,~J.; Graf,~M.; Liu,~K.; Ovchinnikov,~D.; Dumcenco,~D.; Heiranian,~M.; Nandigana,~V.; Aluru,~N.~R.; Kis,~A.; Radenovic,~A. Single-Layer {{MoS2}} Nanopores as Nanopower Generators. \emph{Nature} \textbf{2016}, \emph{536}, 197--200\relax
\mciteBstWouldAddEndPuncttrue
\mciteSetBstMidEndSepPunct{\mcitedefaultmidpunct}
{\mcitedefaultendpunct}{\mcitedefaultseppunct}\relax
\EndOfBibitem
\bibitem[Heiranian \latin{et~al.}(2015)Heiranian, Farimani, and Aluru]{heiranianWaterDesalinationSinglelayer2015c}
Heiranian,~M.; Farimani,~A.~B.; Aluru,~N.~R. Water Desalination with a Single-Layer {{MoS2}} Nanopore. \emph{Nature Communications} \textbf{2015}, \emph{6}, 8616\relax
\mciteBstWouldAddEndPuncttrue
\mciteSetBstMidEndSepPunct{\mcitedefaultmidpunct}
{\mcitedefaultendpunct}{\mcitedefaultseppunct}\relax
\EndOfBibitem
\bibitem[Qian \latin{et~al.}(2020)Qian, Gao, and Pan]{qianNanoconfinementMediatedWaterTreatment2020c}
Qian,~J.; Gao,~X.; Pan,~B. Nanoconfinement-{{Mediated Water Treatment}}: {{From Fundamental}} to {{Application}}. \emph{Environmental Science \& Technology} \textbf{2020}, \emph{54}, 8509--8526\relax
\mciteBstWouldAddEndPuncttrue
\mciteSetBstMidEndSepPunct{\mcitedefaultmidpunct}
{\mcitedefaultendpunct}{\mcitedefaultseppunct}\relax
\EndOfBibitem
\bibitem[Yang \latin{et~al.}(2025)Yang, Li, Yang, Zhang, Chen, Jiao, Wang, Zhang, Zhai, Sun, Xiang, and Gong]{yangUltrafastLithiumIonTransport2025c}
Yang,~Y.; Li,~Z.; Yang,~Z.; Zhang,~Q.; Chen,~Q.; Jiao,~Y.; Wang,~Z.; Zhang,~X.; Zhai,~P.; Sun,~Z.; Xiang,~Y.; Gong,~Y. Ultrafast {{Lithium}}-{{Ion Transport Engineered}} by {{Nanoconfinement Effect}}. \emph{Advanced Materials} \textbf{2025}, \emph{37}, 2416266\relax
\mciteBstWouldAddEndPuncttrue
\mciteSetBstMidEndSepPunct{\mcitedefaultmidpunct}
{\mcitedefaultendpunct}{\mcitedefaultseppunct}\relax
\EndOfBibitem
\bibitem[Jiang \latin{et~al.}(2011)Jiang, Jin, and Wu]{jiangOscillationCapacitanceNanopores2011c}
Jiang,~D.-e.; Jin,~Z.; Wu,~J. Oscillation of {{Capacitance}} inside {{Nanopores}}. \emph{Nano Letters} \textbf{2011}, \emph{11}, 5373--5377\relax
\mciteBstWouldAddEndPuncttrue
\mciteSetBstMidEndSepPunct{\mcitedefaultmidpunct}
{\mcitedefaultendpunct}{\mcitedefaultseppunct}\relax
\EndOfBibitem
\bibitem[Zhang \latin{et~al.}(2021)Zhang, Zhang, Li, Zhang, Li, Ying, and Fu]{zhangNanoconfinementEffectSignal2021c}
Zhang,~J.; Zhang,~L.; Li,~Z.; Zhang,~Q.; Li,~Y.; Ying,~Y.; Fu,~Y. Nanoconfinement {{Effect}} for {{Signal Amplification}} in {{Electrochemical Analysis}} and {{Sensing}}. \emph{Small} \textbf{2021}, \emph{17}, 2101665\relax
\mciteBstWouldAddEndPuncttrue
\mciteSetBstMidEndSepPunct{\mcitedefaultmidpunct}
{\mcitedefaultendpunct}{\mcitedefaultseppunct}\relax
\EndOfBibitem
\bibitem[Chu \latin{et~al.}(2024)Chu, Zhou, Tian, Yu, Lian, Zhang, and Xuan]{chuNanofluidicSensingInspired2024c}
Chu,~T.; Zhou,~Z.; Tian,~P.; Yu,~T.; Lian,~C.; Zhang,~B.; Xuan,~F.-Z. Nanofluidic Sensing Inspired by the Anomalous Water Dynamics in Electrical Angstrom-Scale Channels. \emph{Nature Communications} \textbf{2024}, \emph{15}, 7329\relax
\mciteBstWouldAddEndPuncttrue
\mciteSetBstMidEndSepPunct{\mcitedefaultmidpunct}
{\mcitedefaultendpunct}{\mcitedefaultseppunct}\relax
\EndOfBibitem
\bibitem[Chantipmanee and Xu(2024)Chantipmanee, and Xu]{chantipmaneeNanofluidicManipulationSingle2024c}
Chantipmanee,~N.; Xu,~Y. Nanofluidic {{Manipulation}} of {{Single Nanometric Objects}}: {{Current Progress}}, {{Challenges}}, and {{Future Opportunities}}. \emph{Engineering} \textbf{2024}, \emph{43}, 54--71\relax
\mciteBstWouldAddEndPuncttrue
\mciteSetBstMidEndSepPunct{\mcitedefaultmidpunct}
{\mcitedefaultendpunct}{\mcitedefaultseppunct}\relax
\EndOfBibitem
\bibitem[Aluru \latin{et~al.}(2023)Aluru, Aydin, Bazant, Blankschtein, Brozena, De~Souza, Elimelech, Faucher, Fourkas, Koman, Kuehne, Kulik, Li, Li, Li, Majumdar, Martis, Misra, Noy, Pham, Qu, Rayabharam, Reed, Ritt, Schwegler, Siwy, Strano, Wang, Yao, Zhan, and Zhang]{aluruFluidsElectrolytesConfinement2023g}
Aluru,~N.~R. \latin{et~al.}  Fluids and {{Electrolytes}} under {{Confinement}} in {{Single-Digit Nanopores}}. \emph{Chemical Reviews} \textbf{2023}, \emph{123}, 2737--2831\relax
\mciteBstWouldAddEndPuncttrue
\mciteSetBstMidEndSepPunct{\mcitedefaultmidpunct}
{\mcitedefaultendpunct}{\mcitedefaultseppunct}\relax
\EndOfBibitem
\bibitem[Snook and Henderson(1978)Snook, and Henderson]{snookMonteCarloStudy1978c}
Snook,~I.~K.; Henderson,~D. Monte {{Carlo}} Study of a Hard-Sphere Fluid near a Hard Wall. \emph{The Journal of Chemical Physics} \textbf{1978}, \emph{68}, 2134--2139\relax
\mciteBstWouldAddEndPuncttrue
\mciteSetBstMidEndSepPunct{\mcitedefaultmidpunct}
{\mcitedefaultendpunct}{\mcitedefaultseppunct}\relax
\EndOfBibitem
\bibitem[Meyer and Stanley(1999)Meyer, and Stanley]{meyerLiquidLiquidPhaseTransition1999c}
Meyer,~M.; Stanley,~H.~E. {{Liquid}}-{{Liquid Phase Transition}} in {{Confined Water}}: {{A Monte Carlo Study}}. \emph{The Journal of Physical Chemistry B} \textbf{1999}, \emph{103}, 9728--9730\relax
\mciteBstWouldAddEndPuncttrue
\mciteSetBstMidEndSepPunct{\mcitedefaultmidpunct}
{\mcitedefaultendpunct}{\mcitedefaultseppunct}\relax
\EndOfBibitem
\bibitem[Leng \latin{et~al.}(2013)Leng, Xiang, Lei, and Rao]{lengComparativeStudyGrand2013c}
Leng,~Y.; Xiang,~Y.; Lei,~Y.; Rao,~Q. A Comparative Study by the Grand Canonical {{Monte Carlo}} and Molecular Dynamics Simulations on the Squeezing Behavior of Nanometers Confined Liquid Films. \emph{The Journal of Chemical Physics} \textbf{2013}, \emph{139}, 074704\relax
\mciteBstWouldAddEndPuncttrue
\mciteSetBstMidEndSepPunct{\mcitedefaultmidpunct}
{\mcitedefaultendpunct}{\mcitedefaultseppunct}\relax
\EndOfBibitem
\bibitem[Paquet and Viktor(2015)Paquet, and Viktor]{paquetMolecularDynamicsMonte2015}
Paquet,~E.; Viktor,~H.~L. Molecular {{Dynamics}}, {{Monte Carlo Simulations}}, and {{Langevin Dynamics}}: {{A Computational Review}}. \emph{BioMed Research International} \textbf{2015}, \emph{2015}, 1--18\relax
\mciteBstWouldAddEndPuncttrue
\mciteSetBstMidEndSepPunct{\mcitedefaultmidpunct}
{\mcitedefaultendpunct}{\mcitedefaultseppunct}\relax
\EndOfBibitem
\bibitem[Frenkel and Smit(2002)Frenkel, and Smit]{frenkelUnderstandingMolecularSimulation2002c}
Frenkel,~D.; Smit,~B. \emph{Understanding Molecular Simulation: From Algorithms to Applications}, 2nd ed.; Computational Science Series 1; Academic Press: San Diego, 2002\relax
\mciteBstWouldAddEndPuncttrue
\mciteSetBstMidEndSepPunct{\mcitedefaultmidpunct}
{\mcitedefaultendpunct}{\mcitedefaultseppunct}\relax
\EndOfBibitem
\bibitem[Lynch \latin{et~al.}(2020)Lynch, Rao, and Sansom]{lynchWaterNanoporesBiological2020c}
Lynch,~C.~I.; Rao,~S.; Sansom,~M. S.~P. Water in {{Nanopores}} and {{Biological Channels}}: {{A Molecular Simulation Perspective}}. \emph{Chemical Reviews} \textbf{2020}, \emph{120}, 10298--10335\relax
\mciteBstWouldAddEndPuncttrue
\mciteSetBstMidEndSepPunct{\mcitedefaultmidpunct}
{\mcitedefaultendpunct}{\mcitedefaultseppunct}\relax
\EndOfBibitem
\bibitem[Noid(2023)]{noidPerspectiveAdvancesChallenges2023}
Noid,~W.~G. Perspective: {{Advances}}, {{Challenges}}, and {{Insight}} for {{Predictive Coarse-Grained Models}}. \emph{J. Phys. Chem. B} \textbf{2023}, \emph{127}, 4174--4207\relax
\mciteBstWouldAddEndPuncttrue
\mciteSetBstMidEndSepPunct{\mcitedefaultmidpunct}
{\mcitedefaultendpunct}{\mcitedefaultseppunct}\relax
\EndOfBibitem
\bibitem[Jin \latin{et~al.}(2022)Jin, Pak, Durumeric, Loose, and Voth]{jinBottomupCoarseGrainingPrinciples2022}
Jin,~J.; Pak,~A.~J.; Durumeric,~A. E.~P.; Loose,~T.~D.; Voth,~G.~A. Bottom-up {{Coarse-Graining}}: {{Principles}} and {{Perspectives}}. \emph{J. Chem. Theory Comput.} \textbf{2022}, \emph{18}, 5759--5791\relax
\mciteBstWouldAddEndPuncttrue
\mciteSetBstMidEndSepPunct{\mcitedefaultmidpunct}
{\mcitedefaultendpunct}{\mcitedefaultseppunct}\relax
\EndOfBibitem
\bibitem[Sanghi and Aluru(2012)Sanghi, and Aluru]{sanghiCoarsegrainedPotentialModels2012a}
Sanghi,~T.; Aluru,~N.~R. Coarse-Grained Potential Models for Structural Prediction of Carbon Dioxide ({{CO2}}) in Confined Environments. \emph{The Journal of Chemical Physics} \textbf{2012}, \emph{136}, 024102\relax
\mciteBstWouldAddEndPuncttrue
\mciteSetBstMidEndSepPunct{\mcitedefaultmidpunct}
{\mcitedefaultendpunct}{\mcitedefaultseppunct}\relax
\EndOfBibitem
\bibitem[Motevaselian \latin{et~al.}(2018)Motevaselian, Mashayak, and Aluru]{motevaselianExtendedCoarsegrainedDipole2018a}
Motevaselian,~M.~H.; Mashayak,~S.~Y.; Aluru,~N.~R. Extended Coarse-Grained Dipole Model for Polar Liquids: {{Application}} to Bulk and Confined Water. \emph{Physical Review E} \textbf{2018}, \emph{98}, 052135\relax
\mciteBstWouldAddEndPuncttrue
\mciteSetBstMidEndSepPunct{\mcitedefaultmidpunct}
{\mcitedefaultendpunct}{\mcitedefaultseppunct}\relax
\EndOfBibitem
\bibitem[Mashayak and Aluru(2012)Mashayak, and Aluru]{mashayakCoarseGrainedPotentialModel2012a}
Mashayak,~S.~Y.; Aluru,~N.~R. Coarse-{{Grained Potential Model}} for {{Structural Prediction}} of {{Confined Water}}. \emph{Journal of Chemical Theory and Computation} \textbf{2012}, \emph{8}, 1828--1840\relax
\mciteBstWouldAddEndPuncttrue
\mciteSetBstMidEndSepPunct{\mcitedefaultmidpunct}
{\mcitedefaultendpunct}{\mcitedefaultseppunct}\relax
\EndOfBibitem
\bibitem[Raghunathan \latin{et~al.}(2007)Raghunathan, Park, and Aluru]{raghunathanInteratomicPotentialbasedSemiclassical2007a}
Raghunathan,~A.~V.; Park,~J.~H.; Aluru,~N.~R. Interatomic Potential-Based Semiclassical Theory for {{Lennard-Jones}} Fluids. \emph{The Journal of Chemical Physics} \textbf{2007}, \emph{127}, 174701\relax
\mciteBstWouldAddEndPuncttrue
\mciteSetBstMidEndSepPunct{\mcitedefaultmidpunct}
{\mcitedefaultendpunct}{\mcitedefaultseppunct}\relax
\EndOfBibitem
\bibitem[Mashayak and Aluru(2012)Mashayak, and Aluru]{mashayakThermodynamicStatedependentStructurebased2012a}
Mashayak,~S.~Y.; Aluru,~N.~R. Thermodynamic State-Dependent Structure-Based Coarse-Graining of Confined Water. \emph{The Journal of Chemical Physics} \textbf{2012}, \emph{137}, 214707\relax
\mciteBstWouldAddEndPuncttrue
\mciteSetBstMidEndSepPunct{\mcitedefaultmidpunct}
{\mcitedefaultendpunct}{\mcitedefaultseppunct}\relax
\EndOfBibitem
\bibitem[Stone \latin{et~al.}(2007)Stone, Phillips, Freddolino, Hardy, Trabuco, and Schulten]{stoneAcceleratingMolecularModeling2007c}
Stone,~J.~E.; Phillips,~J.~C.; Freddolino,~L.; Hardy,~D.~J.; Trabuco,~L.~G.; Schulten,~K. Accelerating Molecular Modeling Applications with Graphics Processors. \emph{Journal of Computational Chemistry} \textbf{2007}, \emph{28}, 2618--2640\relax
\mciteBstWouldAddEndPuncttrue
\mciteSetBstMidEndSepPunct{\mcitedefaultmidpunct}
{\mcitedefaultendpunct}{\mcitedefaultseppunct}\relax
\EndOfBibitem
\bibitem[Shaw \latin{et~al.}(2010)Shaw, Maragakis, {Lindorff-Larsen}, Piana, Dror, Eastwood, Bank, Jumper, Salmon, Shan, and Wriggers]{shawAtomicLevelCharacterizationStructural2010c}
Shaw,~D.~E.; Maragakis,~P.; {Lindorff-Larsen},~K.; Piana,~S.; Dror,~R.~O.; Eastwood,~M.~P.; Bank,~J.~A.; Jumper,~J.~M.; Salmon,~J.~K.; Shan,~Y.; Wriggers,~W. Atomic-{{Level Characterization}} of the {{Structural Dynamics}} of {{Proteins}}. \emph{Science} \textbf{2010}, \emph{330}, 341--346\relax
\mciteBstWouldAddEndPuncttrue
\mciteSetBstMidEndSepPunct{\mcitedefaultmidpunct}
{\mcitedefaultendpunct}{\mcitedefaultseppunct}\relax
\EndOfBibitem
\bibitem[Jung \latin{et~al.}(2019)Jung, Nishima, Daniels, Bascom, Kobayashi, Adedoyin, Wall, Lappala, Phillips, Fischer, Tung, Schlick, Sugita, and Sanbonmatsu]{jungScalingMolecularDynamics2019c}
Jung,~J.; Nishima,~W.; Daniels,~M.; Bascom,~G.; Kobayashi,~C.; Adedoyin,~A.; Wall,~M.; Lappala,~A.; Phillips,~D.; Fischer,~W.; Tung,~C.-S.; Schlick,~T.; Sugita,~Y.; Sanbonmatsu,~K.~Y. Scaling Molecular Dynamics beyond 100,000 Processor Cores for Large-scale Biophysical Simulations. \emph{Journal of Computational Chemistry} \textbf{2019}, \emph{40}, 1919--1930\relax
\mciteBstWouldAddEndPuncttrue
\mciteSetBstMidEndSepPunct{\mcitedefaultmidpunct}
{\mcitedefaultendpunct}{\mcitedefaultseppunct}\relax
\EndOfBibitem
\bibitem[Wu \latin{et~al.}(2023)Wu, Liang, Jeong, and Aluru]{wuInitioContinuumLinking2023a}
Wu,~H.; Liang,~C.; Jeong,~J.; Aluru,~N.~R. From {\emph{Ab Initio}} to Continuum: {{Linking}} Multiple Scales Using Deep-Learned Forces. \emph{The Journal of Chemical Physics} \textbf{2023}, \emph{159}, 184108\relax
\mciteBstWouldAddEndPuncttrue
\mciteSetBstMidEndSepPunct{\mcitedefaultmidpunct}
{\mcitedefaultendpunct}{\mcitedefaultseppunct}\relax
\EndOfBibitem
\bibitem[Tuckerman(2002)]{tuckermanInitioMolecularDynamics2002c}
Tuckerman,~M.~E. {\emph{Ab}} Initio Molecular Dynamics: Basic Concepts, Current Trends and Novel Applications. \emph{Journal of Physics: Condensed Matter} \textbf{2002}, \emph{14}, R1297--R1355\relax
\mciteBstWouldAddEndPuncttrue
\mciteSetBstMidEndSepPunct{\mcitedefaultmidpunct}
{\mcitedefaultendpunct}{\mcitedefaultseppunct}\relax
\EndOfBibitem
\bibitem[Liang \latin{et~al.}(2023)Liang, Rayabharam, and Aluru]{liangStructuralDynamicalProperties2023c}
Liang,~C.; Rayabharam,~A.; Aluru,~N.~R. Structural and {{Dynamical Properties}} of {{H}}{$\_2$} {{O}} and {{D}}{$\_2$} {{O}} under {{Confinement}}. \emph{The Journal of Physical Chemistry B} \textbf{2023}, \emph{127}, 6532--6542\relax
\mciteBstWouldAddEndPuncttrue
\mciteSetBstMidEndSepPunct{\mcitedefaultmidpunct}
{\mcitedefaultendpunct}{\mcitedefaultseppunct}\relax
\EndOfBibitem
\bibitem[Berendsen \latin{et~al.}(1987)Berendsen, Grigera, and Straatsma]{berendsenMissingTermEffective1987}
Berendsen,~H. J.~C.; Grigera,~J.~R.; Straatsma,~T.~P. The Missing Term in Effective Pair Potentials. \emph{J. Phys. Chem.} \textbf{1987}, \emph{91}, 6269--6271\relax
\mciteBstWouldAddEndPuncttrue
\mciteSetBstMidEndSepPunct{\mcitedefaultmidpunct}
{\mcitedefaultendpunct}{\mcitedefaultseppunct}\relax
\EndOfBibitem
\bibitem[He \latin{et~al.}(2018)He, Zhu, Epstein, and Mo]{heStatisticalVariancesDiffusional2018c}
He,~X.; Zhu,~Y.; Epstein,~A.; Mo,~Y. Statistical Variances of Diffusional Properties from Ab Initio Molecular Dynamics Simulations. \emph{npj Computational Materials} \textbf{2018}, \emph{4}, 18\relax
\mciteBstWouldAddEndPuncttrue
\mciteSetBstMidEndSepPunct{\mcitedefaultmidpunct}
{\mcitedefaultendpunct}{\mcitedefaultseppunct}\relax
\EndOfBibitem
\bibitem[Choudhary \latin{et~al.}(2023)Choudhary, DeCost, Major, Butler, Thiyagalingam, and Tavazza]{choudharyUnifiedGraphNeural2023}
Choudhary,~K.; DeCost,~B.; Major,~L.; Butler,~K.; Thiyagalingam,~J.; Tavazza,~F. Unified Graph Neural Network Force-Field for the Periodic Table: Solid State Applications. \emph{Digital Discovery} \textbf{2023}, \emph{2}, 346--355\relax
\mciteBstWouldAddEndPuncttrue
\mciteSetBstMidEndSepPunct{\mcitedefaultmidpunct}
{\mcitedefaultendpunct}{\mcitedefaultseppunct}\relax
\EndOfBibitem
\bibitem[Zhang \latin{et~al.}(2018)Zhang, Han, Wang, Car, and E]{zhangDeepPotentialMolecular2018c}
Zhang,~L.; Han,~J.; Wang,~H.; Car,~R.; E,~W. Deep {{Potential Molecular Dynamics}}: {{A Scalable Model}} with the {{Accuracy}} of {{Quantum Mechanics}}. \emph{Physical Review Letters} \textbf{2018}, \emph{120}, 143001\relax
\mciteBstWouldAddEndPuncttrue
\mciteSetBstMidEndSepPunct{\mcitedefaultmidpunct}
{\mcitedefaultendpunct}{\mcitedefaultseppunct}\relax
\EndOfBibitem
\bibitem[Behler and Parrinello(2007)Behler, and Parrinello]{behlerGeneralizedNeuralNetworkRepresentation2007}
Behler,~J.; Parrinello,~M. Generalized {{Neural-Network Representation}} of {{High-Dimensional Potential-Energy Surfaces}}. \emph{Phys. Rev. Lett.} \textbf{2007}, \emph{98}, 146401\relax
\mciteBstWouldAddEndPuncttrue
\mciteSetBstMidEndSepPunct{\mcitedefaultmidpunct}
{\mcitedefaultendpunct}{\mcitedefaultseppunct}\relax
\EndOfBibitem
\bibitem[Sch{\"u}tt \latin{et~al.}(2018)Sch{\"u}tt, Sauceda, Kindermans, Tkatchenko, and M{\"u}ller]{schuttSchNetDeepLearning2018}
Sch{\"u}tt,~K.~T.; Sauceda,~H.~E.; Kindermans,~P.-J.; Tkatchenko,~A.; M{\"u}ller,~K.-R. {{SchNet}} -- {{A}} Deep Learning Architecture for Molecules and Materials. \emph{The Journal of Chemical Physics} \textbf{2018}, \emph{148}\relax
\mciteBstWouldAddEndPuncttrue
\mciteSetBstMidEndSepPunct{\mcitedefaultmidpunct}
{\mcitedefaultendpunct}{\mcitedefaultseppunct}\relax
\EndOfBibitem
\bibitem[Unke and Meuwly(2019)Unke, and Meuwly]{unkePhysNetNeuralNetwork2019}
Unke,~O.~T.; Meuwly,~M. {{PhysNet}}: {{A Neural Network}} for {{Predicting Energies}}, {{Forces}}, {{Dipole Moments}}, and {{Partial Charges}}. \emph{J. Chem. Theory Comput.} \textbf{2019}, \emph{15}, 3678--3693\relax
\mciteBstWouldAddEndPuncttrue
\mciteSetBstMidEndSepPunct{\mcitedefaultmidpunct}
{\mcitedefaultendpunct}{\mcitedefaultseppunct}\relax
\EndOfBibitem
\bibitem[Park \latin{et~al.}(2021)Park, Kornbluth, Vandermause, Wolverton, Kozinsky, and Mailoa]{parkAccurateScalableGraph2021}
Park,~C.~W.; Kornbluth,~M.; Vandermause,~J.; Wolverton,~C.; Kozinsky,~B.; Mailoa,~J.~P. Accurate and Scalable Graph Neural Network Force Field and Molecular Dynamics with Direct Force Architecture. \emph{npj Comput Mater} \textbf{2021}, \emph{7}, 73\relax
\mciteBstWouldAddEndPuncttrue
\mciteSetBstMidEndSepPunct{\mcitedefaultmidpunct}
{\mcitedefaultendpunct}{\mcitedefaultseppunct}\relax
\EndOfBibitem
\bibitem[Batzner \latin{et~al.}(2022)Batzner, Musaelian, Sun, Geiger, Mailoa, Kornbluth, Molinari, Smidt, and Kozinsky]{batznerE3equivariantGraphNeural2022}
Batzner,~S.; Musaelian,~A.; Sun,~L.; Geiger,~M.; Mailoa,~J.~P.; Kornbluth,~M.; Molinari,~N.; Smidt,~T.~E.; Kozinsky,~B. E(3)-Equivariant Graph Neural Networks for Data-Efficient and Accurate Interatomic Potentials. \emph{Nat Commun} \textbf{2022}, \emph{13}, 2453\relax
\mciteBstWouldAddEndPuncttrue
\mciteSetBstMidEndSepPunct{\mcitedefaultmidpunct}
{\mcitedefaultendpunct}{\mcitedefaultseppunct}\relax
\EndOfBibitem
\bibitem[Batatia \latin{et~al.}(2022)Batatia, Kovacs, Simm, Ortner, and Csanyi]{batatiaMACEHigherOrder2022}
Batatia,~I.; Kovacs,~D.~P.; Simm,~G.; Ortner,~C.; Csanyi,~G. {{MACE}}: {{Higher Order Equivariant Message Passing Neural Networks}} for {{Fast}} and {{Accurate Force Fields}}. Advances in {{Neural Information Processing Systems}}. 2022; pp 11423--11436\relax
\mciteBstWouldAddEndPuncttrue
\mciteSetBstMidEndSepPunct{\mcitedefaultmidpunct}
{\mcitedefaultendpunct}{\mcitedefaultseppunct}\relax
\EndOfBibitem
\bibitem[Musaelian \latin{et~al.}(2023)Musaelian, Batzner, Johansson, Sun, Owen, Kornbluth, and Kozinsky]{musaelianLearningLocalEquivariant2023c}
Musaelian,~A.; Batzner,~S.; Johansson,~A.; Sun,~L.; Owen,~C.~J.; Kornbluth,~M.; Kozinsky,~B. Learning Local Equivariant Representations for Large-Scale Atomistic Dynamics. \emph{Nature Communications} \textbf{2023}, \emph{14}, 579\relax
\mciteBstWouldAddEndPuncttrue
\mciteSetBstMidEndSepPunct{\mcitedefaultmidpunct}
{\mcitedefaultendpunct}{\mcitedefaultseppunct}\relax
\EndOfBibitem
\bibitem[Ko and Ong(2023)Ko, and Ong]{koRecentAdvancesOutstanding2023}
Ko,~T.~W.; Ong,~S.~P. Recent Advances and Outstanding Challenges for Machine Learning Interatomic Potentials. \emph{Nat Comput Sci} \textbf{2023}, \emph{3}, 998--1000\relax
\mciteBstWouldAddEndPuncttrue
\mciteSetBstMidEndSepPunct{\mcitedefaultmidpunct}
{\mcitedefaultendpunct}{\mcitedefaultseppunct}\relax
\EndOfBibitem
\bibitem[Li \latin{et~al.}(2022)Li, Meidani, Yadav, and Barati~Farimani]{liGraphNeuralNetworks2022c}
Li,~Z.; Meidani,~K.; Yadav,~P.; Barati~Farimani,~A. Graph Neural Networks Accelerated Molecular Dynamics. \emph{The Journal of Chemical Physics} \textbf{2022}, \emph{156}, 144103\relax
\mciteBstWouldAddEndPuncttrue
\mciteSetBstMidEndSepPunct{\mcitedefaultmidpunct}
{\mcitedefaultendpunct}{\mcitedefaultseppunct}\relax
\EndOfBibitem
\bibitem[Fu \latin{et~al.}(2022)Fu, Xie, Rebello, Olsen, and Jaakkola]{fuSimulateTimeintegratedCoarsegrained2022c}
Fu,~X.; Xie,~T.; Rebello,~N.~J.; Olsen,~B.~D.; Jaakkola,~T. Simulate {{Time-integrated Coarse-grained Molecular Dynamics}} with {{Multi-Scale Graph Networks}}. \textbf{2022}, \relax
\mciteBstWouldAddEndPunctfalse
\mciteSetBstMidEndSepPunct{\mcitedefaultmidpunct}
{}{\mcitedefaultseppunct}\relax
\EndOfBibitem
\bibitem[Li \latin{et~al.}(2023)Li, Persaud, Choudhary, DeCost, Greenwood, and {Hattrick-Simpers}]{liExploitingRedundancyLarge2023c}
Li,~K.; Persaud,~D.; Choudhary,~K.; DeCost,~B.; Greenwood,~M.; {Hattrick-Simpers},~J. Exploiting Redundancy in Large Materials Datasets for Efficient Machine Learning with Less Data. \emph{Nature Communications} \textbf{2023}, \emph{14}, 7283\relax
\mciteBstWouldAddEndPuncttrue
\mciteSetBstMidEndSepPunct{\mcitedefaultmidpunct}
{\mcitedefaultendpunct}{\mcitedefaultseppunct}\relax
\EndOfBibitem
\bibitem[Nadkarni \latin{et~al.}(2023)Nadkarni, Wu, and Aluru]{nadkarniDataDrivenApproachCoarseGraining2023a}
Nadkarni,~I.; Wu,~H.; Aluru,~N.~R. Data-{{Driven Approach}} to {{Coarse-Graining Simple Liquids}} in {{Confinement}}. \emph{Journal of Chemical Theory and Computation} \textbf{2023}, \emph{19}, 7358--7370\relax
\mciteBstWouldAddEndPuncttrue
\mciteSetBstMidEndSepPunct{\mcitedefaultmidpunct}
{\mcitedefaultendpunct}{\mcitedefaultseppunct}\relax
\EndOfBibitem
\bibitem[Wu and Aluru(2022)Wu, and Aluru]{wuDeepLearningbasedQuasicontinuum2022}
Wu,~H.; Aluru,~N.~R. Deep Learning-Based Quasi-Continuum Theory for Structure of Confined Fluids. \emph{The Journal of Chemical Physics} \textbf{2022}, \emph{157}, 084121\relax
\mciteBstWouldAddEndPuncttrue
\mciteSetBstMidEndSepPunct{\mcitedefaultmidpunct}
{\mcitedefaultendpunct}{\mcitedefaultseppunct}\relax
\EndOfBibitem
\bibitem[Deng and Cahill(1993)Deng, and Cahill]{dengAdaptiveGaussianFilter1993}
Deng,~G.; Cahill,~L. An Adaptive {{Gaussian}} Filter for Noise Reduction and Edge Detection. 1993 {{IEEE Conference Record Nuclear Science Symposium}} and {{Medical Imaging Conference}}. San Francisco, CA, USA, 1993; pp 1615--1619\relax
\mciteBstWouldAddEndPuncttrue
\mciteSetBstMidEndSepPunct{\mcitedefaultmidpunct}
{\mcitedefaultendpunct}{\mcitedefaultseppunct}\relax
\EndOfBibitem
\bibitem[Nichol and Dhariwal(2021)Nichol, and Dhariwal]{pmlr-v139-nichol21a}
Nichol,~A.~Q.; Dhariwal,~P. Improved Denoising Diffusion Probabilistic Models. Proceedings of the 38th International Conference on Machine Learning. 2021; pp 8162--8171\relax
\mciteBstWouldAddEndPuncttrue
\mciteSetBstMidEndSepPunct{\mcitedefaultmidpunct}
{\mcitedefaultendpunct}{\mcitedefaultseppunct}\relax
\EndOfBibitem
\bibitem[Ho \latin{et~al.}(2020)Ho, Jain, and Abbeel]{hoDenoisingDiffusionProbabilistic2020c}
Ho,~J.; Jain,~A.; Abbeel,~P. Denoising {{Diffusion Probabilistic Models}}. 2020\relax
\mciteBstWouldAddEndPuncttrue
\mciteSetBstMidEndSepPunct{\mcitedefaultmidpunct}
{\mcitedefaultendpunct}{\mcitedefaultseppunct}\relax
\EndOfBibitem
\bibitem[{Sohl-Dickstein} \latin{et~al.}(2015){Sohl-Dickstein}, Weiss, Maheswaranathan, and Ganguli]{sohl-dicksteinDeepUnsupervisedLearning2015}
{Sohl-Dickstein},~J.; Weiss,~E.~A.; Maheswaranathan,~N.; Ganguli,~S. Deep {{Unsupervised Learning}} Using {{Nonequilibrium Thermodynamics}}. 2015\relax
\mciteBstWouldAddEndPuncttrue
\mciteSetBstMidEndSepPunct{\mcitedefaultmidpunct}
{\mcitedefaultendpunct}{\mcitedefaultseppunct}\relax
\EndOfBibitem
\bibitem[Teorell(1953)]{teorellTransportProcessesElectrical1953}
Teorell,~T. Transport {{Processes}} and {{Electrical Phenomena}} in {{Ionic Membranes}}. \emph{Progress in Biophysics and Biophysical Chemistry} \textbf{1953}, \emph{3}, 305--369\relax
\mciteBstWouldAddEndPuncttrue
\mciteSetBstMidEndSepPunct{\mcitedefaultmidpunct}
{\mcitedefaultendpunct}{\mcitedefaultseppunct}\relax
\EndOfBibitem
\bibitem[Thompson \latin{et~al.}(2022)Thompson, Aktulga, Berger, Bolintineanu, Brown, Crozier, In~'T~Veld, Kohlmeyer, Moore, Nguyen, Shan, Stevens, Tranchida, Trott, and Plimpton]{thompsonLAMMPSFlexibleSimulation2022}
Thompson,~A.~P.; Aktulga,~H.~M.; Berger,~R.; Bolintineanu,~D.~S.; Brown,~W.~M.; Crozier,~P.~S.; In~'T~Veld,~P.~J.; Kohlmeyer,~A.; Moore,~S.~G.; Nguyen,~T.~D.; Shan,~R.; Stevens,~M.~J.; Tranchida,~J.; Trott,~C.; Plimpton,~S.~J. {{LAMMPS}} - a Flexible Simulation Tool for Particle-Based Materials Modeling at the Atomic, Meso, and Continuum Scales. \emph{Computer Physics Communications} \textbf{2022}, \emph{271}, 108171\relax
\mciteBstWouldAddEndPuncttrue
\mciteSetBstMidEndSepPunct{\mcitedefaultmidpunct}
{\mcitedefaultendpunct}{\mcitedefaultseppunct}\relax
\EndOfBibitem
\bibitem[Fu \latin{et~al.}(2022)Fu, Wu, Wang, Xie, Keten, {Gomez-Bombarelli}, and Jaakkola]{fuForcesAreNot2022}
Fu,~X.; Wu,~Z.; Wang,~W.; Xie,~T.; Keten,~S.; {Gomez-Bombarelli},~R.; Jaakkola,~T. Forces Are Not {{Enough}}: {{Benchmark}} and {{Critical Evaluation}} for {{Machine Learning Force Fields}} with {{Molecular Simulations}}. \textbf{2022}, \relax
\mciteBstWouldAddEndPunctfalse
\mciteSetBstMidEndSepPunct{\mcitedefaultmidpunct}
{}{\mcitedefaultseppunct}\relax
\EndOfBibitem
\bibitem[Zhang \latin{et~al.}(2020)Zhang, Wang, Chen, Zeng, Zhang, Wang, and E]{zhangDPGENConcurrentLearning2020}
Zhang,~Y.; Wang,~H.; Chen,~W.; Zeng,~J.; Zhang,~L.; Wang,~H.; E,~W. {{DP-GEN}}: {{A}} Concurrent Learning Platform for the Generation of Reliable Deep Learning Based Potential Energy Models. \emph{Computer Physics Communications} \textbf{2020}, \emph{253}, 107206\relax
\mciteBstWouldAddEndPuncttrue
\mciteSetBstMidEndSepPunct{\mcitedefaultmidpunct}
{\mcitedefaultendpunct}{\mcitedefaultseppunct}\relax
\EndOfBibitem
\bibitem[Raja \latin{et~al.}(2024)Raja, Amin, Pedregosa, and Krishnapriyan]{rajaStabilityAwareTrainingMachine2024}
Raja,~S.; Amin,~I.; Pedregosa,~F.; Krishnapriyan,~A.~S. Stability-{{Aware Training}} of {{Machine Learning Force Fields}} with {{Differentiable Boltzmann Estimators}}. \textbf{2024}, \relax
\mciteBstWouldAddEndPunctfalse
\mciteSetBstMidEndSepPunct{\mcitedefaultmidpunct}
{}{\mcitedefaultseppunct}\relax
\EndOfBibitem
\bibitem[Silverman(2018)]{silvermanDensityEstimationStatistics2018}
Silverman,~B. \emph{Density {{Estimation}} for {{Statistics}} and {{Data Analysis}}}, 1st ed.; Routledge, 2018\relax
\mciteBstWouldAddEndPuncttrue
\mciteSetBstMidEndSepPunct{\mcitedefaultmidpunct}
{\mcitedefaultendpunct}{\mcitedefaultseppunct}\relax
\EndOfBibitem
\bibitem[Rizzi \latin{et~al.}(2012)Rizzi, Najm, Debusschere, Sargsyan, Salloum, Adalsteinsson, and Knio]{rizziUncertaintyQuantificationMD2012}
Rizzi,~F.; Najm,~H.~N.; Debusschere,~B.~J.; Sargsyan,~K.; Salloum,~M.; Adalsteinsson,~H.; Knio,~O.~M. Uncertainty {{Quantification}} in {{MD Simulations}}. {{Part I}}: {{Forward Propagation}}. \emph{Multiscale Model. Simul.} \textbf{2012}, \emph{10}, 1428--1459\relax
\mciteBstWouldAddEndPuncttrue
\mciteSetBstMidEndSepPunct{\mcitedefaultmidpunct}
{\mcitedefaultendpunct}{\mcitedefaultseppunct}\relax
\EndOfBibitem
\bibitem[Wand and Jones(1994)Wand, and Jones]{wandKernelSmoothing1994}
Wand,~M.; Jones,~M. \emph{Kernel {{Smoothing}}}, 0th ed.; {Chapman and Hall/CRC}, 1994\relax
\mciteBstWouldAddEndPuncttrue
\mciteSetBstMidEndSepPunct{\mcitedefaultmidpunct}
{\mcitedefaultendpunct}{\mcitedefaultseppunct}\relax
\EndOfBibitem
\bibitem[Opsomer \latin{et~al.}(2001)Opsomer, Wang, and Yang]{opsomerNonparametricRegressinCorrelated2001}
Opsomer,~J.; Wang,~Y.; Yang,~Y. Nonparametric {{Regressin}} with {{Correlated Errors}}. \emph{Statist. Sci.} \textbf{2001}, \emph{16}\relax
\mciteBstWouldAddEndPuncttrue
\mciteSetBstMidEndSepPunct{\mcitedefaultmidpunct}
{\mcitedefaultendpunct}{\mcitedefaultseppunct}\relax
\EndOfBibitem
\bibitem[Rosenblatt(1956)]{rosenblattRemarksNonparametricEstimates1956}
Rosenblatt,~M. Remarks on {{Some Nonparametric Estimates}} of a {{Density Function}}. \emph{Ann. Math. Statist.} \textbf{1956}, \emph{27}, 832--837\relax
\mciteBstWouldAddEndPuncttrue
\mciteSetBstMidEndSepPunct{\mcitedefaultmidpunct}
{\mcitedefaultendpunct}{\mcitedefaultseppunct}\relax
\EndOfBibitem
\bibitem[Fix and Hodges(1989)Fix, and Hodges]{fixDiscriminatoryAnalysisNonparametric1989}
Fix,~E.; Hodges,~J.~L. Discriminatory {{Analysis}}. {{Nonparametric Discrimination}}: {{Consistency Properties}}. \emph{International Statistical Review / Revue Internationale de Statistique} \textbf{1989}, \emph{57}, 238\relax
\mciteBstWouldAddEndPuncttrue
\mciteSetBstMidEndSepPunct{\mcitedefaultmidpunct}
{\mcitedefaultendpunct}{\mcitedefaultseppunct}\relax
\EndOfBibitem
\bibitem[Oppenheim and Verghese(2016)Oppenheim, and Verghese]{oppenheimSignalsSystemsInference2016}
Oppenheim,~A.~V.; Verghese,~G.~C. \emph{Signals, Systems \& Inference}; Pearson: Boston, 2016\relax
\mciteBstWouldAddEndPuncttrue
\mciteSetBstMidEndSepPunct{\mcitedefaultmidpunct}
{\mcitedefaultendpunct}{\mcitedefaultseppunct}\relax
\EndOfBibitem
\bibitem[Wiener(1949)]{wienerExtrapolationInterpolationSmoothing1949}
Wiener,~N. \emph{Extrapolation, {{Interpolation}}, and {{Smoothing}} of {{Stationary Time Series}}: {{With Engineering Applications}}}; The MIT Press, 1949\relax
\mciteBstWouldAddEndPuncttrue
\mciteSetBstMidEndSepPunct{\mcitedefaultmidpunct}
{\mcitedefaultendpunct}{\mcitedefaultseppunct}\relax
\EndOfBibitem
\bibitem[Savitzky and Golay(1964)Savitzky, and Golay]{savitzkySmoothingDifferentiationData1964}
Savitzky,~{\relax Abraham}.; Golay,~M. J.~E. Smoothing and {{Differentiation}} of {{Data}} by {{Simplified Least Squares Procedures}}. \emph{Anal. Chem.} \textbf{1964}, \emph{36}, 1627--1639\relax
\mciteBstWouldAddEndPuncttrue
\mciteSetBstMidEndSepPunct{\mcitedefaultmidpunct}
{\mcitedefaultendpunct}{\mcitedefaultseppunct}\relax
\EndOfBibitem
\bibitem[Num(1992)]{NumericalRecipesArt1992}
\emph{Numerical Recipes in {{C}}: The Art of Scientific Computing}, 2nd ed.; Cambridge university press: Cambridge, 1992\relax
\mciteBstWouldAddEndPuncttrue
\mciteSetBstMidEndSepPunct{\mcitedefaultmidpunct}
{\mcitedefaultendpunct}{\mcitedefaultseppunct}\relax
\EndOfBibitem
\bibitem[Breiman \latin{et~al.}(1977)Breiman, Meisel, and Purcell]{breimanVariableKernelEstimates1977}
Breiman,~L.; Meisel,~W.; Purcell,~E. Variable {{Kernel Estimates}} of {{Multivariate Densities}}. \emph{Technometrics} \textbf{1977}, \emph{19}, 135--144\relax
\mciteBstWouldAddEndPuncttrue
\mciteSetBstMidEndSepPunct{\mcitedefaultmidpunct}
{\mcitedefaultendpunct}{\mcitedefaultseppunct}\relax
\EndOfBibitem
\bibitem[Flyvbjerg and Petersen(1989)Flyvbjerg, and Petersen]{flyvbjergErrorEstimatesAverages1989}
Flyvbjerg,~H.; Petersen,~H.~G. Error Estimates on Averages of Correlated Data. \emph{The Journal of Chemical Physics} \textbf{1989}, \emph{91}, 461--466\relax
\mciteBstWouldAddEndPuncttrue
\mciteSetBstMidEndSepPunct{\mcitedefaultmidpunct}
{\mcitedefaultendpunct}{\mcitedefaultseppunct}\relax
\EndOfBibitem
\bibitem[Hansen(1992)]{hansenAnalysisDiscreteIllPosed1992}
Hansen,~P.~C. Analysis of {{Discrete Ill-Posed Problems}} by {{Means}} of the {{L-Curve}}. \emph{SIAM Rev.} \textbf{1992}, \emph{34}, 561--580\relax
\mciteBstWouldAddEndPuncttrue
\mciteSetBstMidEndSepPunct{\mcitedefaultmidpunct}
{\mcitedefaultendpunct}{\mcitedefaultseppunct}\relax
\EndOfBibitem
\bibitem[Hengartner and {Matzner-L{\o}ber}(2009)Hengartner, and {Matzner-L{\o}ber}]{hengartnerAsymptoticUnbiasedDensity2009}
Hengartner,~N.~W.; {Matzner-L{\o}ber},~{\'E}. Asymptotic Unbiased Density Estimators. \emph{ESAIM: PS} \textbf{2009}, \emph{13}, 1--14\relax
\mciteBstWouldAddEndPuncttrue
\mciteSetBstMidEndSepPunct{\mcitedefaultmidpunct}
{\mcitedefaultendpunct}{\mcitedefaultseppunct}\relax
\EndOfBibitem
\bibitem[Klein(2021)]{kleinUncertaintyPropagationGaussian2021}
Klein,~R. Uncertainty {{Propagation}} in ({{Gaussian}}) {{Convolution}}. \emph{Res. Notes AAS} \textbf{2021}, \emph{5}, 39\relax
\mciteBstWouldAddEndPuncttrue
\mciteSetBstMidEndSepPunct{\mcitedefaultmidpunct}
{\mcitedefaultendpunct}{\mcitedefaultseppunct}\relax
\EndOfBibitem
\bibitem[Nadkarni \latin{et~al.}(2025)Nadkarni, Mart{\'i}nez~Cordeiro, and Aluru]{nadkarniMolecularDenoisingUsing2025}
Nadkarni,~I.; Mart{\'i}nez~Cordeiro,~J.; Aluru,~N.~R. Molecular {{Denoising Using Diffusion Models}} with {{Physics-Informed Priors}}. \emph{J. Phys. Chem. Lett.} \textbf{2025}, \emph{16}, 3078--3085\relax
\mciteBstWouldAddEndPuncttrue
\mciteSetBstMidEndSepPunct{\mcitedefaultmidpunct}
{\mcitedefaultendpunct}{\mcitedefaultseppunct}\relax
\EndOfBibitem
\bibitem[Saharia \latin{et~al.}(2022)Saharia, Ho, Chan, Salimans, Fleet, and Norouzi]{sahariaImageSuperResolutionIterative2022}
Saharia,~C.; Ho,~J.; Chan,~W.; Salimans,~T.; Fleet,~D.~J.; Norouzi,~M. Image {{Super-Resolution Via Iterative Refinement}}. \emph{IEEE Trans. Pattern Anal. Mach. Intell.} \textbf{2022}, 1--14\relax
\mciteBstWouldAddEndPuncttrue
\mciteSetBstMidEndSepPunct{\mcitedefaultmidpunct}
{\mcitedefaultendpunct}{\mcitedefaultseppunct}\relax
\EndOfBibitem
\bibitem[Li \latin{et~al.}(2024)Li, Ditzler, Roveda, and Li]{liDeScoDECGDeepScoreBased2024}
Li,~H.; Ditzler,~G.; Roveda,~J.; Li,~A. {{DeScoD-ECG}}: {{Deep Score-Based Diffusion Model}} for {{ECG Baseline Wander}} and {{Noise Removal}}. \emph{IEEE J. Biomed. Health Inform.} \textbf{2024}, \emph{28}, 5081--5091\relax
\mciteBstWouldAddEndPuncttrue
\mciteSetBstMidEndSepPunct{\mcitedefaultmidpunct}
{\mcitedefaultendpunct}{\mcitedefaultseppunct}\relax
\EndOfBibitem
\bibitem[Virtanen \latin{et~al.}(2020)Virtanen, Gommers, Oliphant, Haberland, Reddy, Cournapeau, Burovski, Peterson, Weckesser, Bright, {van der Walt}, Brett, Wilson, Millman, Mayorov, Nelson, Jones, Kern, Larson, Carey, Polat, Feng, Moore, {VanderPlas}, Laxalde, Perktold, Cimrman, Henriksen, Quintero, Harris, Archibald, Ribeiro, Pedregosa, {van Mulbregt}, and {SciPy 1.0 Contributors}]{2020SciPy-NMeth}
Virtanen,~P. \latin{et~al.}  {{SciPy} 1.0: Fundamental Algorithms for Scientific Computing in Python}. \emph{Nature Methods} \textbf{2020}, \emph{17}, 261--272\relax
\mciteBstWouldAddEndPuncttrue
\mciteSetBstMidEndSepPunct{\mcitedefaultmidpunct}
{\mcitedefaultendpunct}{\mcitedefaultseppunct}\relax
\EndOfBibitem
\bibitem[Sanghi and Aluru(2014)Sanghi, and Aluru]{sanghiThermalNoiseConfined2014}
Sanghi,~T.; Aluru,~N.~R. Thermal Noise in Confined Fluids. \emph{The Journal of Chemical Physics} \textbf{2014}, \emph{141}, 174707\relax
\mciteBstWouldAddEndPuncttrue
\mciteSetBstMidEndSepPunct{\mcitedefaultmidpunct}
{\mcitedefaultendpunct}{\mcitedefaultseppunct}\relax
\EndOfBibitem
\bibitem[Kingma and Ba(2014)Kingma, and Ba]{kingmaAdamMethodStochastic2014}
Kingma,~D.~P.; Ba,~J. Adam: {{A Method}} for {{Stochastic Optimization}}. 2014\relax
\mciteBstWouldAddEndPuncttrue
\mciteSetBstMidEndSepPunct{\mcitedefaultmidpunct}
{\mcitedefaultendpunct}{\mcitedefaultseppunct}\relax
\EndOfBibitem
\bibitem[Kresse and Furthm{\"u}ller(1996)Kresse, and Furthm{\"u}ller]{kresseEfficiencyAbinitioTotal1996}
Kresse,~G.; Furthm{\"u}ller,~J. Efficiency of Ab-Initio Total Energy Calculations for Metals and Semiconductors Using a Plane-Wave Basis Set. \emph{Computational Materials Science} \textbf{1996}, \emph{6}, 15--50\relax
\mciteBstWouldAddEndPuncttrue
\mciteSetBstMidEndSepPunct{\mcitedefaultmidpunct}
{\mcitedefaultendpunct}{\mcitedefaultseppunct}\relax
\EndOfBibitem
\bibitem[Kresse and Furthm{\"u}ller(1996)Kresse, and Furthm{\"u}ller]{kresseEfficientIterativeSchemes1996}
Kresse,~G.; Furthm{\"u}ller,~J. Efficient Iterative Schemes for {\emph{Ab Initio}} Total-Energy Calculations Using a Plane-Wave Basis Set. \emph{Phys. Rev. B} \textbf{1996}, \emph{54}, 11169--11186\relax
\mciteBstWouldAddEndPuncttrue
\mciteSetBstMidEndSepPunct{\mcitedefaultmidpunct}
{\mcitedefaultendpunct}{\mcitedefaultseppunct}\relax
\EndOfBibitem
\end{mcitethebibliography}
\clearpage
\appendix

% Reset counters at the start of the appendix
\renewcommand{\thesection}{\Alph{section}} 
\renewcommand{\theequation}{\thesection\arabic{equation}} 
\renewcommand{\thefigure}{\thesection\arabic{figure}}     
\renewcommand{\thetable}{\thesection\arabic{table}}      

\setcounter{section}{0}  

\setcounter{equation}{0}
\setcounter{figure}{0}
\setcounter{table}{0}

\section{Numerical Solution of the Nernst–Planck Equation}
We solve the steady-state, 1D Nernst–Planck equation. The second-order differential equation is given as:

\begin{equation}
\frac{d}{dx} \left( \frac{d\rho}{dx} - \frac{\rho}{RT} f(x) \right) = 0
\label{eq:np_steady}
\end{equation}

\noindent where:
\begin{itemize}
    \item $\rho(x)$ is the density profile,
    \item $R$ is the gas constant, $T$ is the temperature,
    \item $f(x)$ is the force profile acting on the particles.
\end{itemize}

The equation is subject to Dirichlet boundary conditions:

\begin{align}
\rho(x = 0) &= 0  \\
\rho(x = L) &= 0 
\end{align}

and the constraint that ensures the channel is in thermodynamic equilibrium with the reservoir:

\begin{equation}
\frac{1}{L} \int_0^L \rho(z)\,dz = \rho_{\text{ave}}
\end{equation}

\noindent where $\rho_{\text{ave}}$ is the average density in the channel.
\subsection*{Numerical Implementation via Shooting Method}
\label{sec:AppendixA}
To numerically solve the boundary value problem defined by the steady-state Nernst–Planck equation, we employ the shooting method. Although other numerical methods, such as finite-volume schemes or collocation-based solvers, are also viable, we adopt the shooting method due to its conceptual simplicity and ease of implementation for smooth, one-dimensional problems. The shooting method involves converting the second-order differential equation into a first-order form through a single integration step:

\begin{equation}
\frac{d\rho}{dx} - \frac{\rho f(x)}{RT} = C
\label{eq:first_order_np}
\end{equation}

Here, $C$ is an integration constant that must be determined such that the resulting solution satisfies the boundary conditions and average density constraint.

\paragraph{}Solving the Nernst–Planck equation from one boundary alone (e.g., left to right) can lead to accumulated integration bias. To mitigate this, we adopt an averaging approach, where the equation is solved forward (from $x = 0$ to $x = L$) and backward (from $x = L$ to $x = 0$), and the resulting density profiles are averaged. This strategy improves numerical stability and helps reduce artifacts introduced by local integration errors. In addition, to aid convergence and enforce the correct asymptotic behavior, we apply a simple boundary force modification scheme. Specifically, we fix the force to a constant value at either end of the channel, equal to the maximum and minimum observed values respectively, so that the solution decays to zero at the boundaries. This ensures that the resulting density profile satisfies the Dirichlet conditions and helps stabilize the shooting procedure.

\paragraph{}The unknown integration constant $C$ is calibrated by minimizing the discrepancy between the computed density profile and the constraint imposed by the average density:

\begin{equation}
\frac{1}{L} \int_0^L \rho(z)\,dz = \rho_{\text{ave}}
\end{equation}

This is done by iteratively adjusting $C$ using a bounded scalar minimization routine until the constraint is satisfied to within numerical tolerance.

\paragraph{Code Implementation.} 
The numerical solution is implemented in \texttt{Python}, using the \texttt{solve\_ivp} function from the \texttt{SciPy} \cite{2020SciPy-NMeth} library for integrating the first-order differential equation, and \texttt{minimize\_scalar} from the same library to optimize the integration constant $C$ under the average density constraint.

\paragraph{Robustness of the solver to noise.}
To evaluate the robustness of the numerical Nernst–Planck solver, we conduct a systematic noise sensitivity analysis by introducing perturbations into the input force profile. Let $f_{\mathrm{clean}}(x)$ denote the original (smooth) force profile, evaluated at discrete spatial bins $\{x_i\}_{i=1}^N$. Noise is added to generate perturbed profiles $f_{\mathrm{noisy}}(x_i)$, and the corresponding density profiles $\rho(x)$ are obtained using the solver. The deviation from the ground-truth density $\rho_{\mathrm{clean}}(x)$ is quantified using the root mean squared error (RMSE), averaged across 5 samples.

Three representative noise models are tested:

\begin{itemize}
  \item \textbf{White noise:} Independent Gaussian perturbations are added to each bin,
  \[
  f_{\mathrm{noisy}}(x_i) = f_{\mathrm{clean}}(x_i) + \epsilon F \cdot \xi_i, \quad \xi_i \sim \mathcal{N}(0, 1)
  \]
  where $F = \max_i |f_{\mathrm{clean}}(x_i)|$ is the force scale, and $\epsilon$ controls the relative noise magnitude.

  \item \textbf{Colored noise:} Spatially correlated noise is generated by convolving white noise with a Gaussian kernel:
  \[
  \eta_i = \sum_{j} g_j \xi_{i-j}, \quad g_j = \frac{1}{Z} \exp\left(-\frac{1}{2} \left(\frac{j}{\sigma}\right)^2 \right)
  \]
  where $\sigma$ controls the spatial correlation length and $Z$ normalizes the kernel. The noise is rescaled to have standard deviation $\epsilon F$, and applied as
  \[
  f_{\mathrm{noisy}}(x_i) = f_{\mathrm{clean}}(x_i) + \epsilon F \cdot \frac{\eta_i}{\sqrt{\operatorname{Var}[\eta]}}
  \]

  \item \textbf{Systematic noise:} Low-frequency trends are added to model systematic drift:
  \[
  f_{\mathrm{noisy}}(x_i) = f_{\mathrm{clean}}(x_i) + \delta_0 + \delta_1 \cdot i + \zeta_i
  \]
  where $\delta_0 \sim \mathcal{U}(-\epsilon F, \epsilon F)$ (offset), $\delta_1 \sim \mathcal{U}(-\epsilon F/N, \epsilon F/N)$ (slope), and $\zeta_i \sim \mathcal{N}(0, (0.2\epsilon F)^2)$ represents small random fluctuations.

\end{itemize}

For each noise type and magnitude $\epsilon$, multiple realizations are generated, and the average and standard deviation of RMSE are reported in Figure 6. 
\begin{figure}[H]
    \centering
    \includegraphics[width=0.8\linewidth]{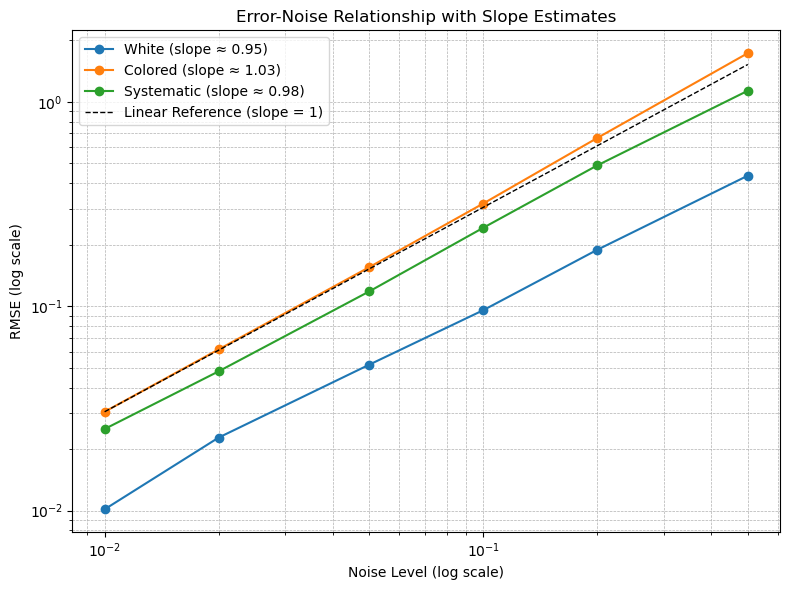}
    \caption{RMSE in predicted density profiles as a function of noise level introduced in the force profile. Three noise models are considered: (1) white noise (uncorrelated Gaussian noise), (2) colored noise (spatially correlated via Gaussian convolution), and (3) systematic noise (low-frequency drift with random offset and slope). Each curve shows the root mean squared error (RMSE) averaged over 5 random samples, with slopes computed in log-log space. A dashed black line indicates the linear reference trend ($\propto \epsilon$)}
    \label{fig:noise_robustness}
\end{figure}
\paragraph{}The log-log plot (Figure~\ref{fig:noise_robustness}) shows that the error scales approximately linearly with noise level (slope $\approx 1$) for all three noise types, indicating that the solver is both robust and predictable. 
\clearpage
\section{Neural Network Potential Training Details}
\label{sec:AppendixB}
\addcontentsline{toc}{section}{Neural Network Potential Training Details}

We provide additional implementation details related to training the neural network potential used in the molecular dynamics simulations. 
\paragraph{Neural Network Architecture.}

The input to the network is defined by atomic positions $X = \{q_1, q_2, \dots, q_n\}$, where $q_i \in \mathbb{R}^3$ corresponds to the Cartesian coordinates of atom $i$, and atomic species $Z = \{z_1, z_2, \dots, z_n\}$, where $z_i \in \{0,1,2\}$ encodes Carbon (C), Oxygen (O), and Hydrogen (H), respectively.

Atomic neighborhood graphs are constructed using a 6.0~\AA~cutoff radius. For flexible distance encoding, radial basis functions are represented by trainable Bessel functions with a polynomial cutoff of 6.0~\AA. For each atomic pair, 32-dimensional embeddings are generated via a single-layer multilayer perceptron (MLP). These embeddings are passed through a three-layer MLP that expands representations from 32 to 64 to 128 dimensions. The 128-dimensional representation is then processed by two Allegro layers that preserve $\mathcal{O}(3)$ equivariance through spherical harmonics, using a maximum rotation order of $\ell_\text{max} = 2$.

Following the Allegro layers, another three-layer MLP with hidden dimensions [128, 128, 128] and SiLU activation is used. A residual connection is applied between the final MLP output and the Allegro layers. For energy prediction, an Edge Energy MLP with a 32-node hidden layer transforms the pairwise features into per-edge energy contributions. These are aggregated into per-atom energies, which sum to give the total energy $E$ of the system. Forces are computed via the energy gradient: $F_i = -\frac{\partial E}{\partial q_i}$.

The model is trained using a combined loss function:

\begin{equation}
\mathcal{L}_\text{total} = 1000 \cdot \mathcal{L}_F + \mathcal{L}_E
\label{eq:loss}
\end{equation}

where $\mathcal{L}_F$ and $\mathcal{L}_E$ denote the mean squared errors in force and per-atom energy predictions, respectively. The force component is up-weighted by a factor of 1000 to prioritize force accuracy.

Training is performed on an NVIDIA A100 GPU using the Adam optimizer with an initial learning rate of $1 \times 10^{-3}$. If validation performance plateaus for 5 epochs, the learning rate is decayed by a factor of 0.8. Training is terminated once the learning rate drops below $1 \times 10^{-6}$. To improve generalization, Exponential Moving Average (EMA) with a decay factor of 0.99 is applied to model weights. The total training process took 23.04 hours in wall-clock time.

\paragraph{Training Data.}
The neural network was trained using a random 30\% subsample of AIMD configurations derived from simulations of water confined between graphene sheets at various channel widths $L$. Table~B1 summarizes the original AIMD dataset, including the number of water molecules, simulation temperature, and trajectory length for each confinement case.

\begin{table}[H]
    \centering
    \begin{tabular}{cccc}
        \toprule
        \textbf{L (\AA)} & \textbf{\# Water Molecules} & \textbf{Temperature (K)} & \textbf{AIMD Length (ps)} \\
        \midrule
        10.0\phantom{\textsuperscript{*}} & 33  & 400 & 20.0 \\
        11.0\textsuperscript{*}           & 35  & 400 & 24.0 \\
        12.0\phantom{\textsuperscript{*}} & 39  & 400 & 20.0 \\
        13.0\phantom{\textsuperscript{*}} & 46  & 400 & 17.0 \\
        14.0\phantom{\textsuperscript{*}} & 48  & 400 & 16.0 \\
        15.0\textsuperscript{*}           & 54  & 400 & 16.0 \\
        16.0\phantom{\textsuperscript{*}} & 59  & 400 & 12.0 \\
        17.0\phantom{\textsuperscript{*}} & 66  & 400 & 11.5 \\
        18.0\phantom{\textsuperscript{*}} & 71  & 400 & 8.5 \\
        19.0\phantom{\textsuperscript{*}} & 75  & 400 & 8.5 \\
        20.0\phantom{\textsuperscript{*}} & 81  & 400 & 8.5 \\
        30.0\phantom{\textsuperscript{*}} & 130 & 400 & 5.0 \\
        40.0\phantom{\textsuperscript{*}} & 188 & 400 & 3.0 \\
        \bottomrule
    \end{tabular}
    \caption{Overview of the AIMD datasets used for training and evaluation.}
    \vspace{0.5em}
    {\footnotesize \textsuperscript{*} Not included in training.}
    \label{tab:aimd_dataset}
\end{table}

\paragraph{}Below we report the training results for the neural network potential using the initial dataset augmented with data obtained from two rounds of active learning. The final set consists of 2793 AIMD samples, with an additional 300 samples reserved for validation.

\vspace{0.5em}

\begin{figure}[H]
    \centering
    \begin{minipage}{0.48\linewidth}
        \centering
        \includegraphics[width=\linewidth]{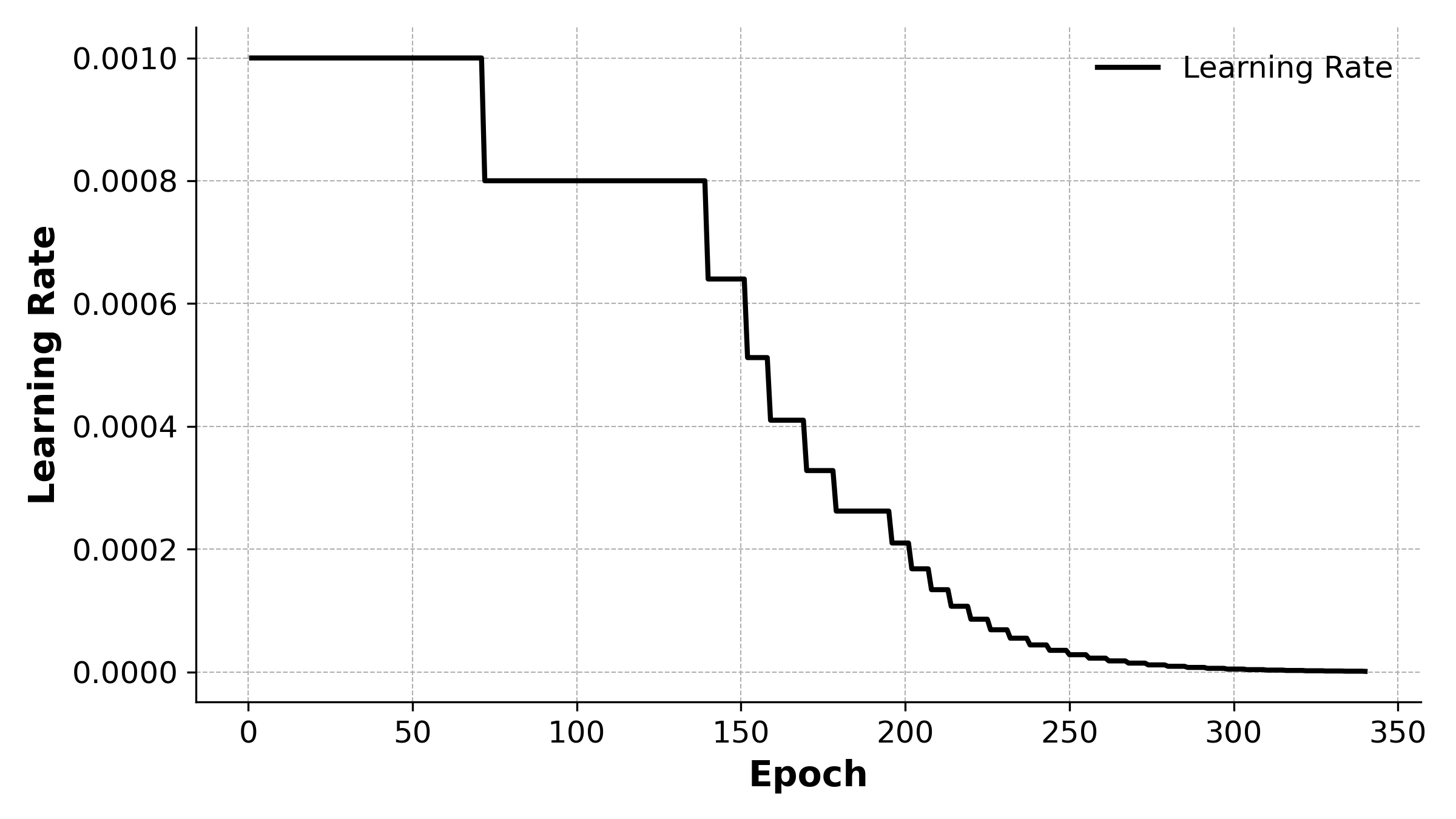}
        \caption*{(a) Learning rate schedule.}
    \end{minipage}
    \hfill
    \begin{minipage}{0.48\linewidth}
        \centering
        \includegraphics[width=\linewidth]{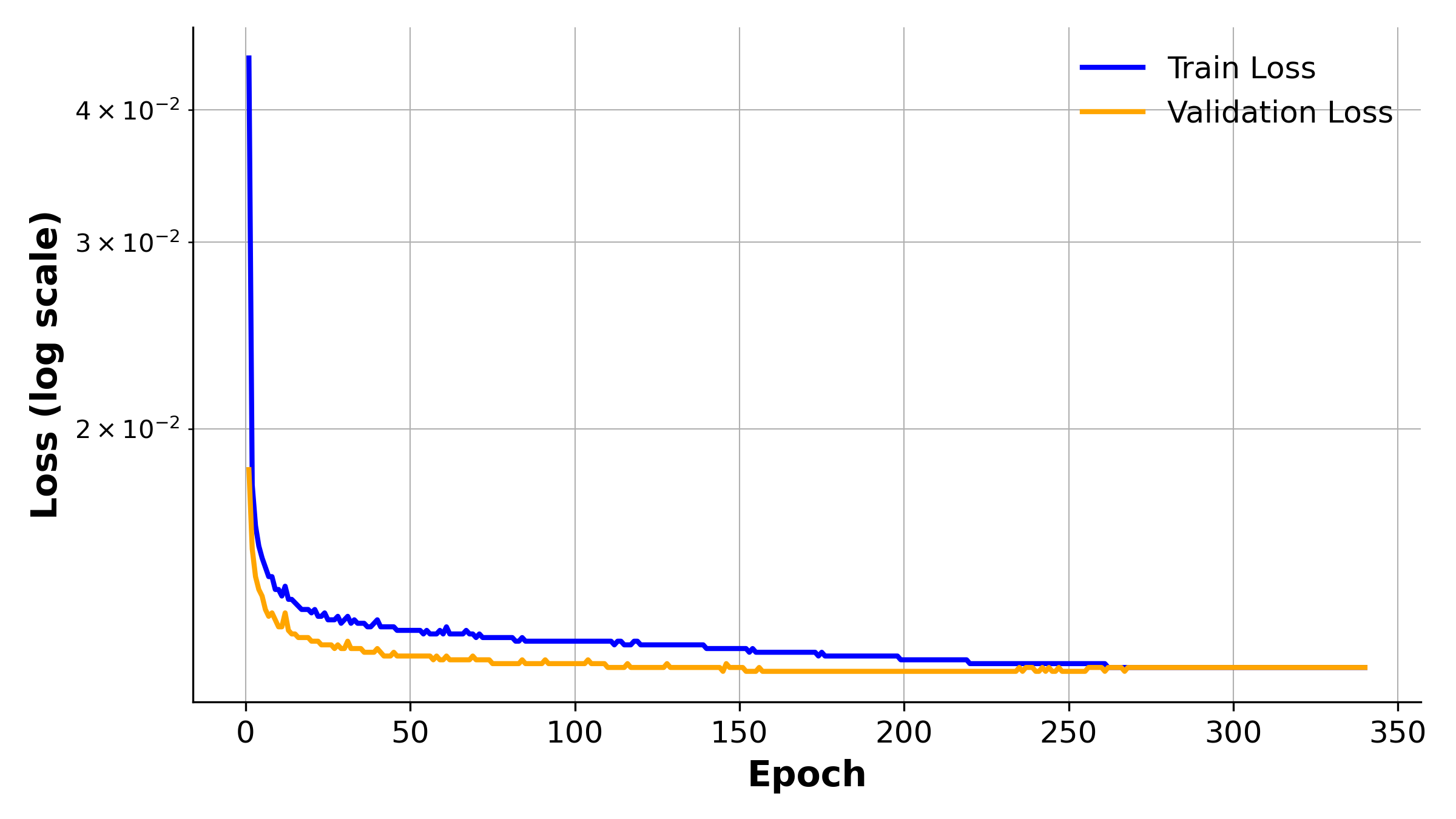}
        \caption*{(b) Training and validation loss curves.}
    \end{minipage}
    \caption{Training curves for the neural network potential.(a) shows the dynamic learning rate schedule during training.   (b) shows the training and validation loss curves across epochs.}
    \label{fig:training_curves}
\end{figure}

Figure~\ref{fig:training_curves} presents the learning rate schedule alongside the training and validation loss curves. Training is terminated at epoch 340 once the learning rate reaches the threshold value of $1\times10^{-6}$. The final training and validation losses, along with the corresponding performance metrics for force and energy prediction, are summarized in Table~\ref{tab:training_metrics}.

\begin{table}[H]
\centering
\small  % Reduce font size slightly
\setlength{\tabcolsep}{4pt}  % Reduce horizontal padding between columns
\caption{Training and validation performance metrics at the end of training.}
\label{tab:training_metrics}
\begin{tabular}{lcccccccc}
\toprule
\textbf{Split} 
& $\mathcal{L}_{\text{total}}$ 
& $\mathcal{L}_{F}$ 
& $\mathcal{L}_{E}$ 
& $\text{F}_{\text{MAE}}$ (eV/\AA) 
& $\text{F}_{\text{RMSE}}$ (eV/\AA) 
& $\text{E}_{\text{MAE}}$ (eV) 
& $\text{E}/N_{\text{MAE}}$ (eV/atom) \\
\midrule
Training   & 11.90 & 0.0119 & 0.00033  & 0.0375 & 0.0720 & 2.25 & 0.00747 \\
Validation & 11.90 & 0.0119 & 0.000332 & 0.0382 & 0.0731 & 2.34 & 0.00769 \\
\bottomrule
\end{tabular}
\end{table}

We assess the generalizability of the trained diffusion model on two channel widths, $L = 11$~\AA{} and $L = 15$~\AA{}, which were excluded from training (see Table~\ref{tab:aimd_dataset}). For each case, we compare the predicted molecular structure via MLMD against AIMD ground truth by evaluating the O--H bond length and H--O--H bond angle distributions. The MLMD distributions are obtained from 1.5~ns long simulations.

\begin{figure}[H]
    \centering
    \begin{subfigure}{0.48\textwidth}
        \centering
        \includegraphics[width=\linewidth]{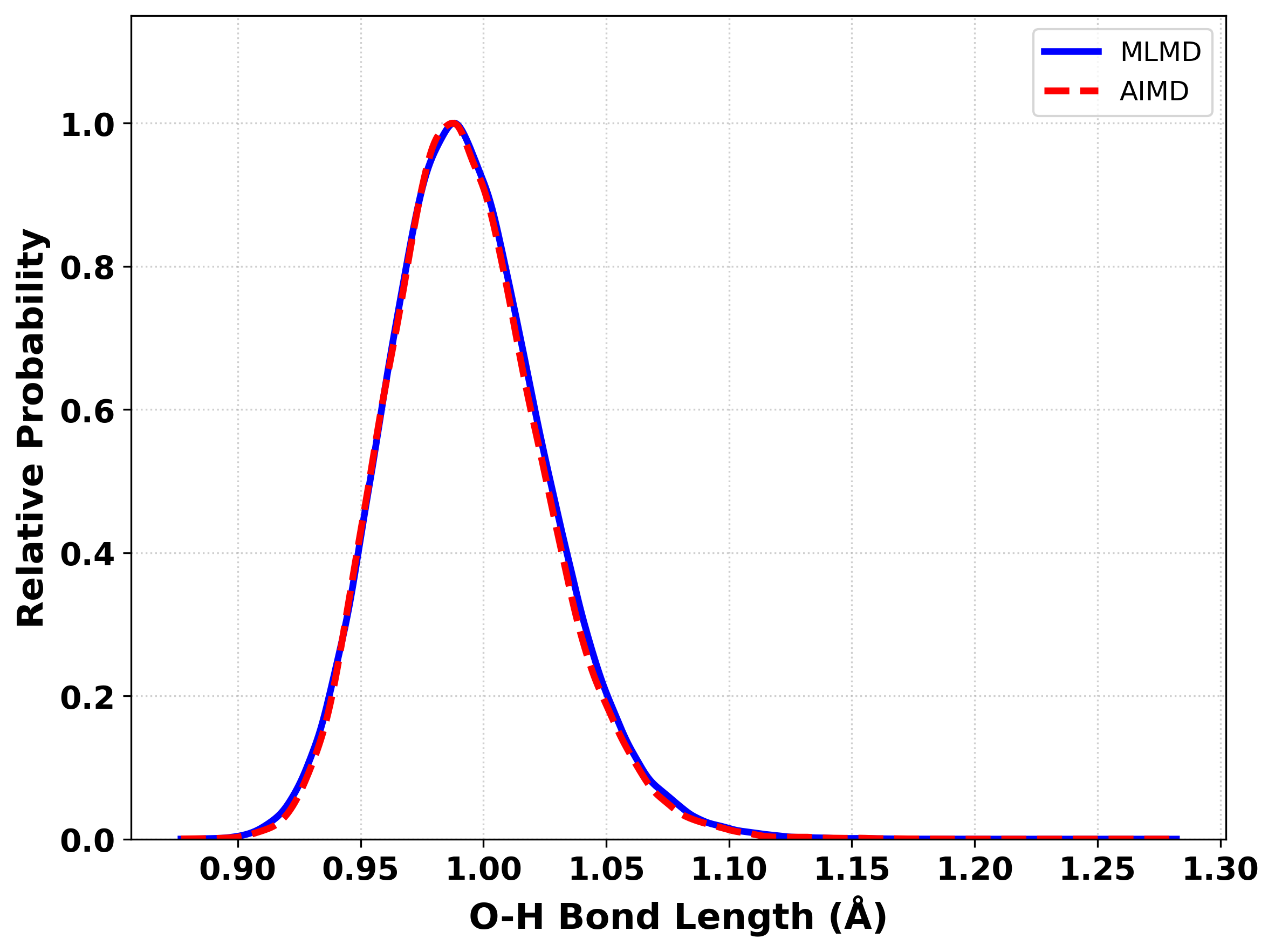}
        \caption{$L = 15$~\AA{}: O--H bond length}
    \end{subfigure}
    \hfill
    \begin{subfigure}{0.48\textwidth}
        \centering
        \includegraphics[width=\linewidth]{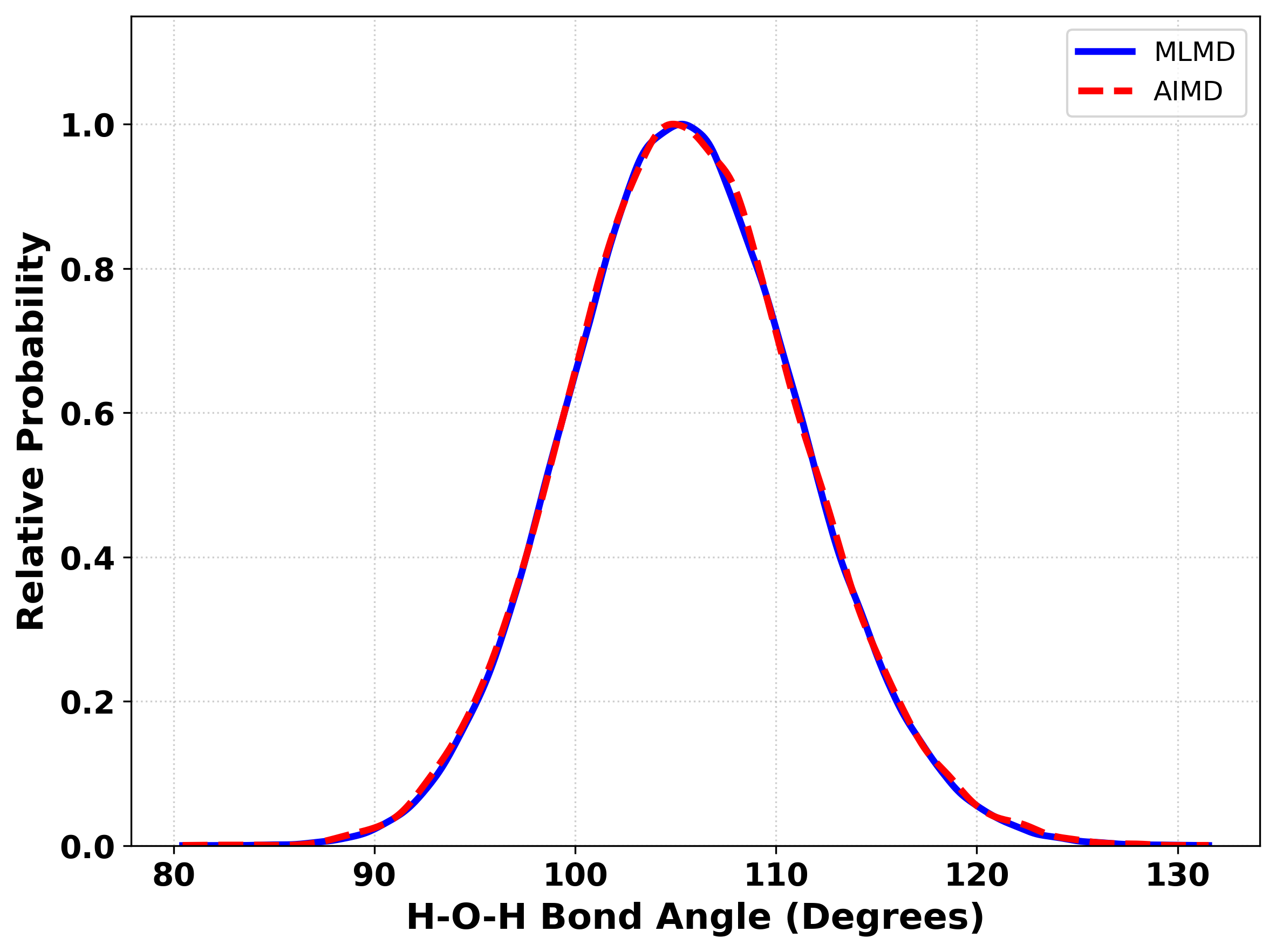}
        \caption{$L = 15$~\AA{}: H--O--H bond angle}
    \end{subfigure}

    \vspace{1em}

    \begin{subfigure}{0.48\textwidth}
        \centering
        \includegraphics[width=\linewidth]{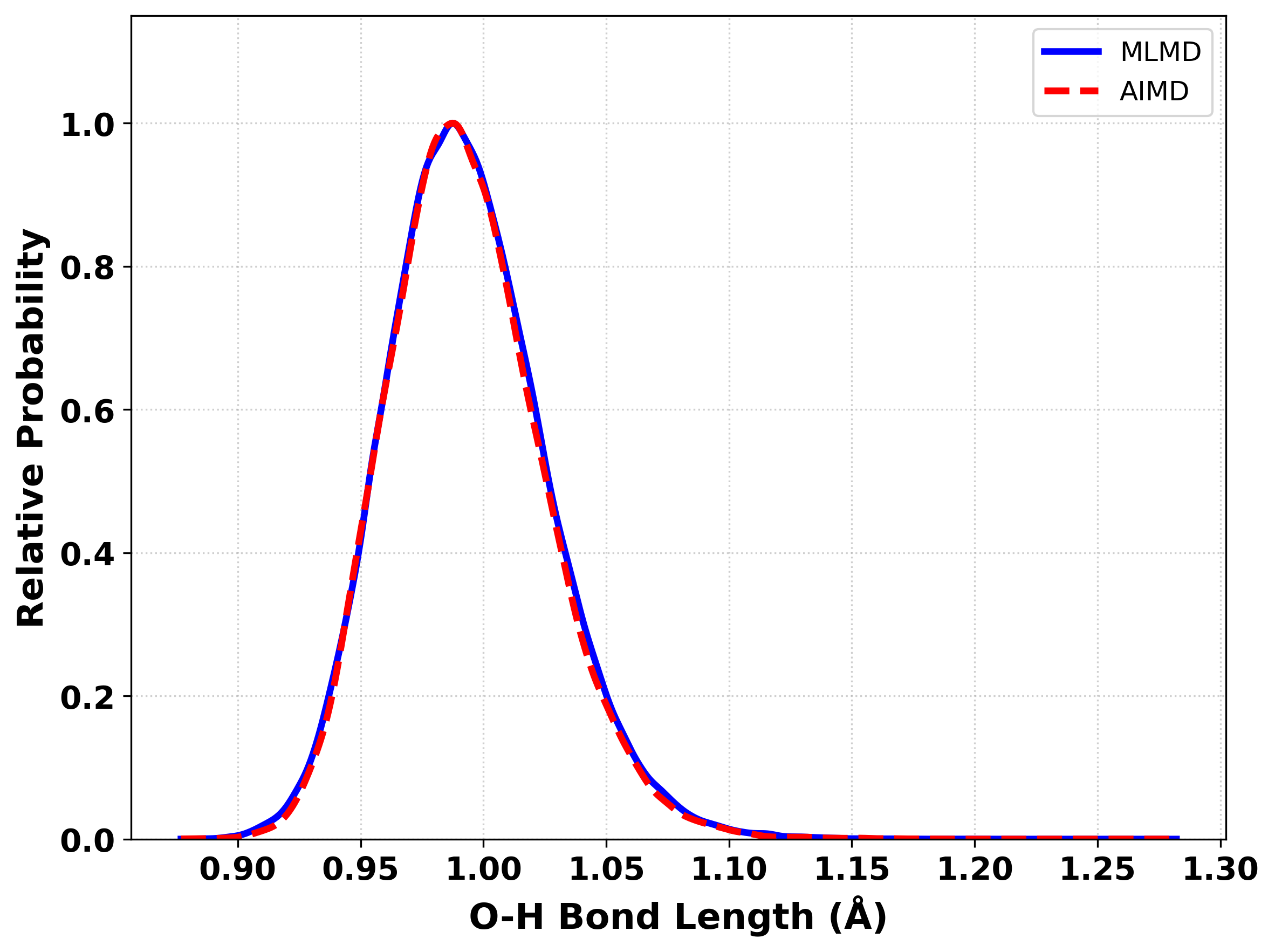}
        \caption{$L = 11$~\AA{}: O--H bond length}
    \end{subfigure}
    \hfill
    \begin{subfigure}{0.48\textwidth}
        \centering
        \includegraphics[width=\linewidth]{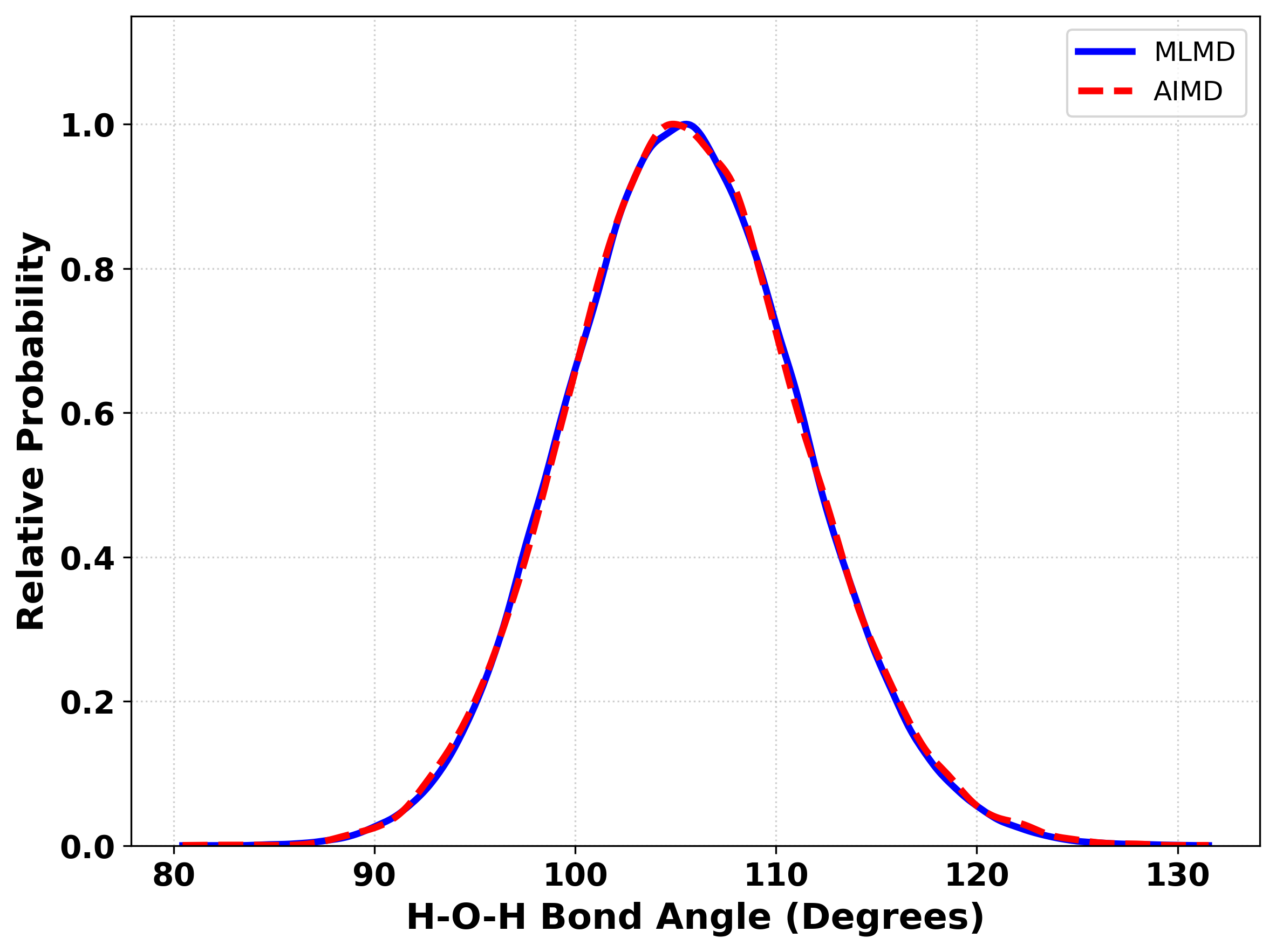}
        \caption{$L = 11$~\AA{}: H--O--H bond angle}
    \end{subfigure}
    
    \caption{Comparison of O--H bond length and H--O--H bond angle distributions between AIMD and MLMD for two unseen channel widths. MLMD distributions are computed from 1.5~ns simulations.}
    \label{fig:unseen_channel_validation}
\end{figure}
\clearpage
\section{Details for Local Force Approximation}
\label{sec:AppendixC}
\textbf{Block Averaging for Sampling Uncertainty.}
A block-averaging strategy is employed to quantify the uncertainty in observables such as force and density. The trajectory is divided into non-overlapping blocks of fixed length \( B \), chosen to exceed the autocorrelation time of the thermal noise, ensuring block-wise independence. We set \( B = 500 \) frames (2.5 ps) following \cite{sanghiThermalNoiseConfined2014}. For a given observable \( O(x) \), the average over block \( j \) is computed as:
\begin{equation}
\langle O \rangle_j^B(x) = \frac{1}{B} \sum_{i = (j - 1)B + 1}^{jB} O_i(x)
\label{eq:block_avg}
\end{equation}

The trajectory-wide average across all \( n_B \) blocks is:
\begin{equation}
\langle O \rangle^B(x) = \frac{1}{n_B} \sum_{j = 1}^{n_B} \langle O \rangle_j^B(x)
\label{eq:mean_block_avg}
\end{equation}

The uncertainty is estimated using the standard error of the mean (SEM):
\begin{equation}
\sigma_{\langle O \rangle}(x) = \frac{1}{\sqrt{n_B}} 
\sqrt{ \frac{1}{n_B - 1} \sum_{j=1}^{n_B} \left( \langle O \rangle_j^B(x) - \langle O \rangle^B(x) \right)^2 }
\label{eq:sem}
\end{equation}

\begin{figure}[htbp]
\centering
\begin{tabular}{cccc}
\multicolumn{1}{c}{\textbf{Convergence}} &
\multicolumn{1}{c}{\textbf{Covariance}} &
\multicolumn{1}{c}{\textbf{Autocorrelation}} &
\multicolumn{1}{c}{\textbf{L-curve}} \\

\includegraphics[width=0.22\textwidth]{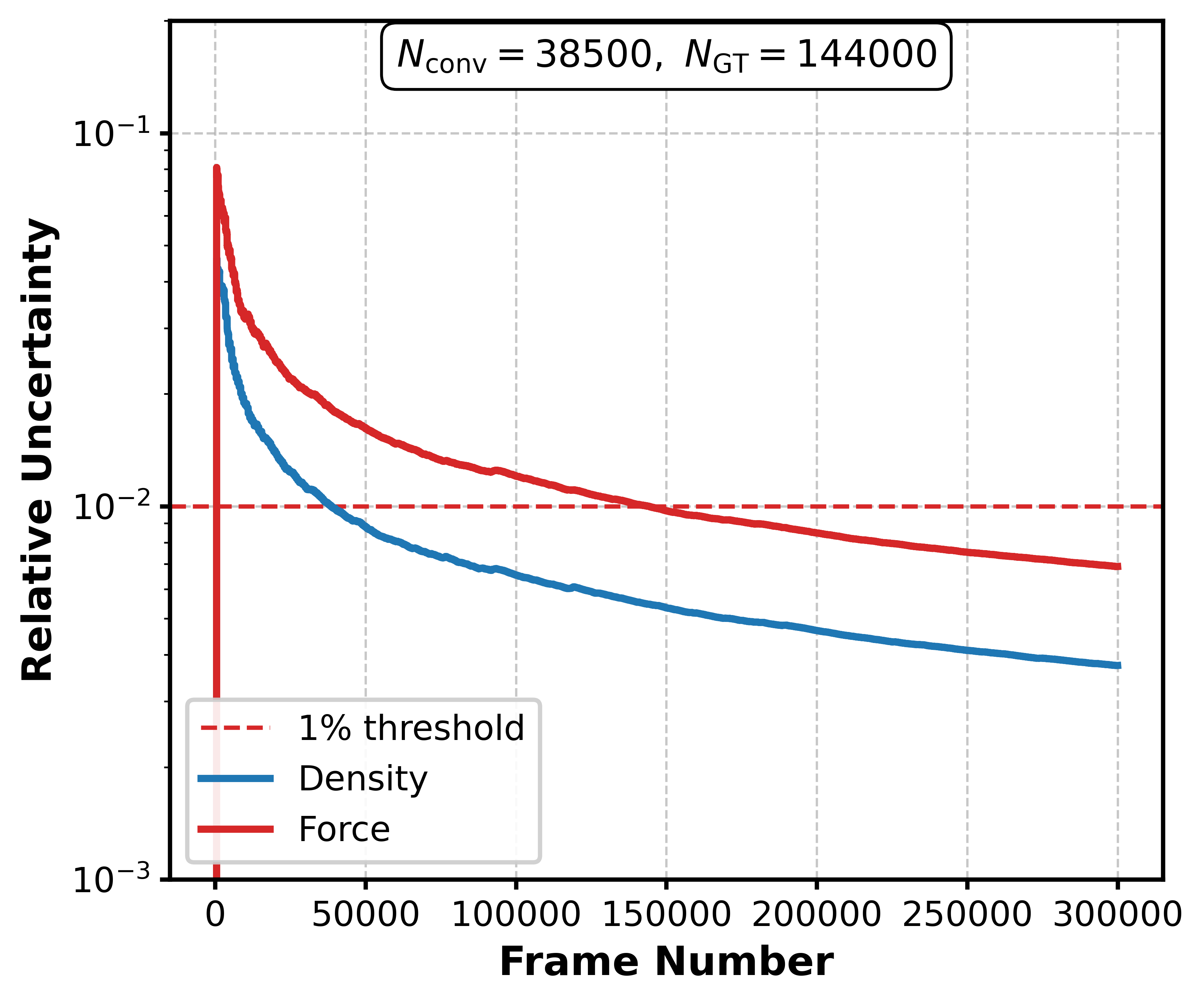} &
\includegraphics[width=0.22\textwidth]{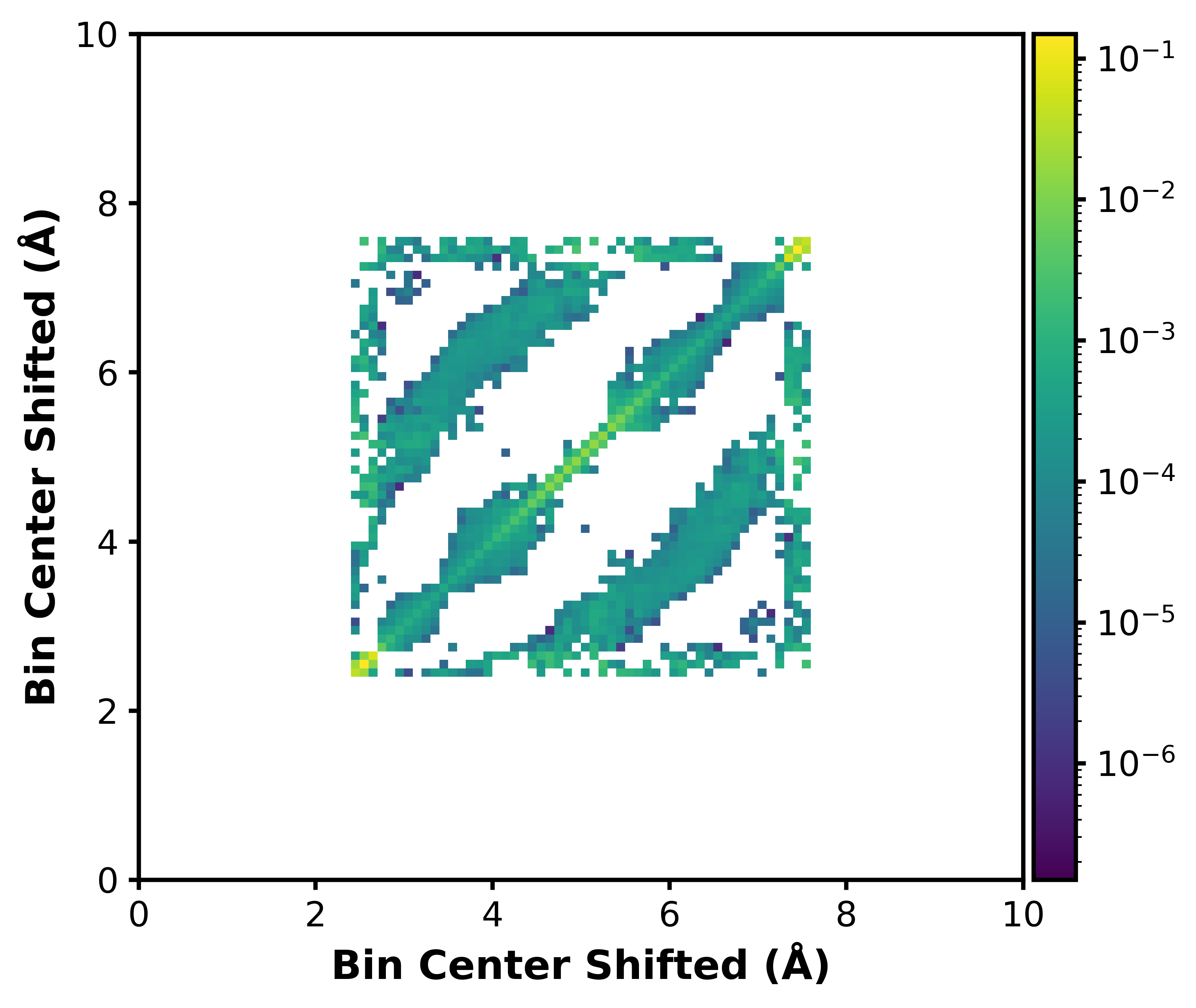} &
\includegraphics[width=0.22\textwidth]{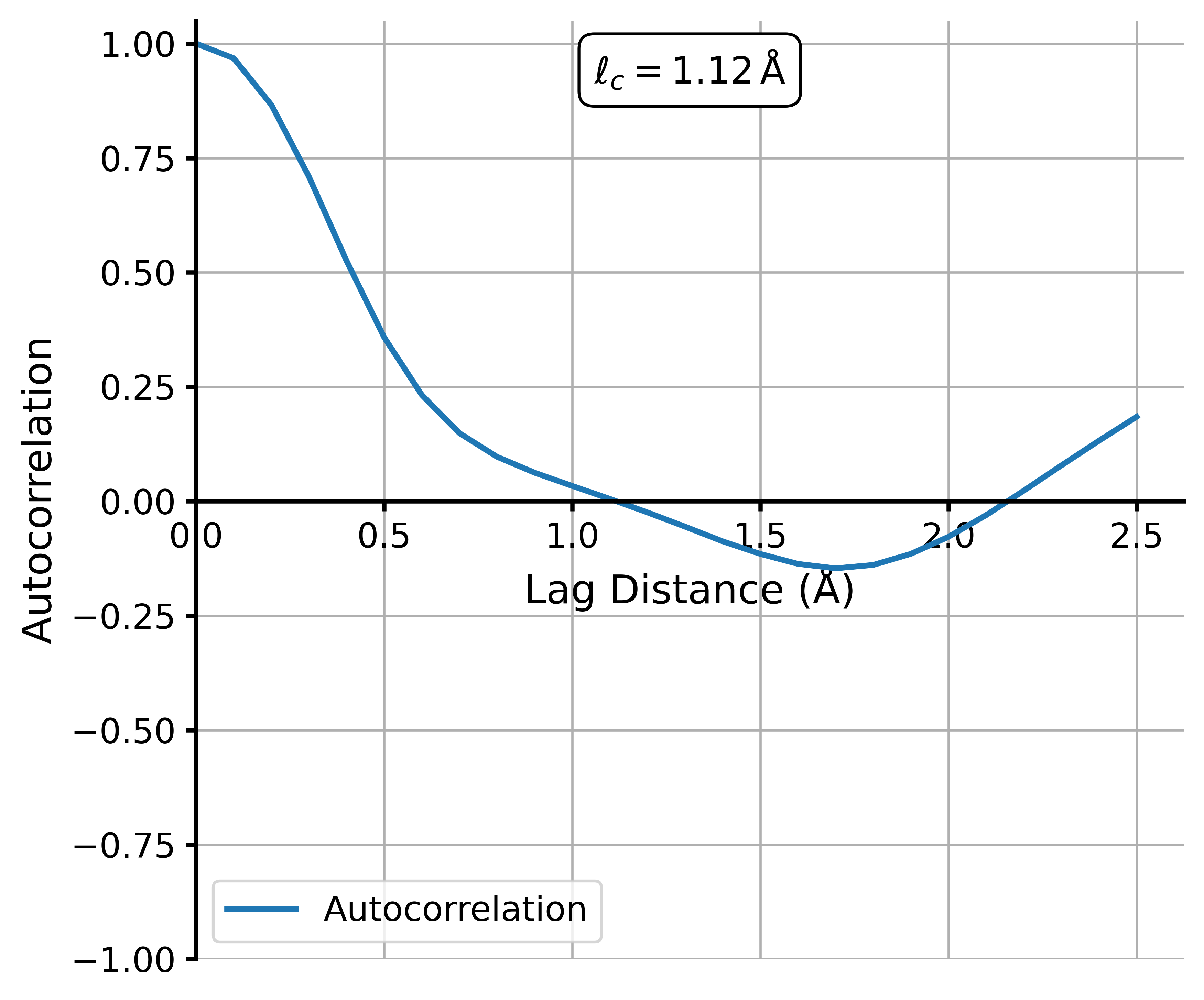} &
\includegraphics[width=0.22\textwidth]{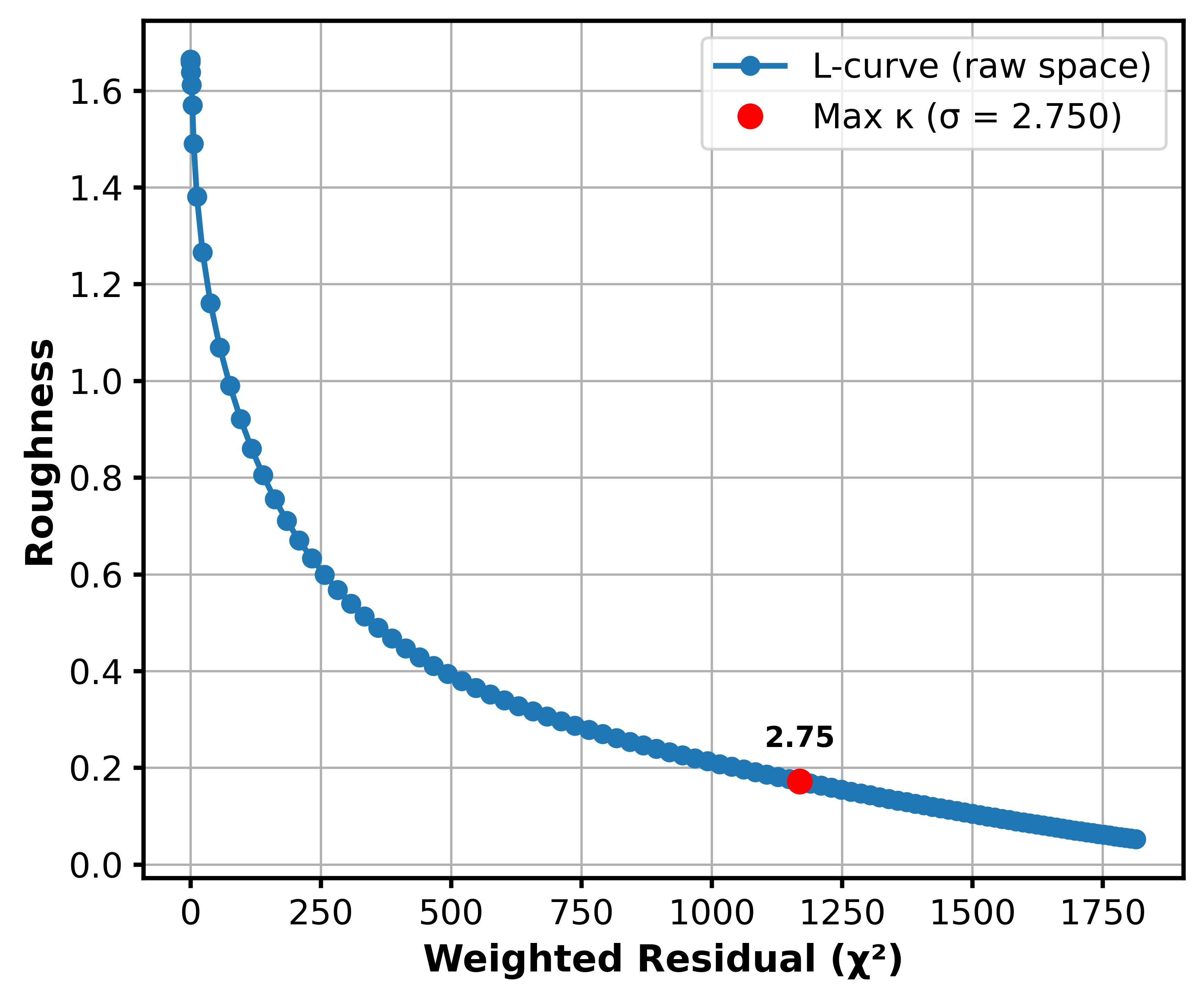} \\

\includegraphics[width=0.22\textwidth]{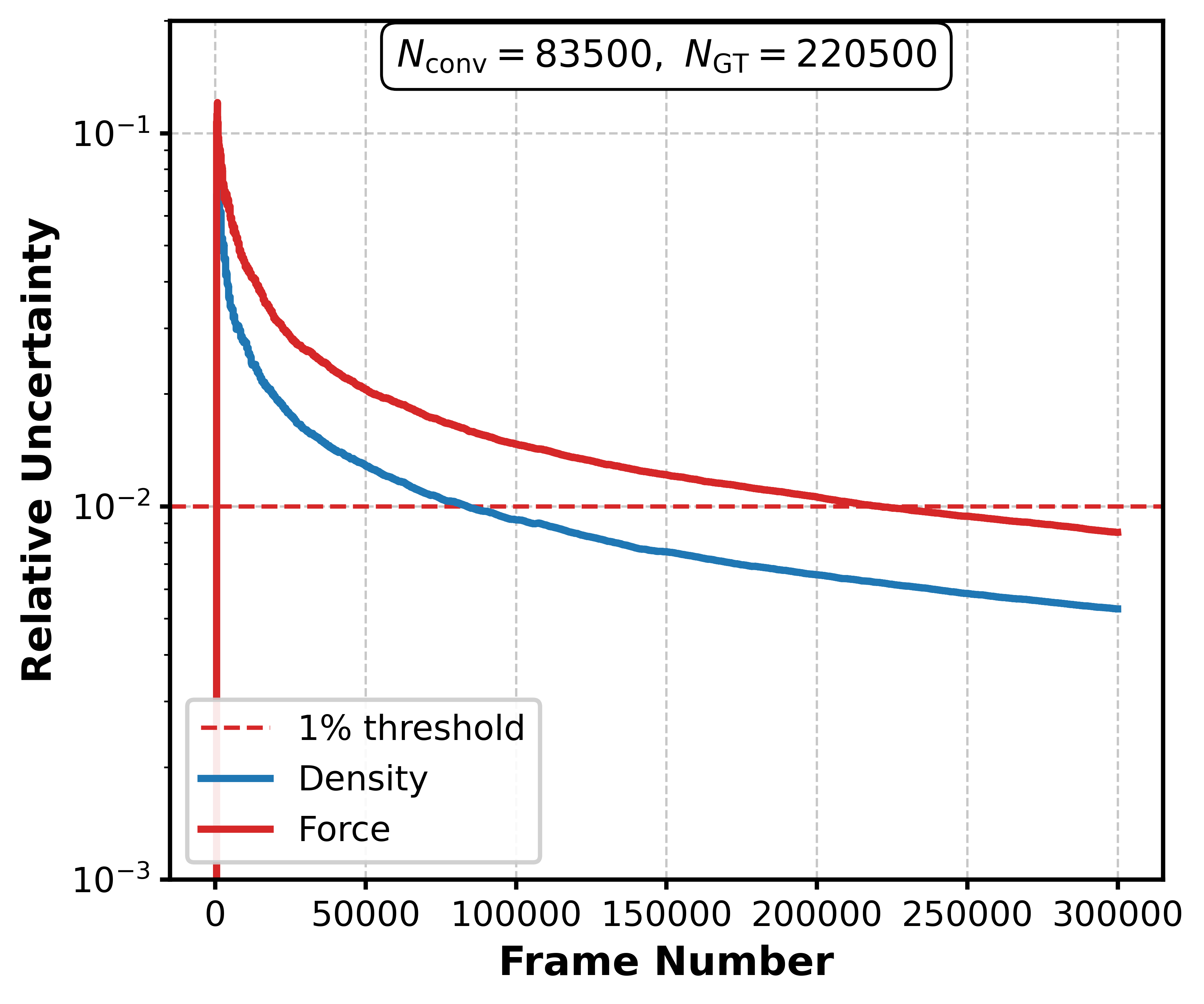} &
\includegraphics[width=0.22\textwidth]{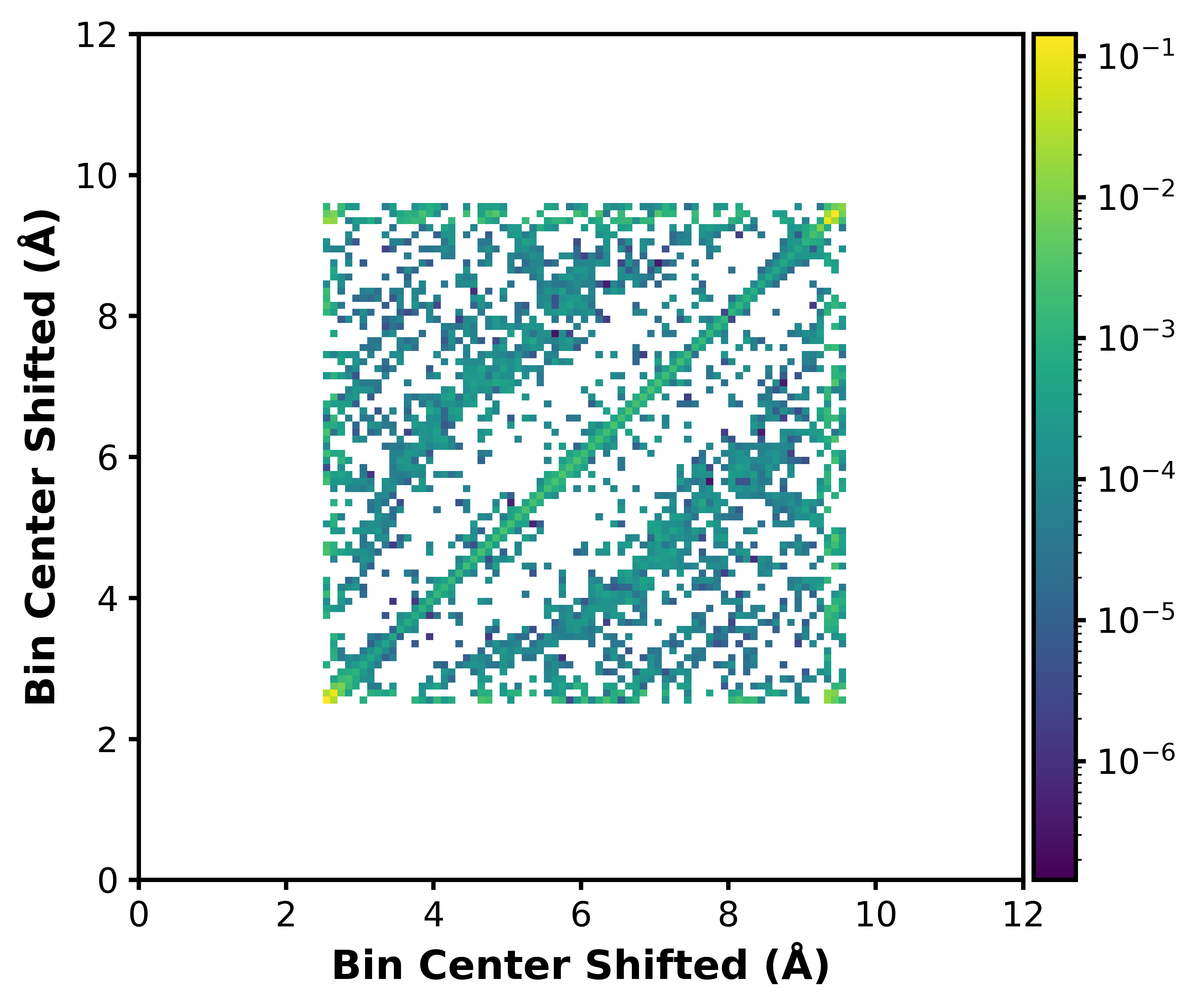} &
\includegraphics[width=0.22\textwidth]{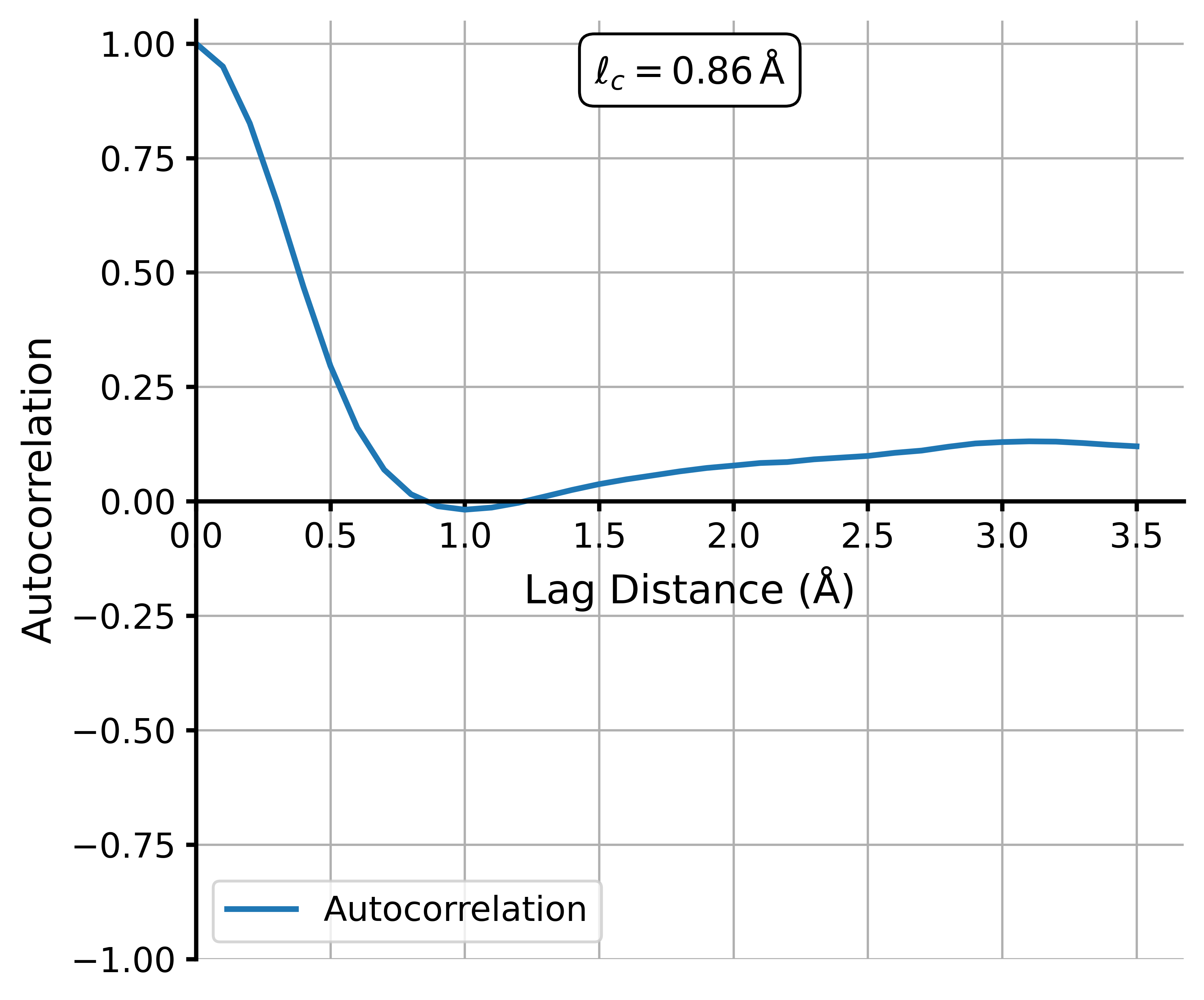} &
\includegraphics[width=0.22\textwidth]{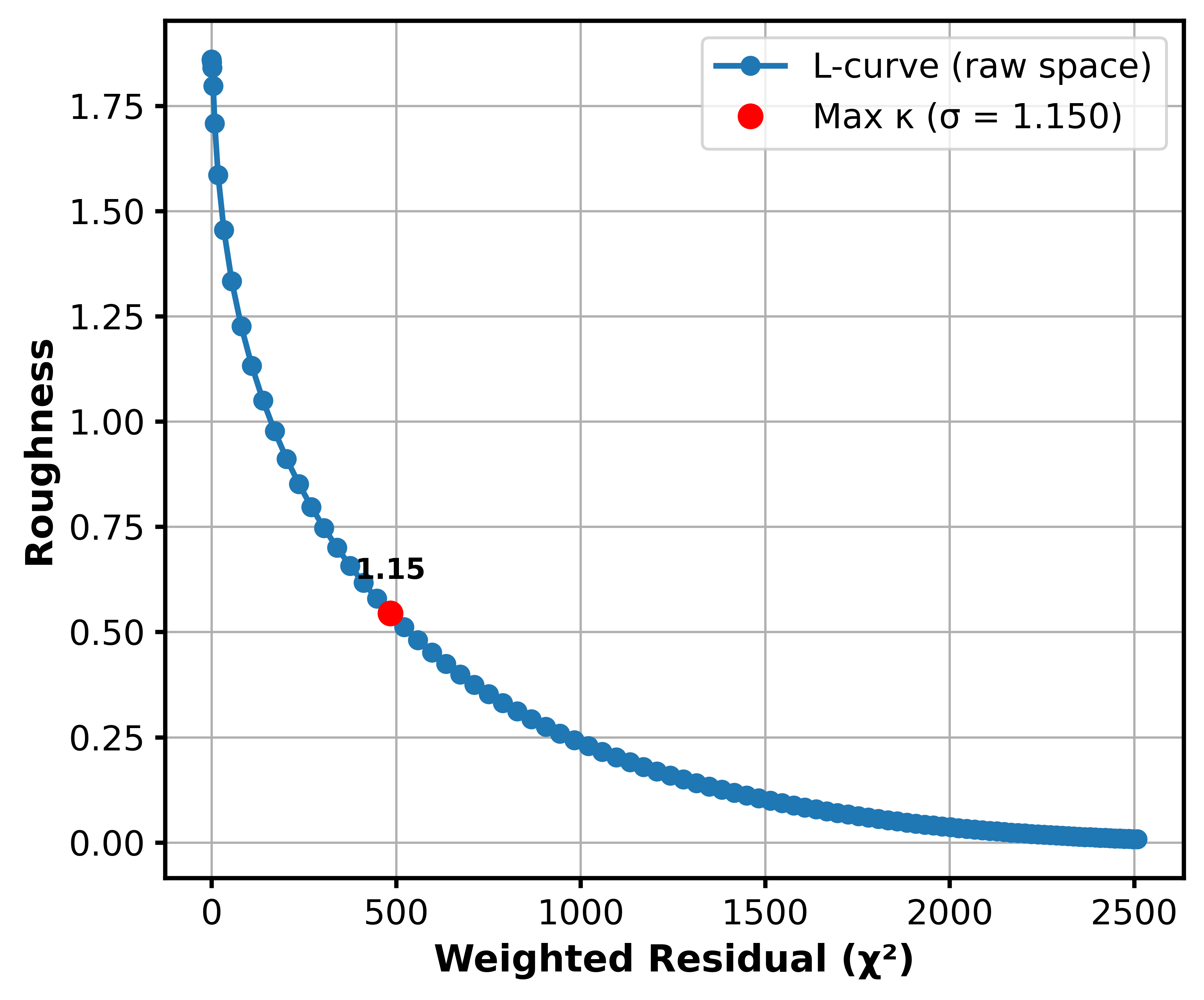} \\

\includegraphics[width=0.22\textwidth]{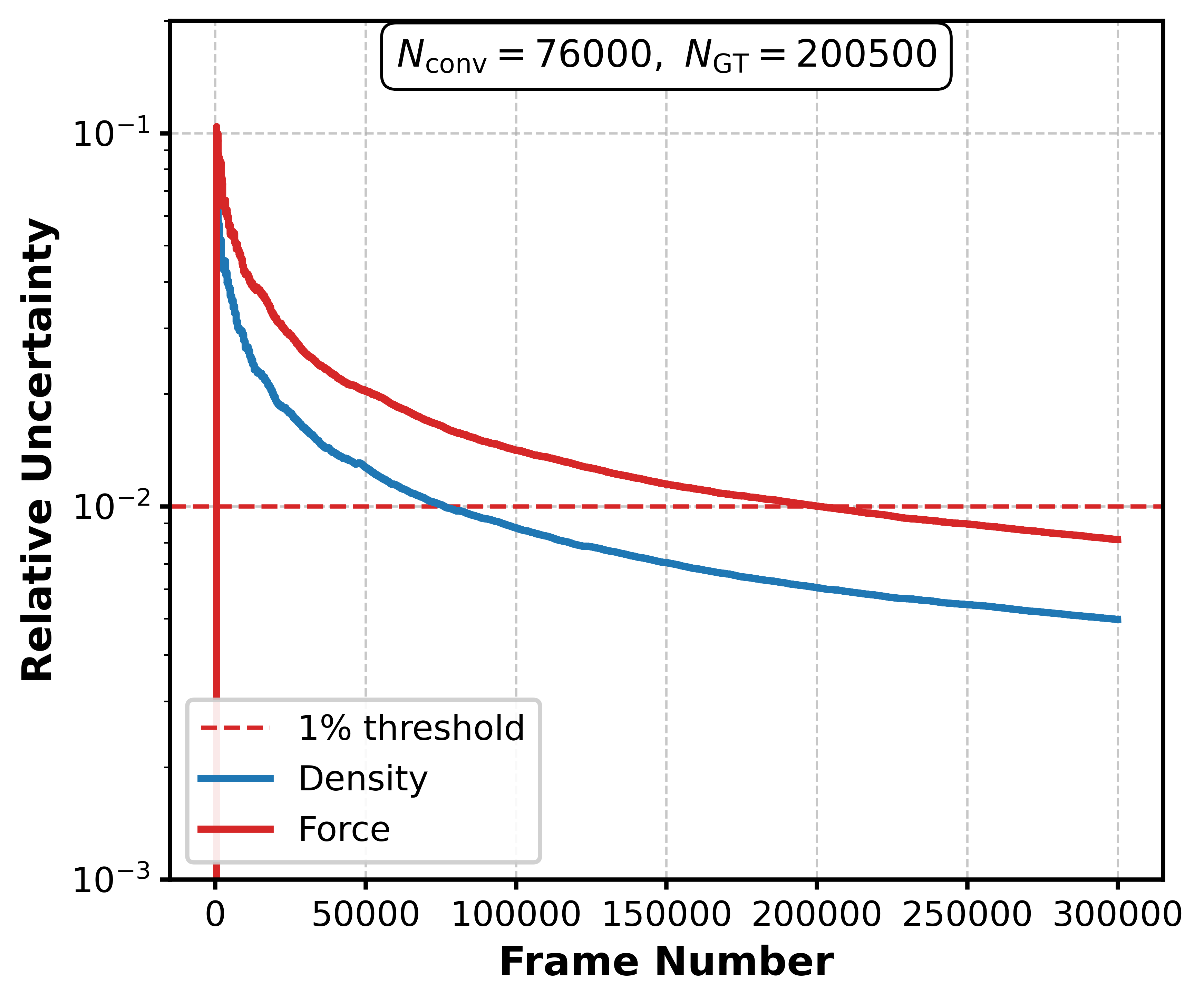} &
\includegraphics[width=0.22\textwidth]{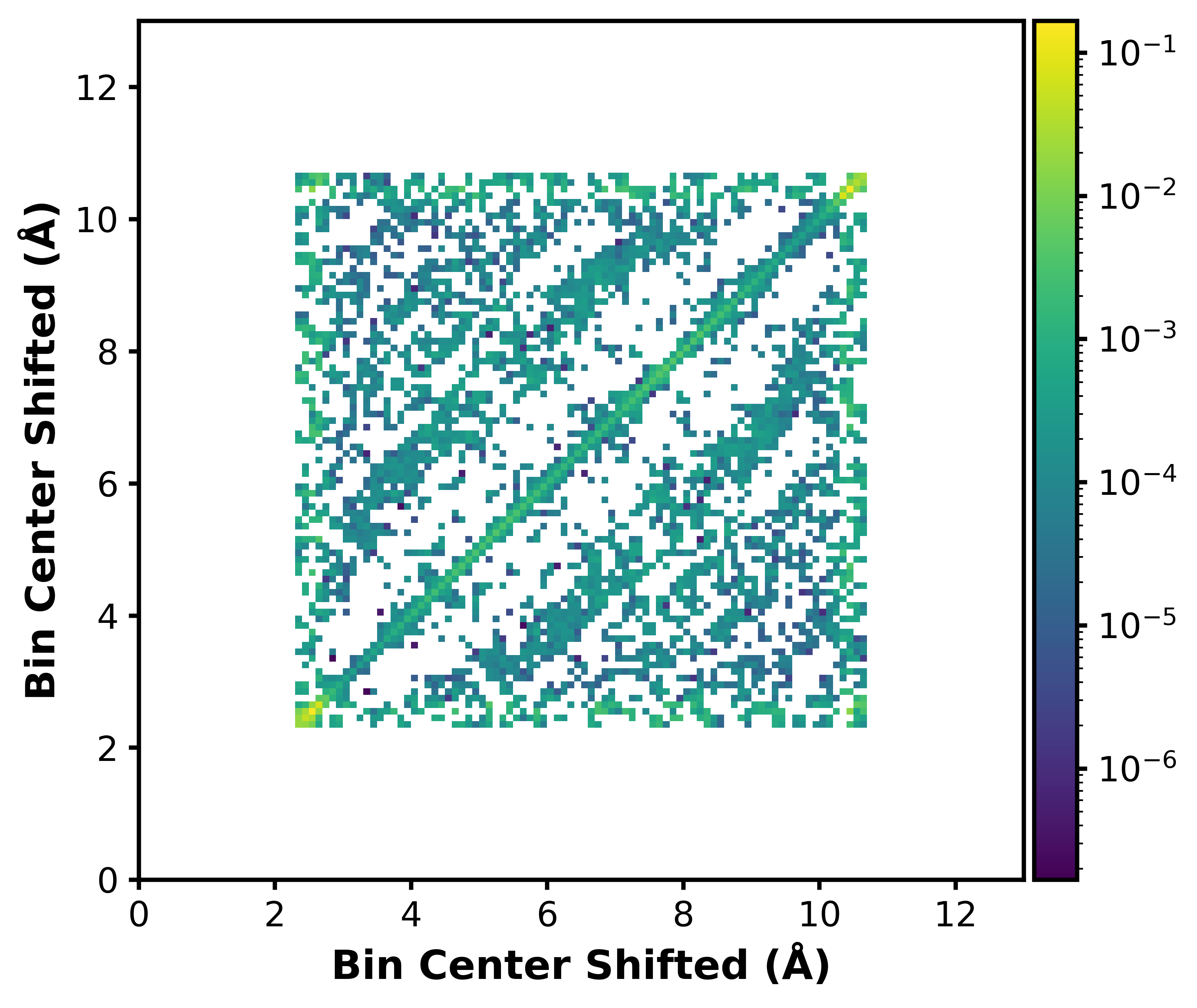} &
\includegraphics[width=0.22\textwidth]{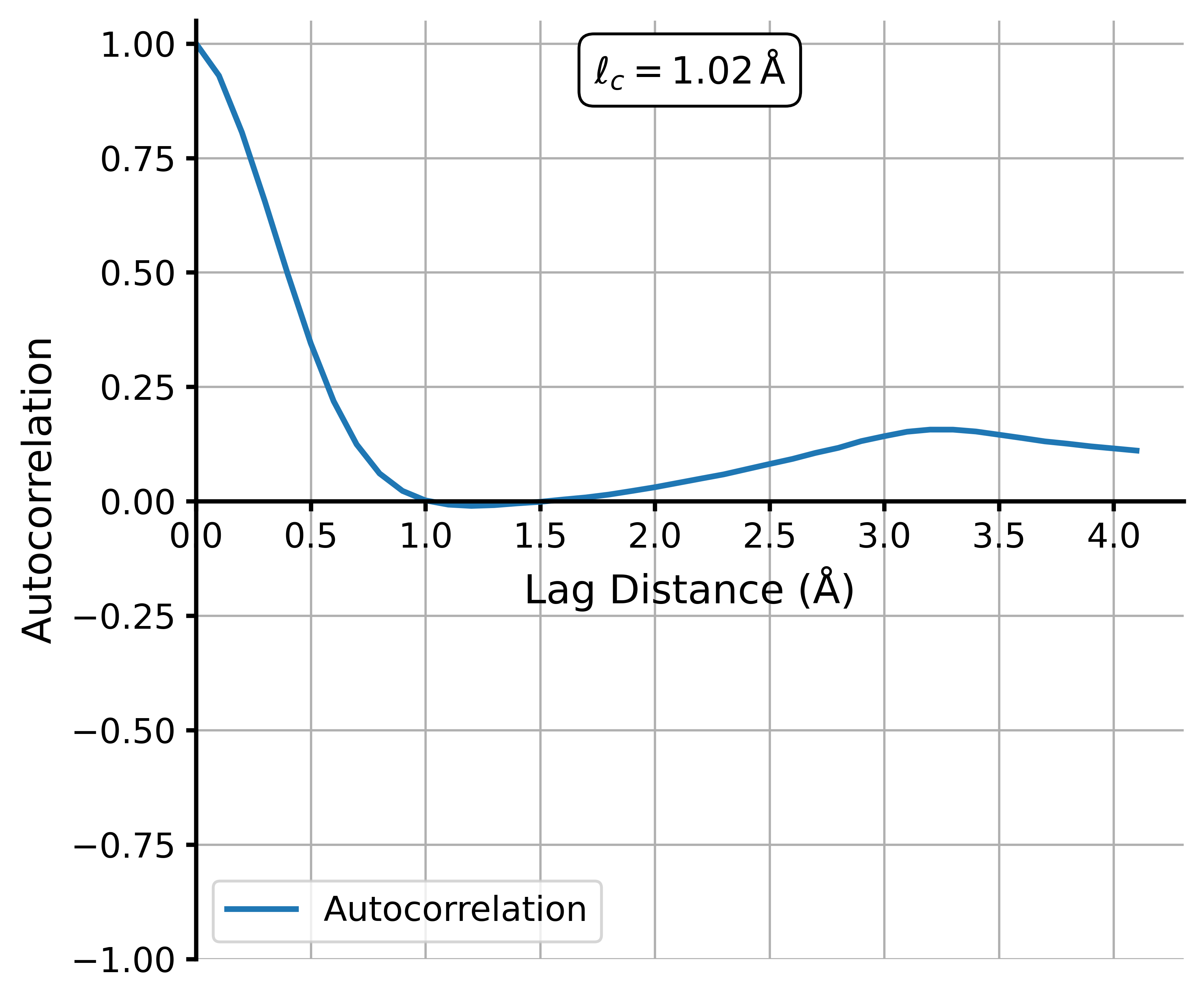} &
\includegraphics[width=0.22\textwidth]{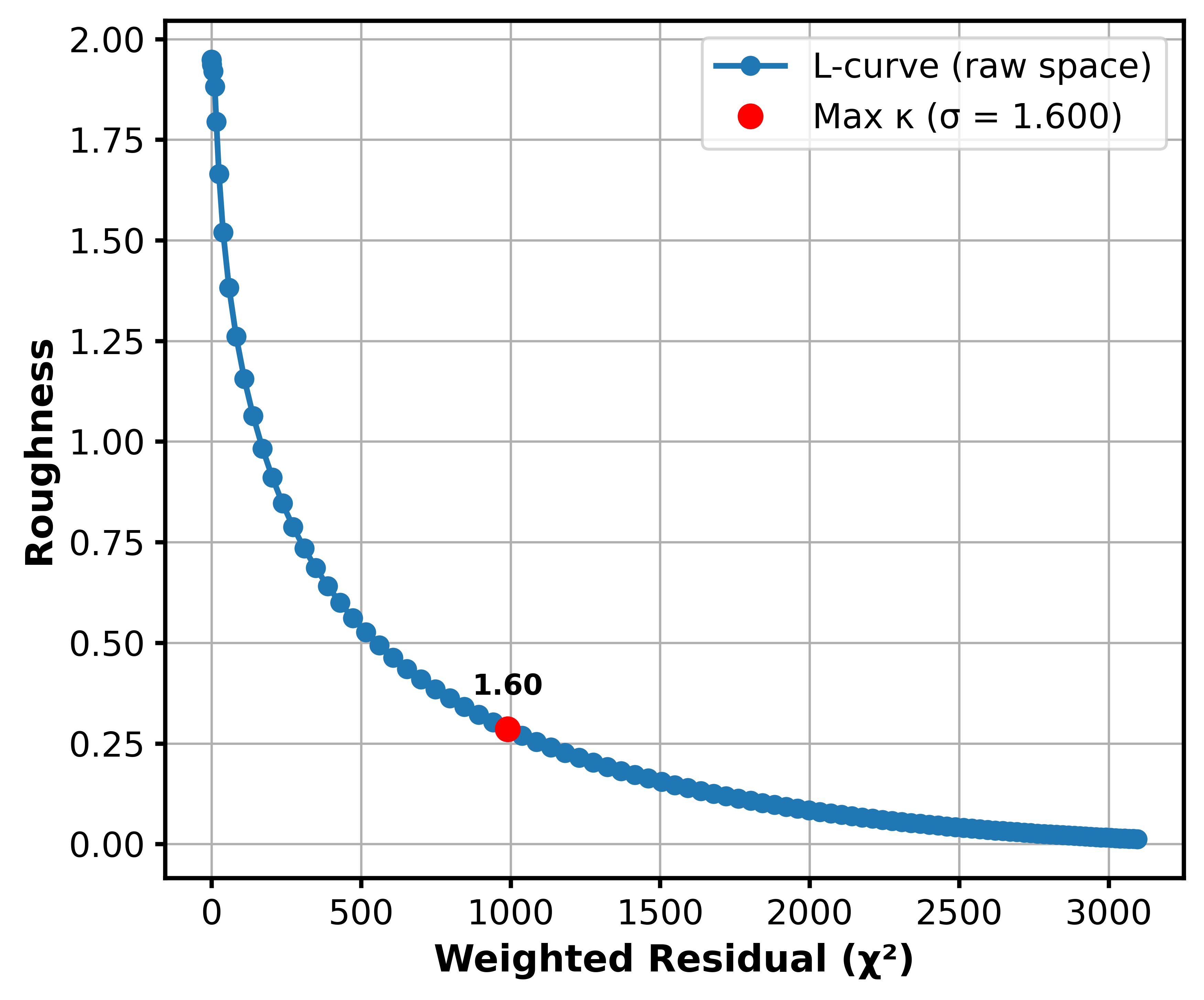} \\

\includegraphics[width=0.22\textwidth]{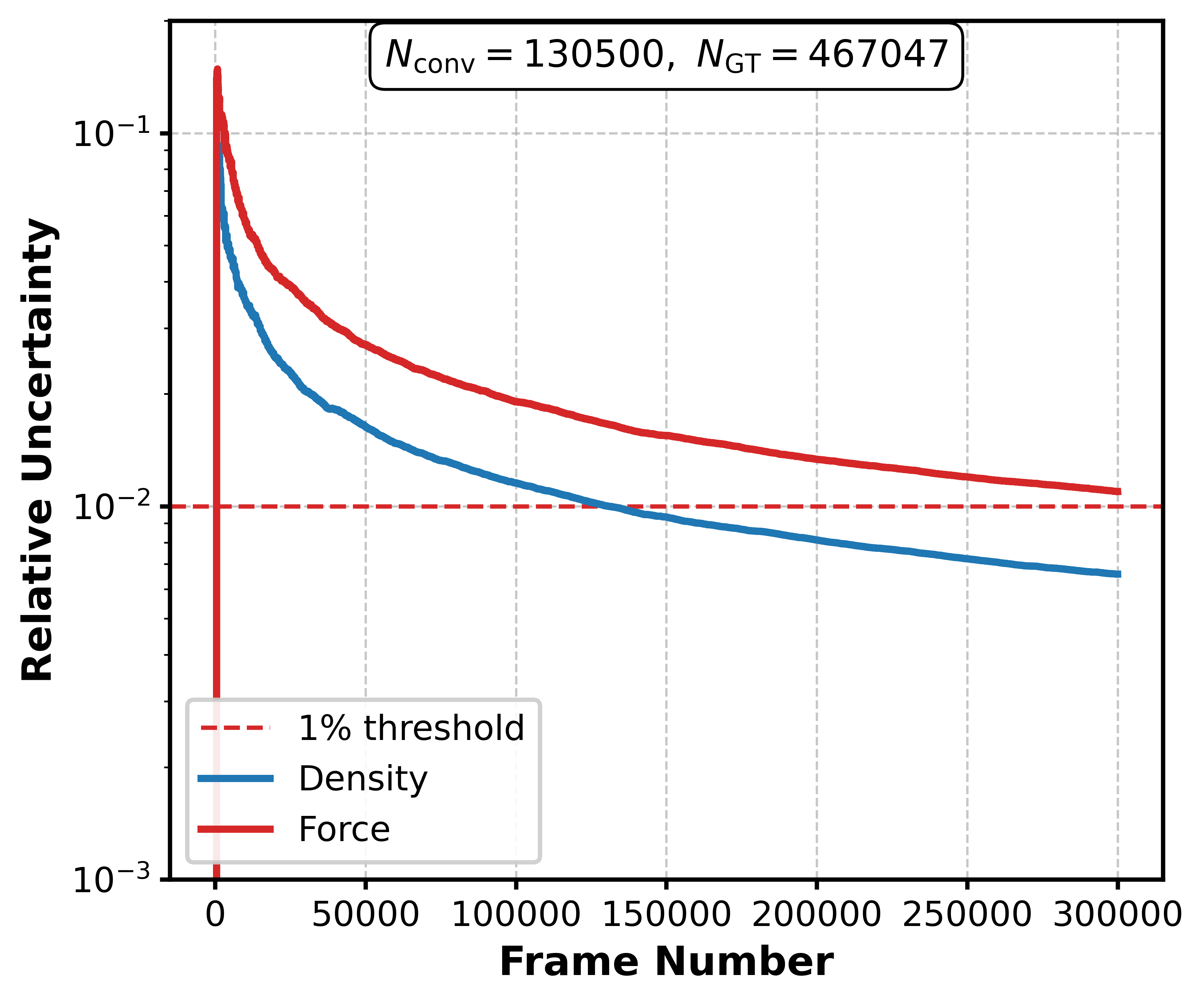} &
\includegraphics[width=0.22\textwidth]{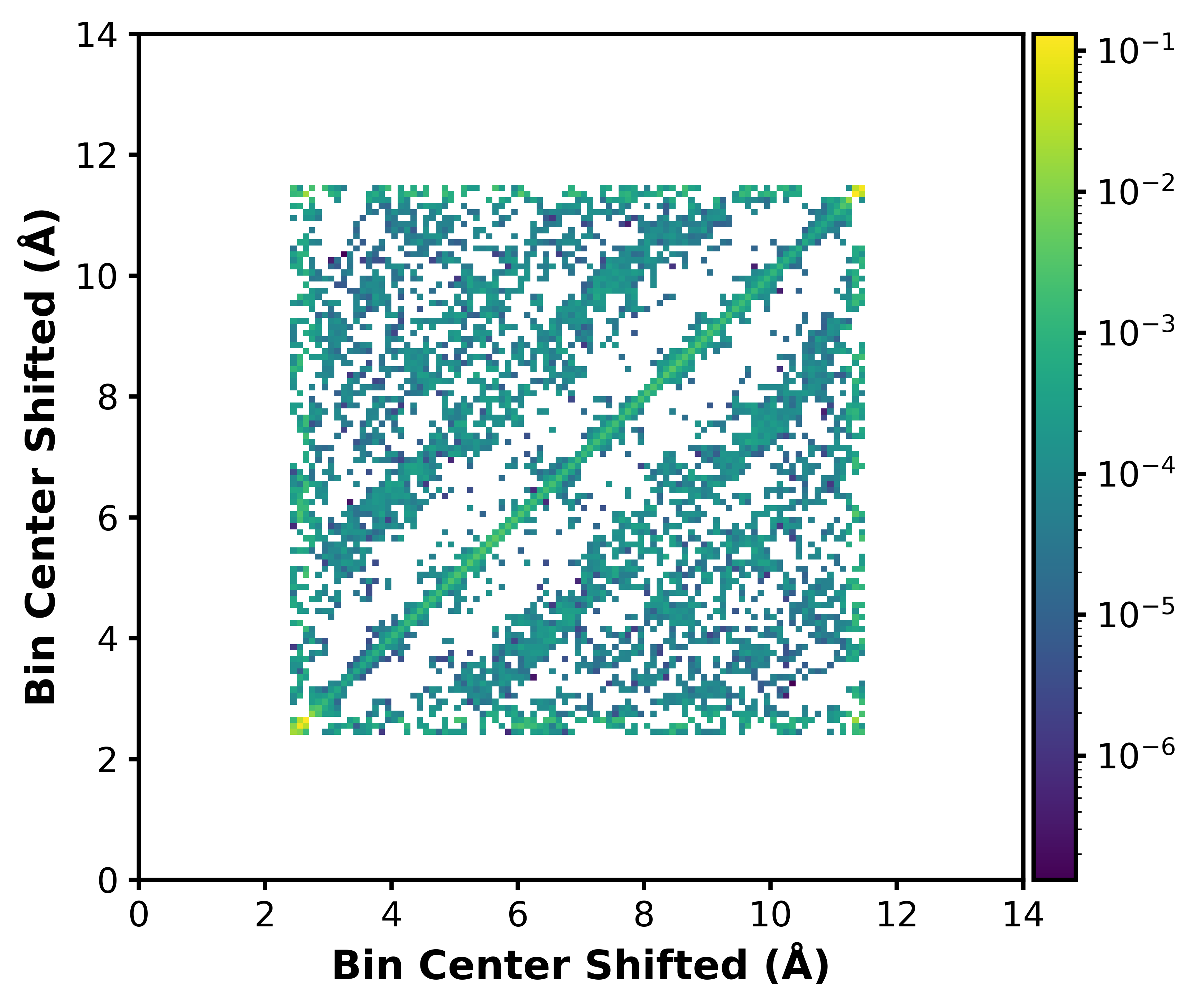} &
\includegraphics[width=0.22\textwidth]{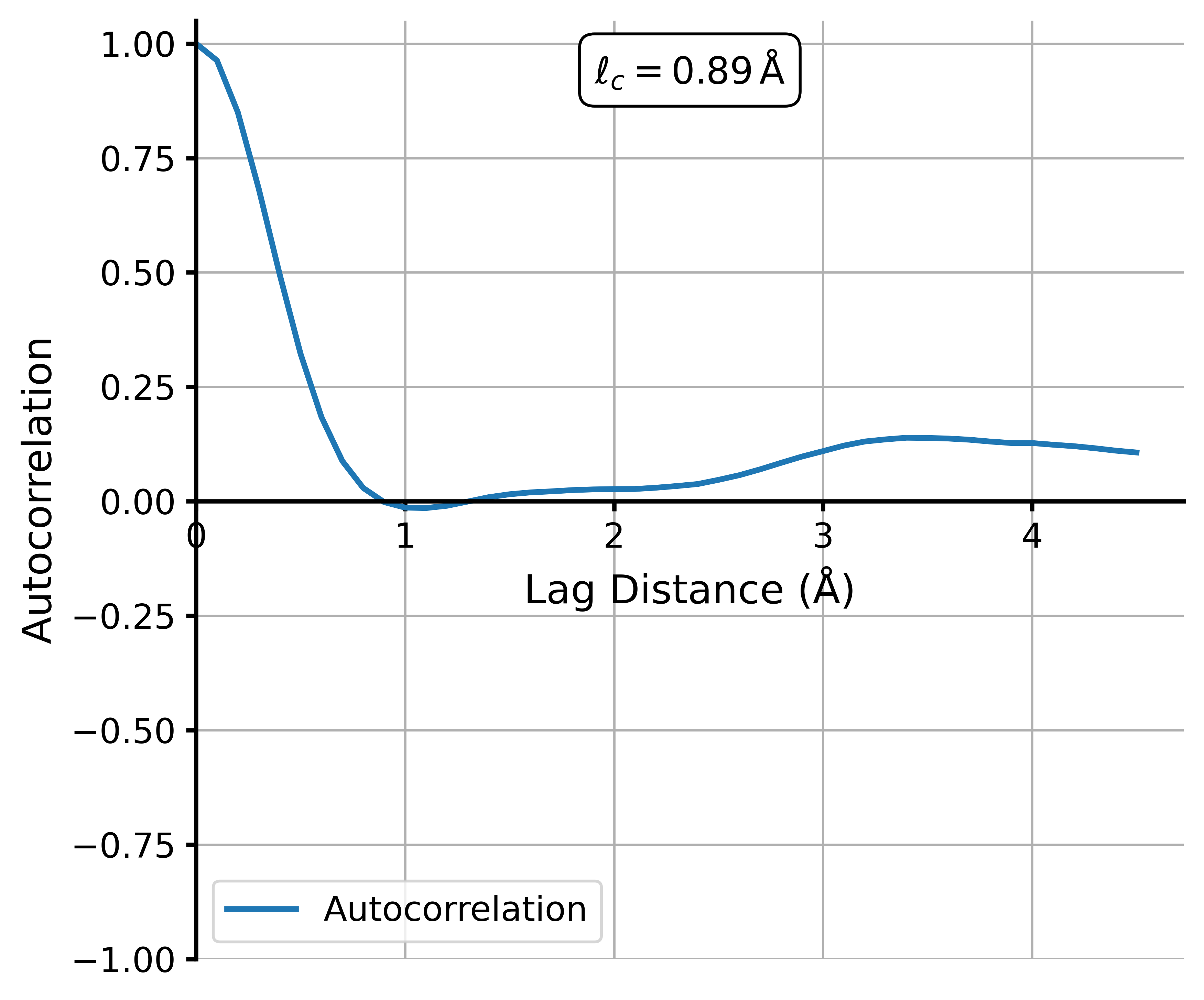} &
\includegraphics[width=0.22\textwidth]{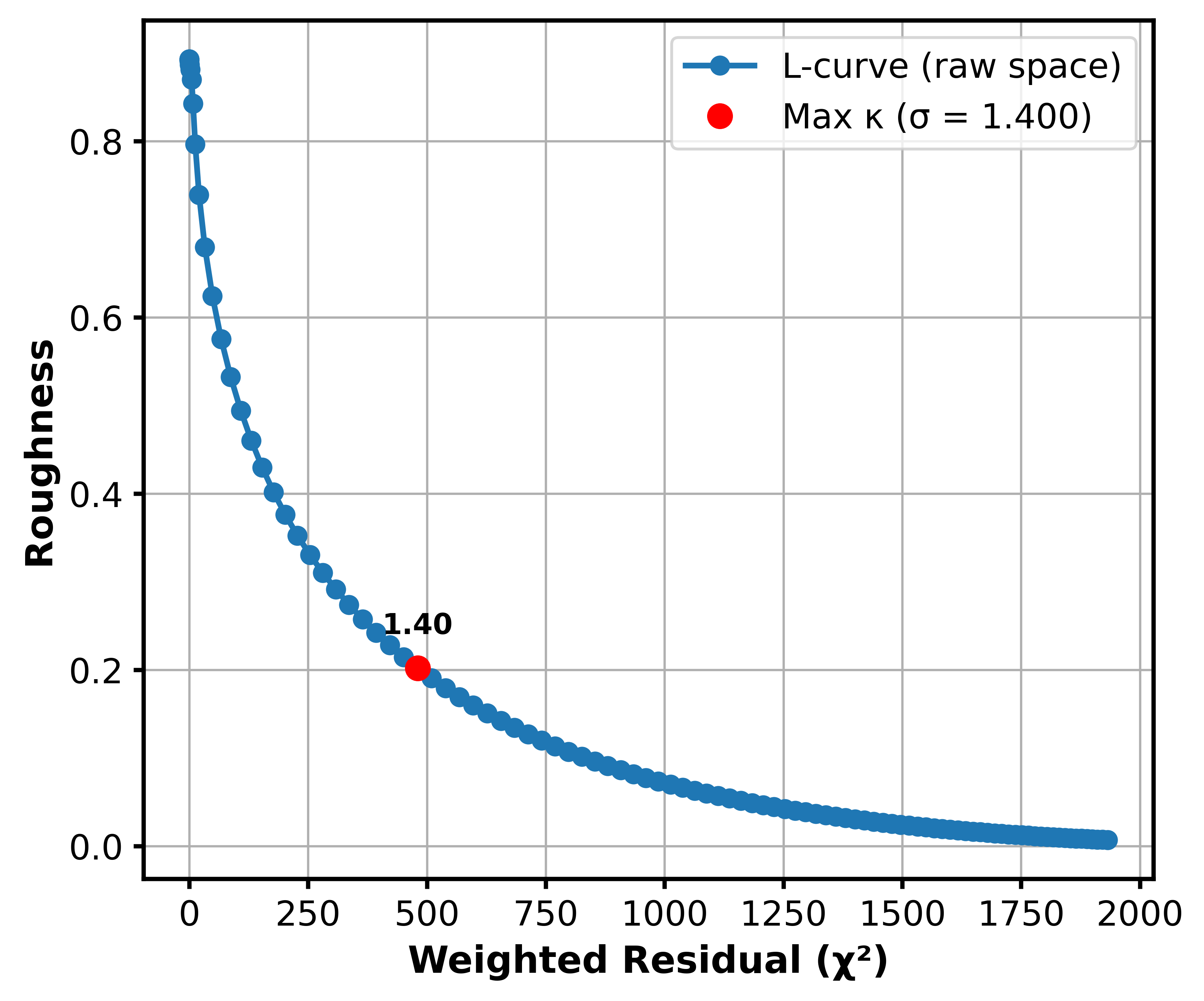} \\

\includegraphics[width=0.22\textwidth]{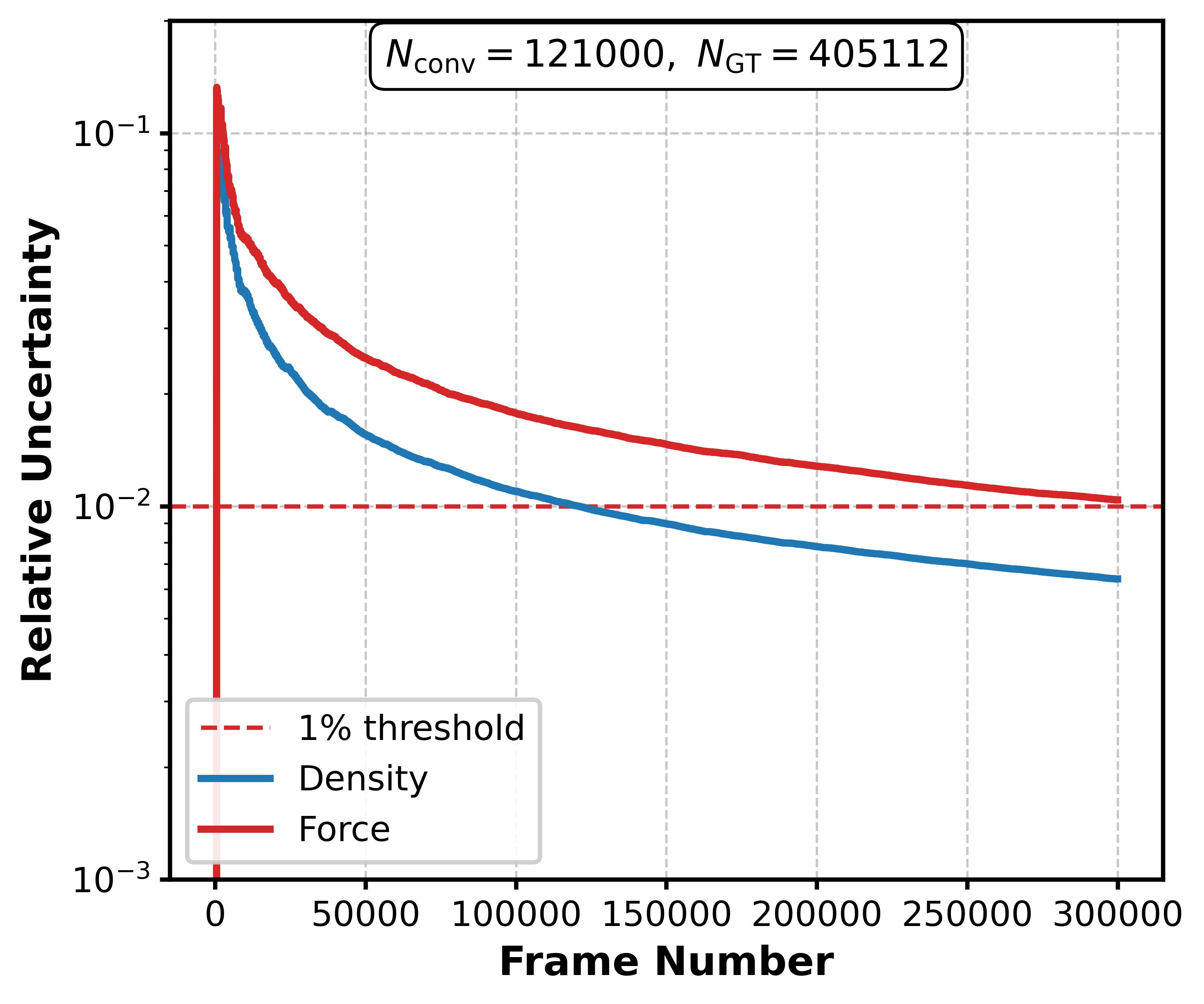} &
\includegraphics[width=0.22\textwidth]{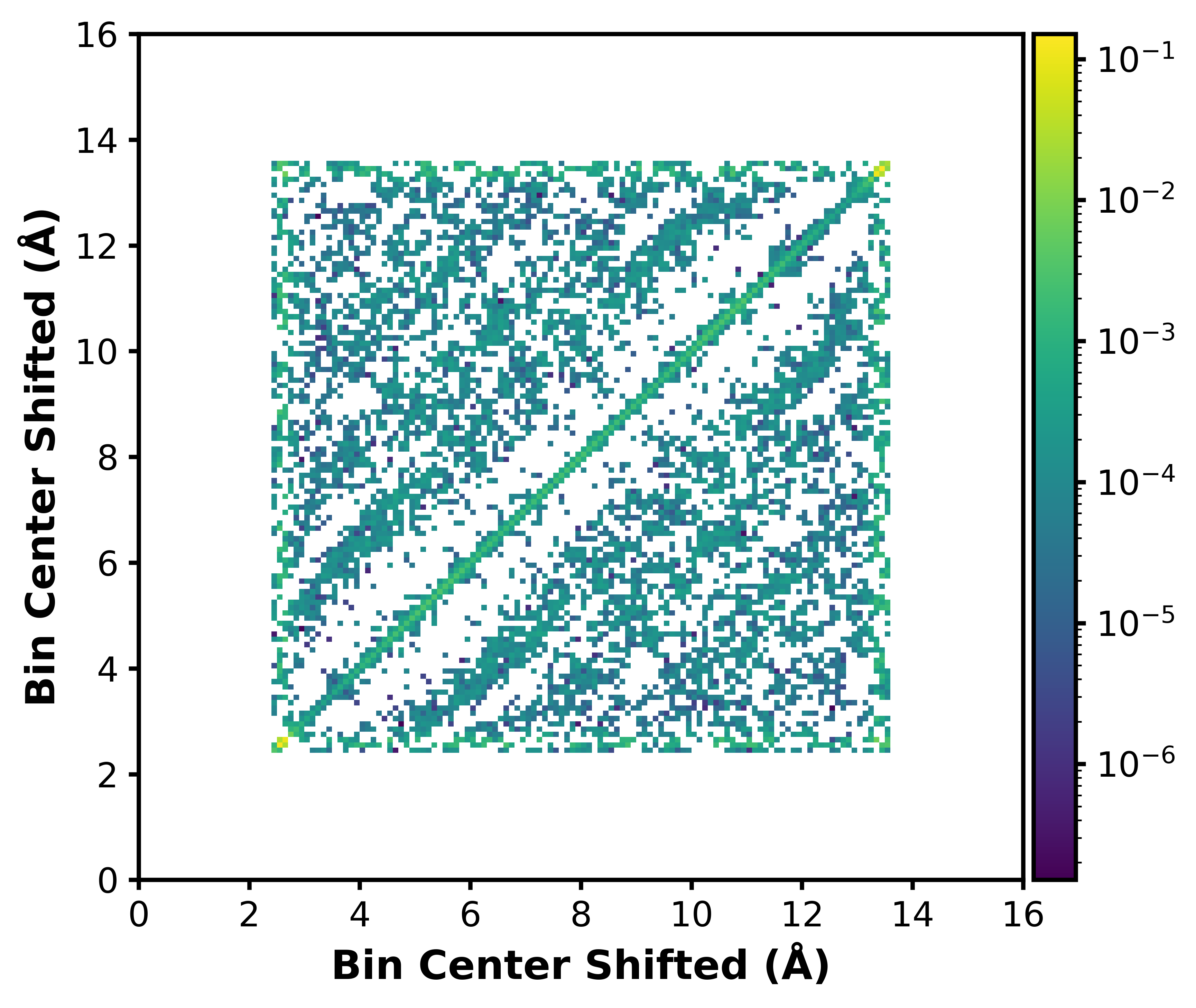} &
\includegraphics[width=0.22\textwidth]{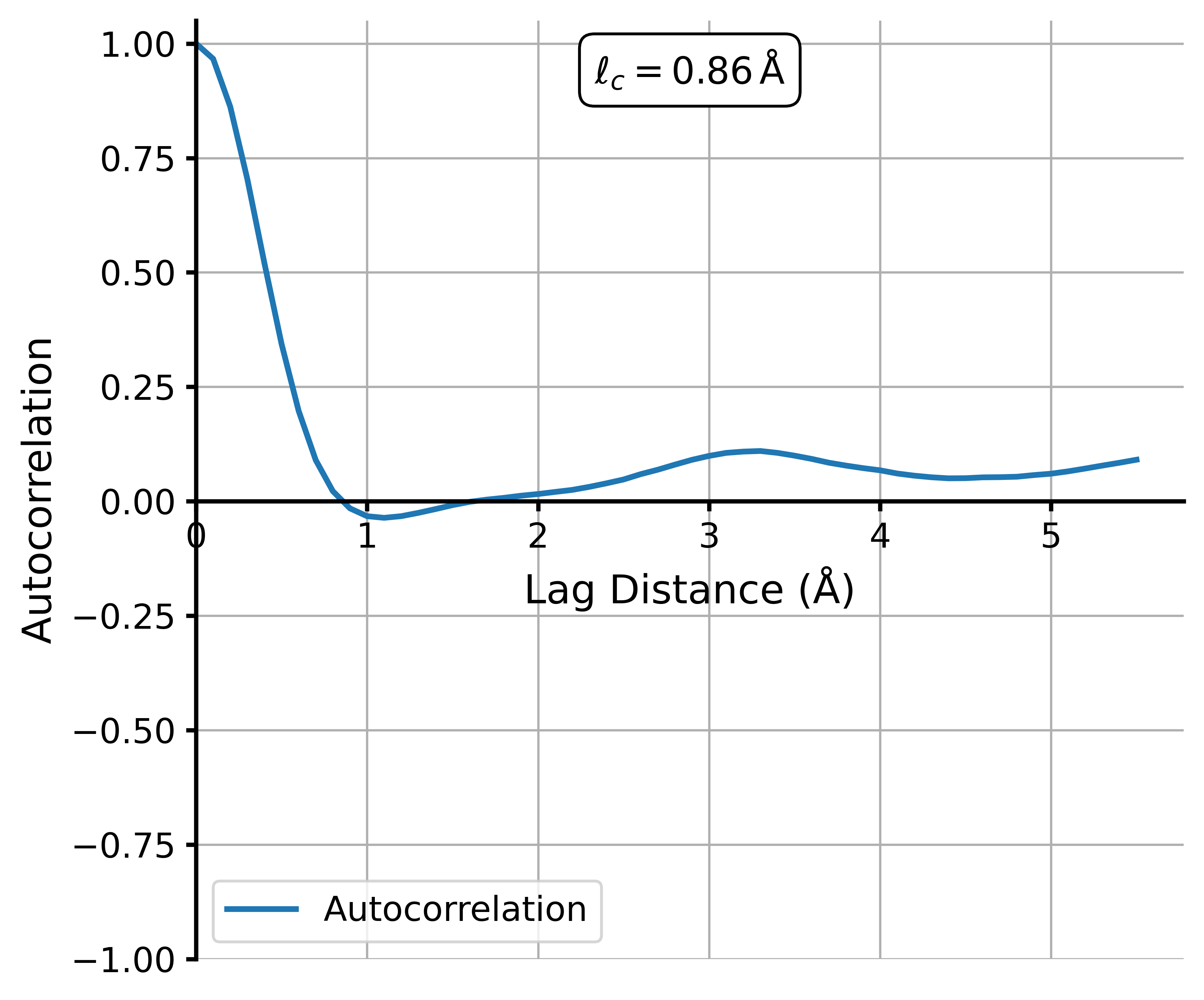} &
\includegraphics[width=0.22\textwidth]{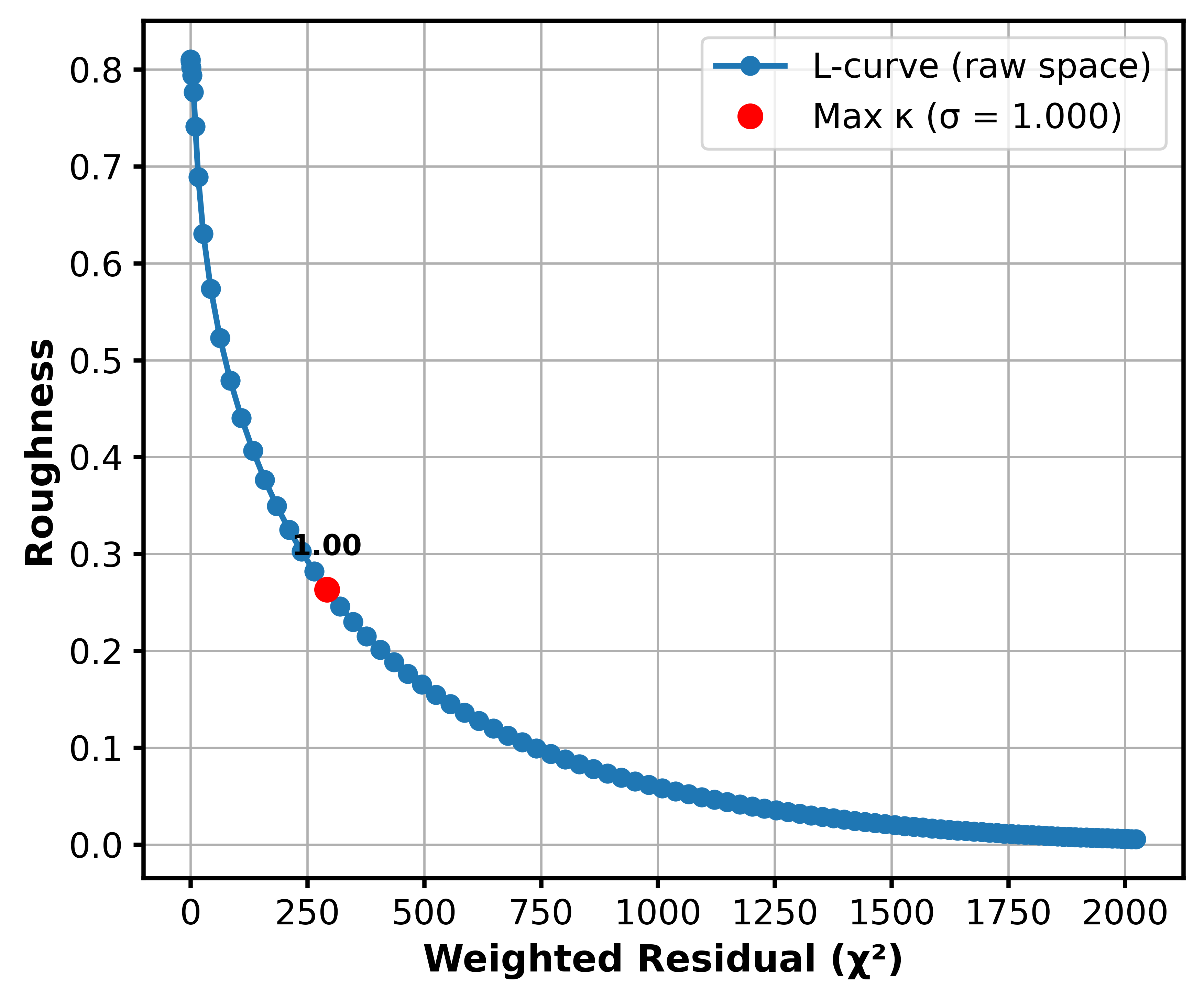} \\

\includegraphics[width=0.22\textwidth]{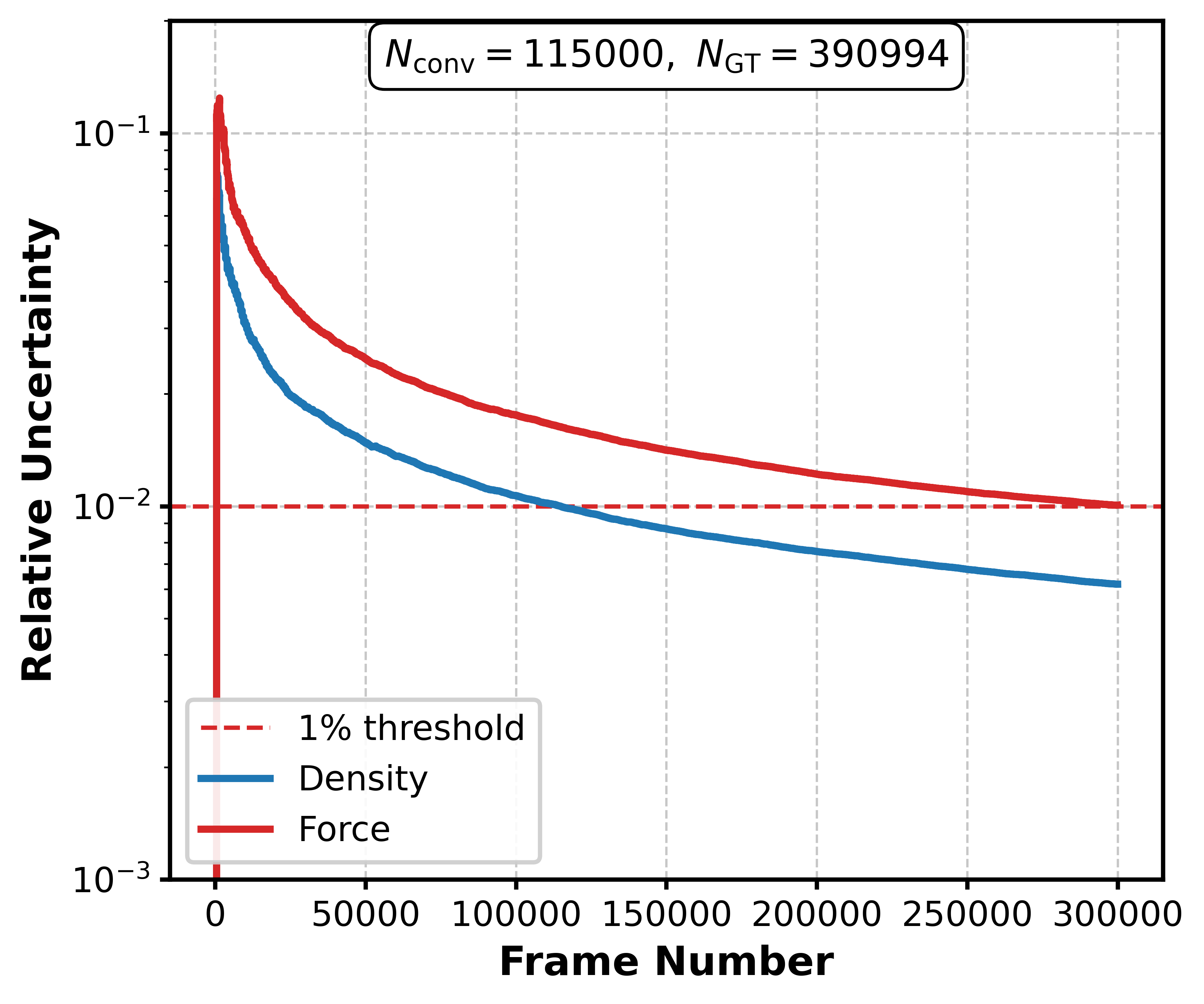} &
\includegraphics[width=0.22\textwidth]{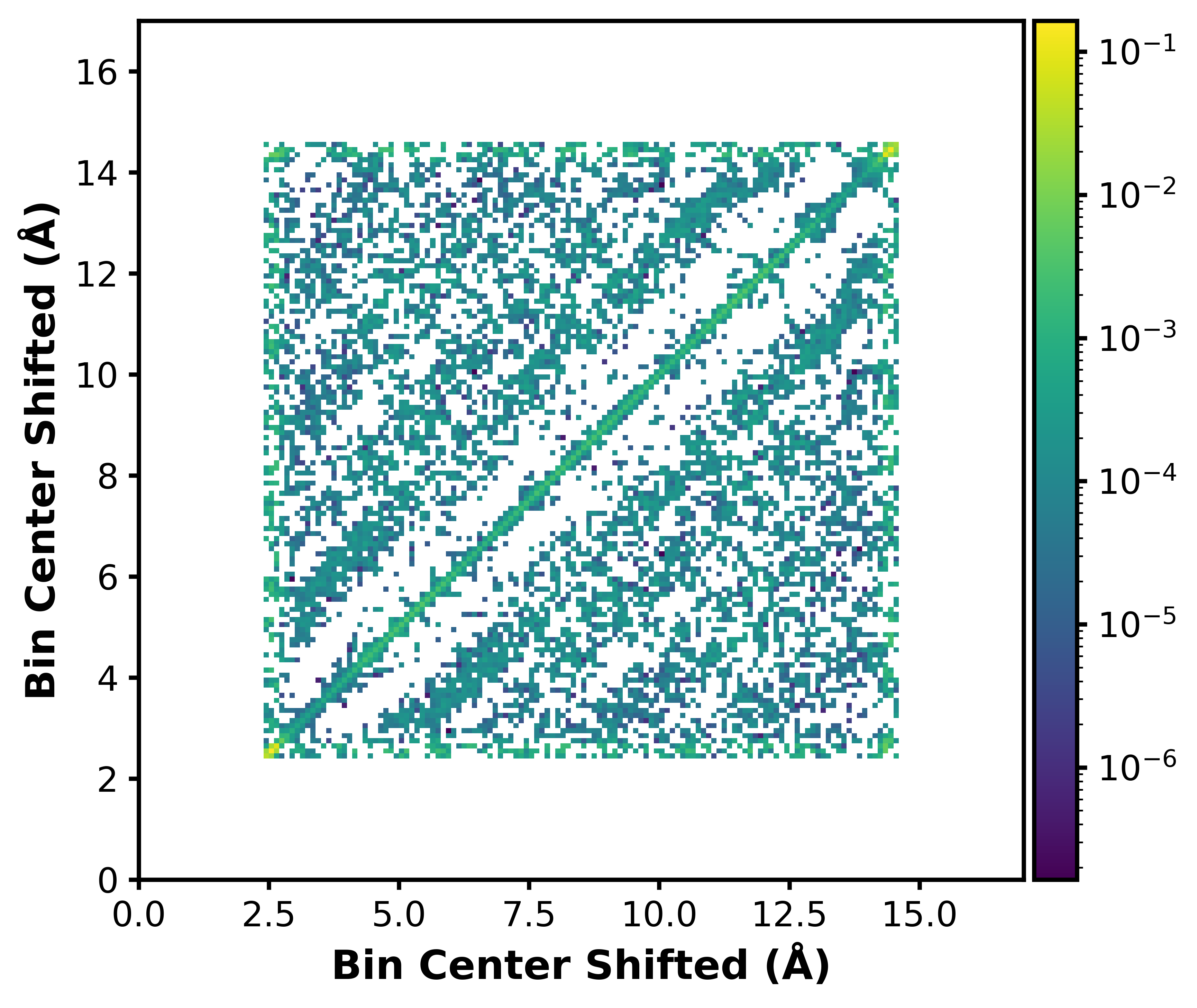} &
\includegraphics[width=0.22\textwidth]{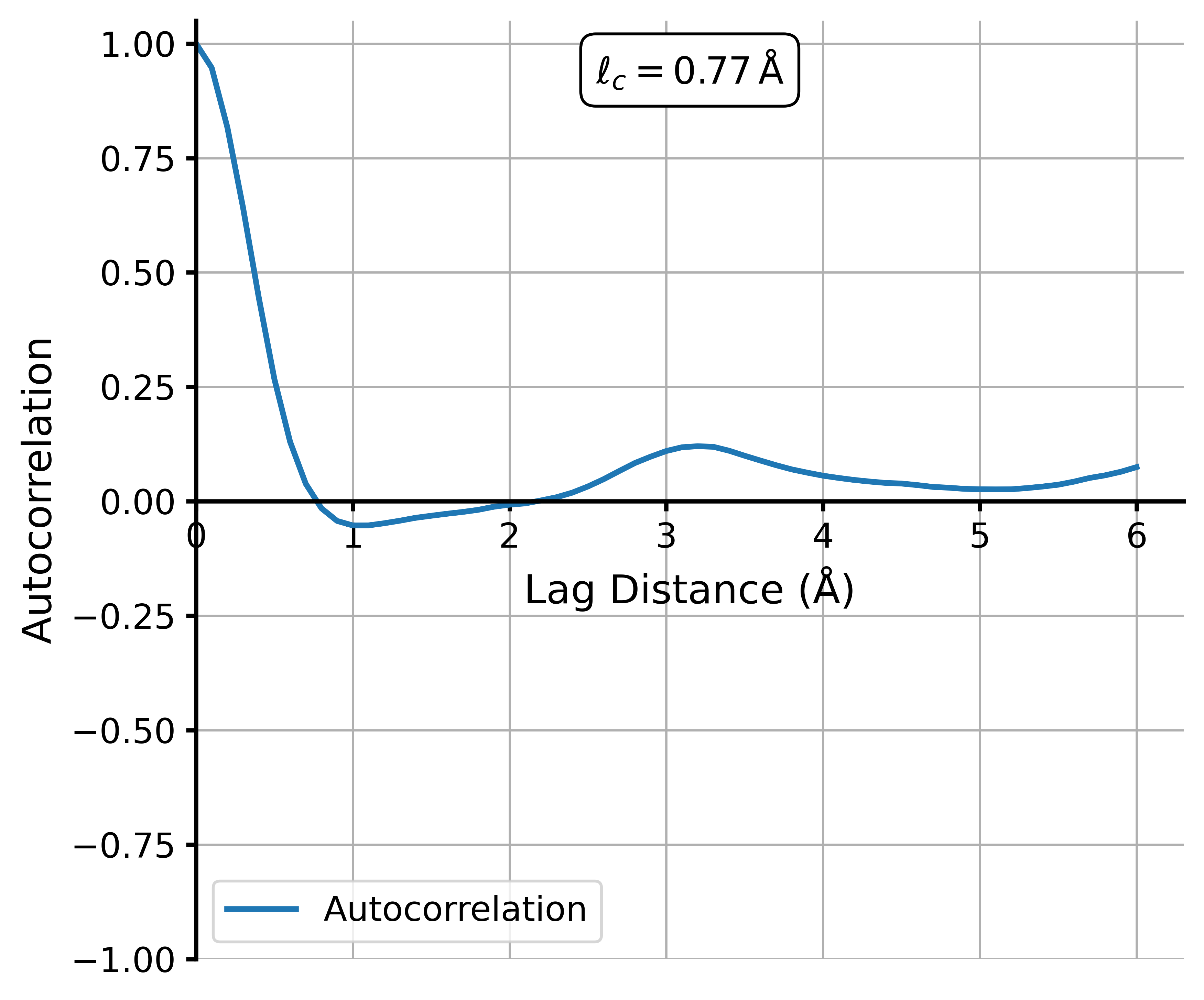} &
\includegraphics[width=0.22\textwidth]{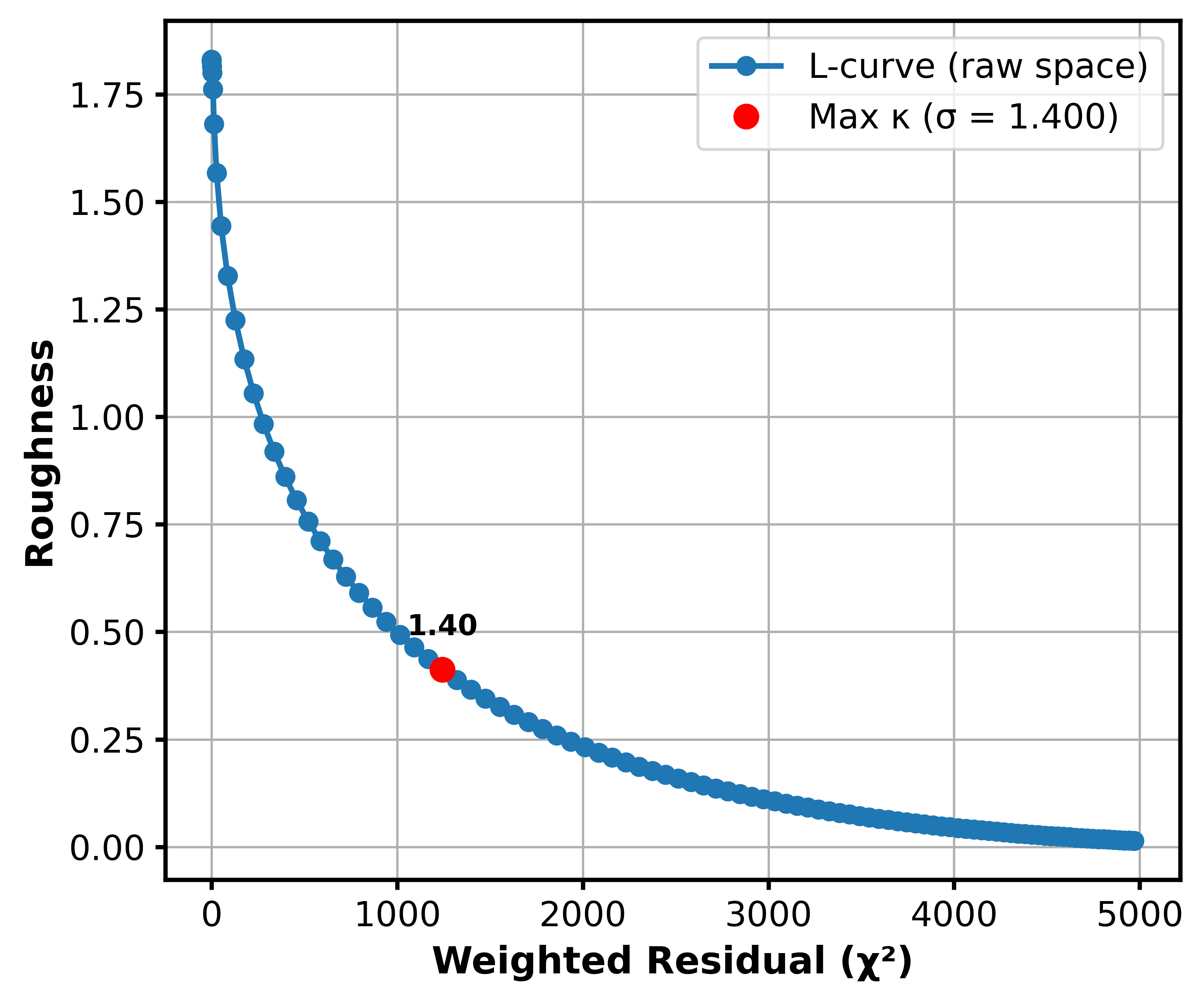} \\
\end{tabular}
\caption{Convergence, covariance, autocorrelation, and L-curve plots for channels 1\,nm (top) to 1.7\,nm (bottom).}
\label{fig:summary_06_13}
\label{fig:lfa_convergence}
\end{figure}

\vspace{1em}

\begin{figure}[htbp]
\centering
\begin{tabular}{cccc}
\multicolumn{1}{c}{\textbf{Convergence}} &
\multicolumn{1}{c}{\textbf{Covariance}} &
\multicolumn{1}{c}{\textbf{Autocorrelation}} &
\multicolumn{1}{c}{\textbf{L-curve}} \\

\includegraphics[width=0.22\textwidth]{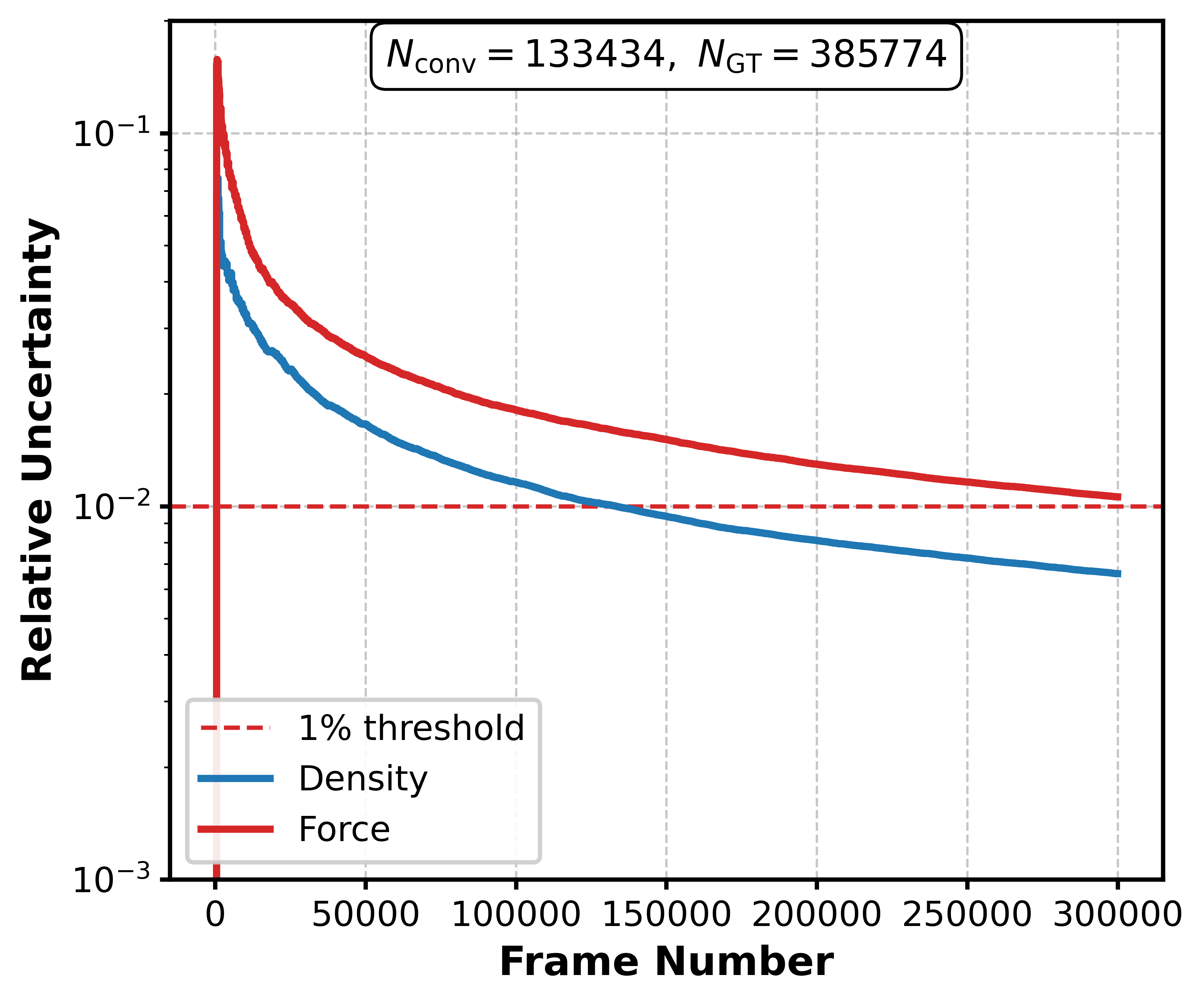} &
\includegraphics[width=0.22\textwidth]{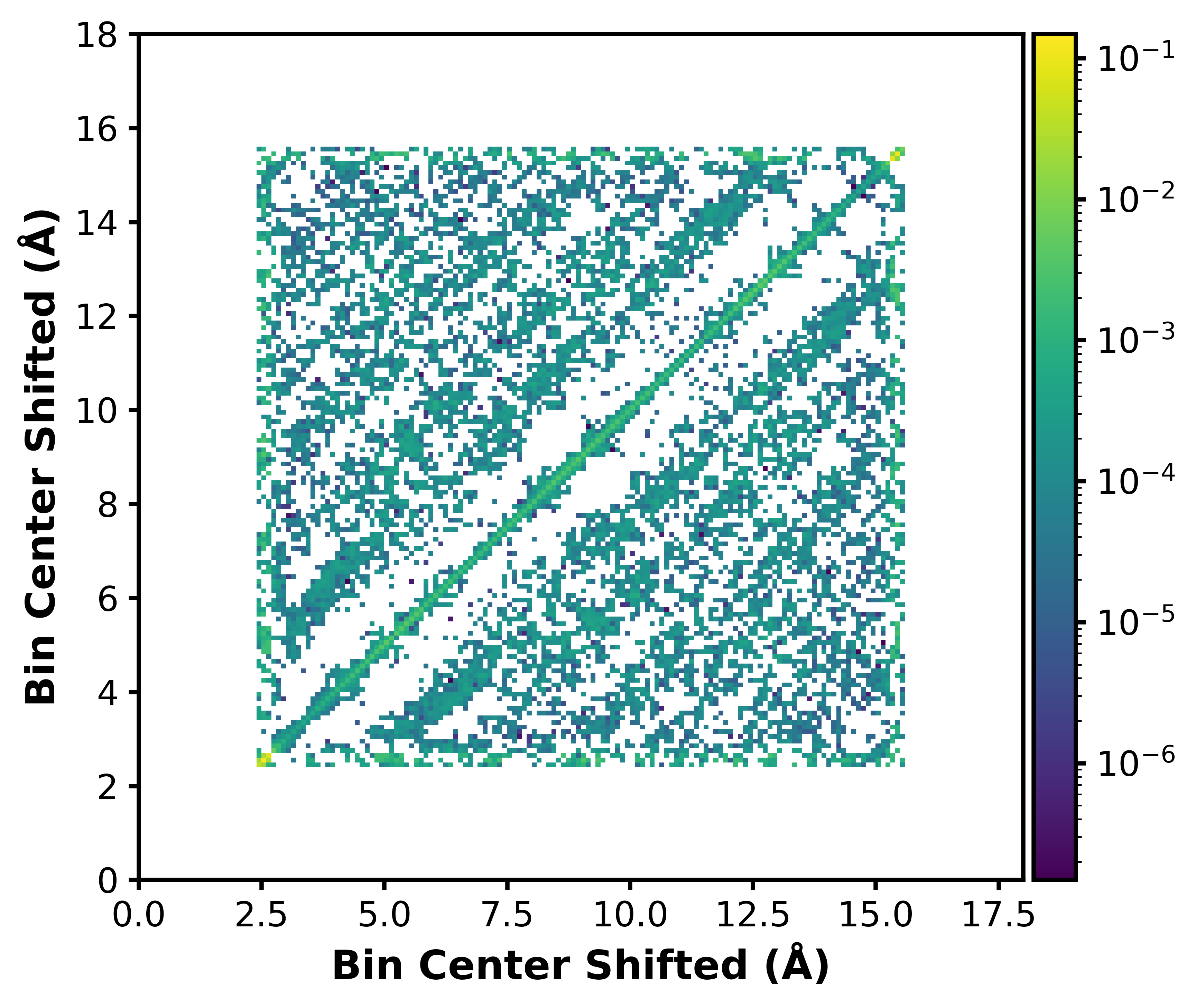} &
\includegraphics[width=0.22\textwidth]{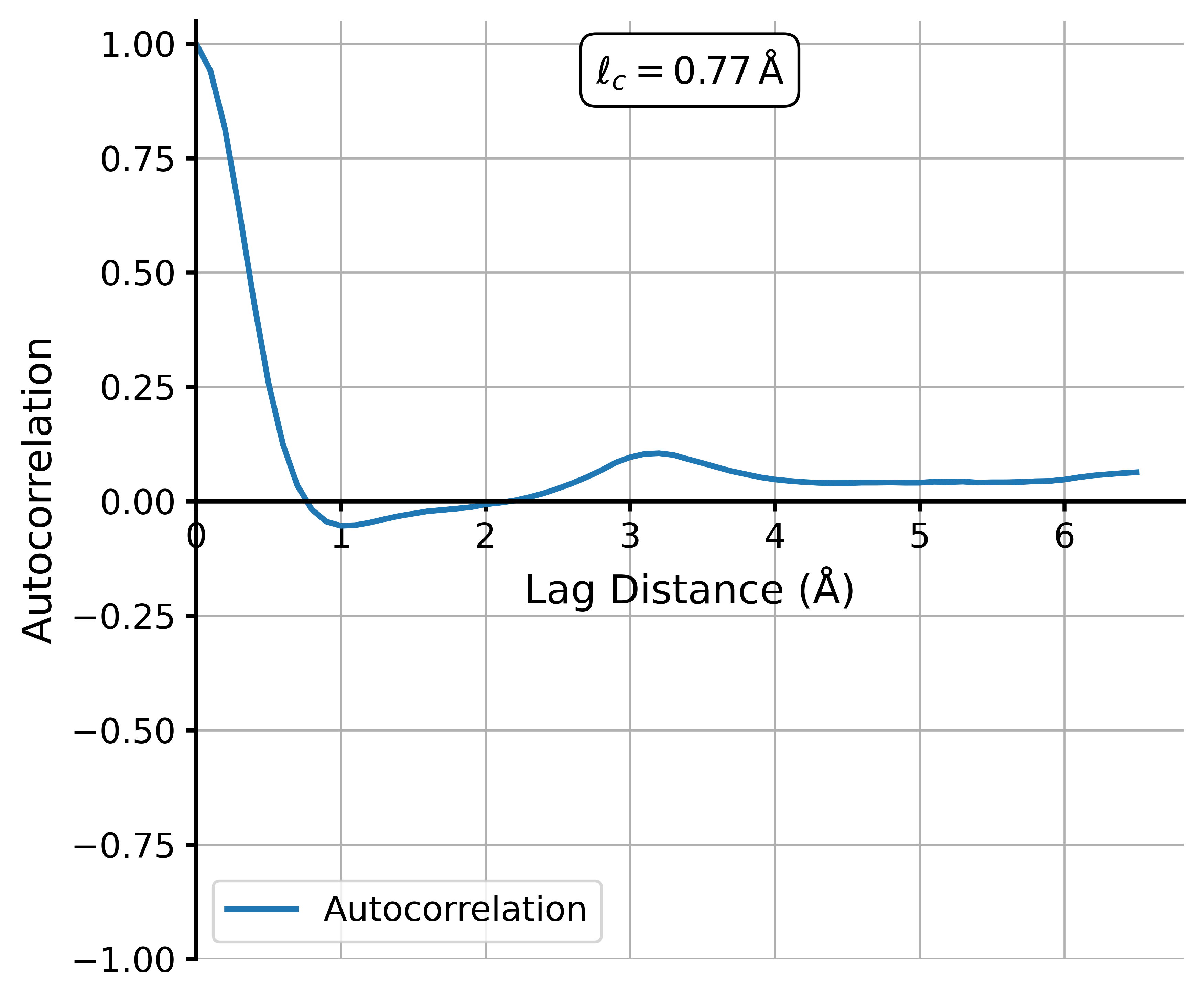} &
\includegraphics[width=0.22\textwidth]{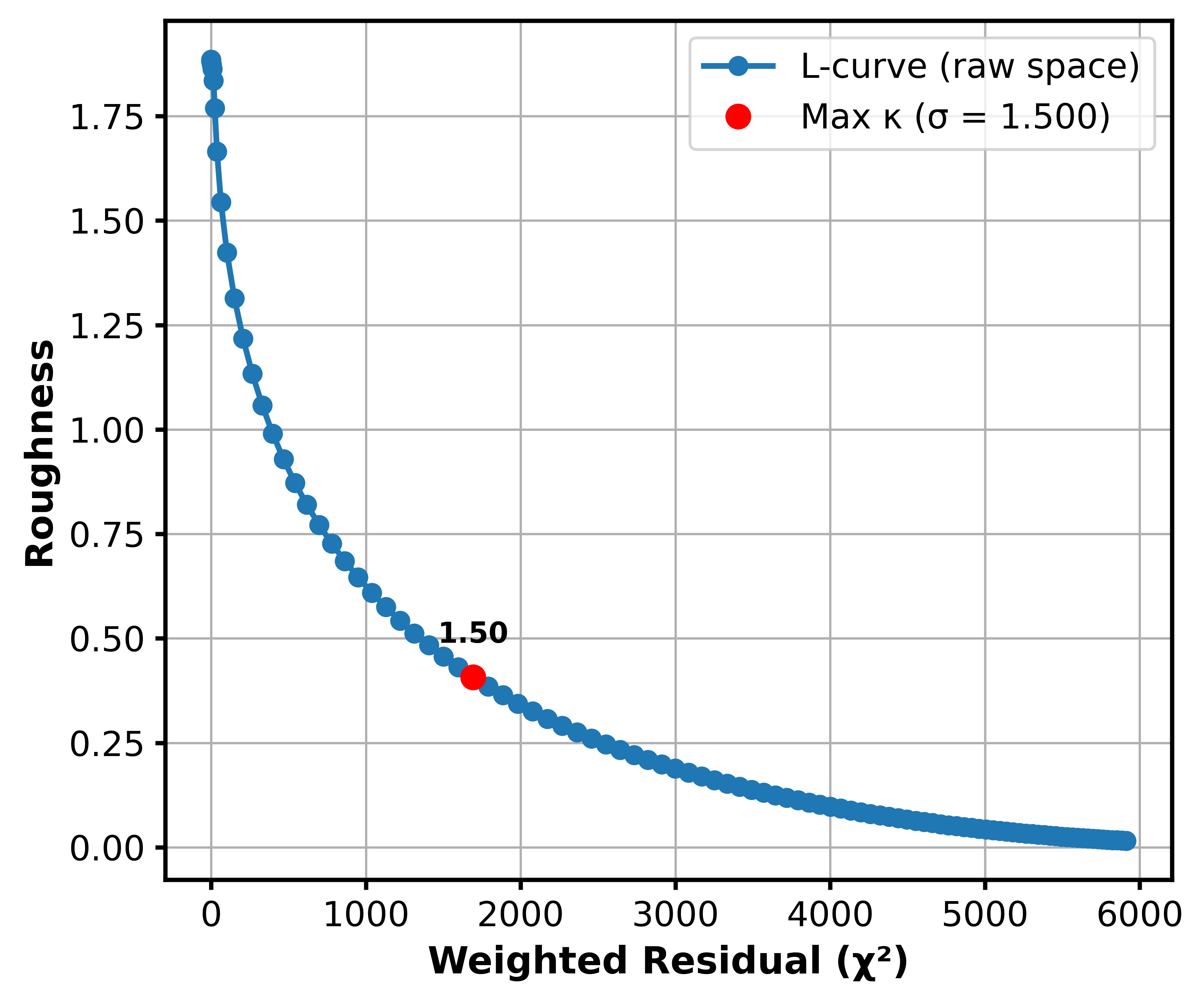} \\

\includegraphics[width=0.22\textwidth]{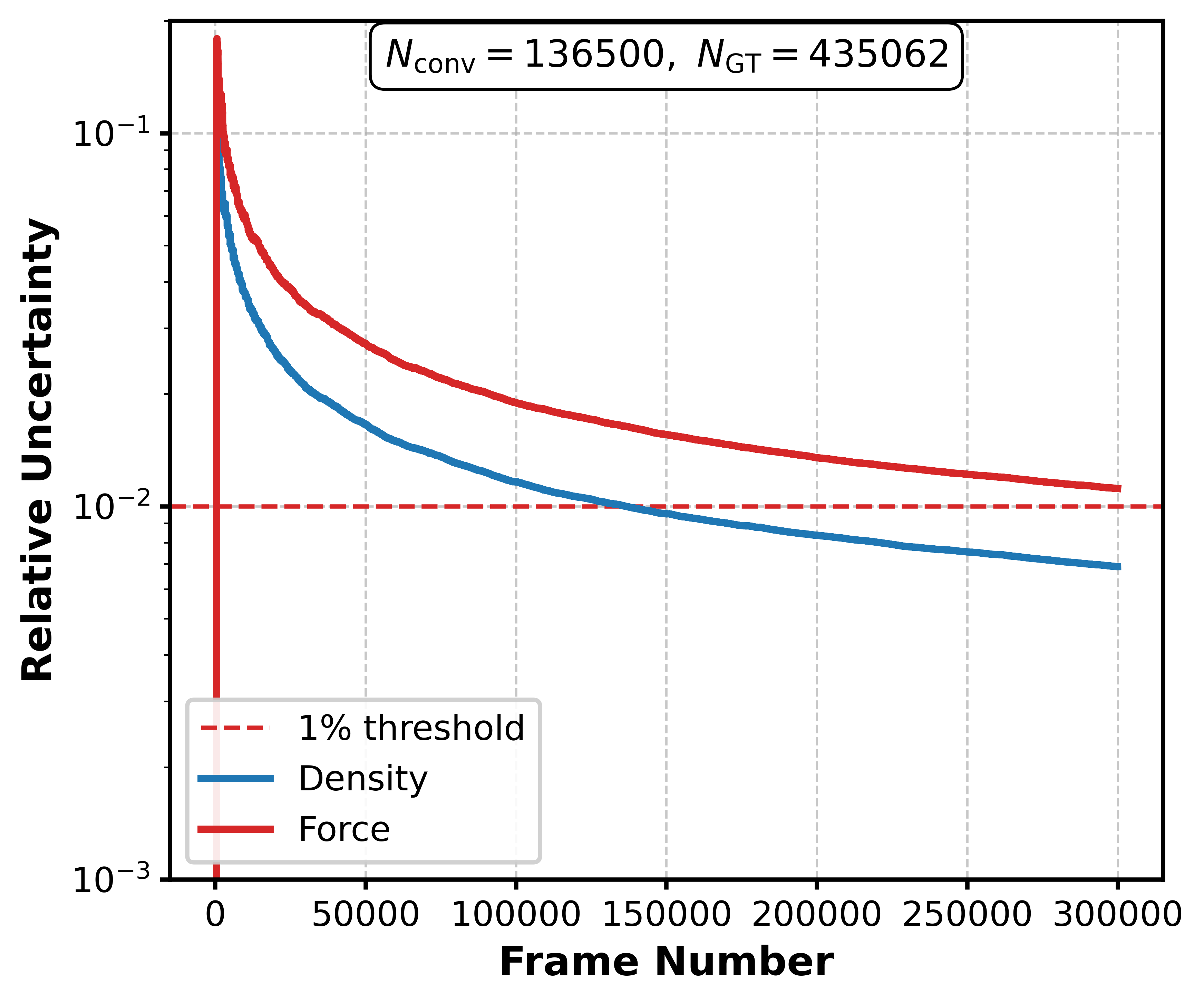} &
\includegraphics[width=0.22\textwidth]{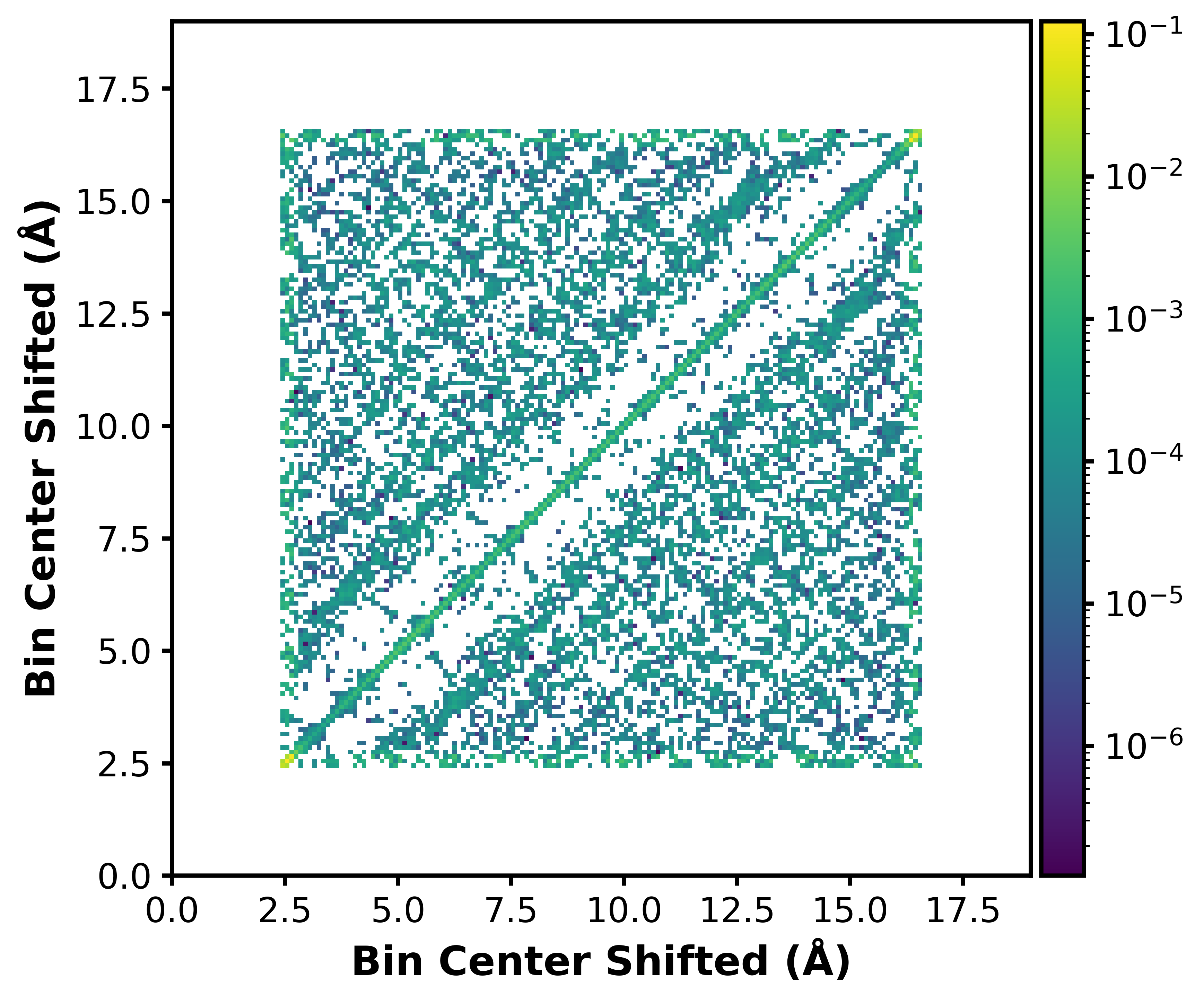} &
\includegraphics[width=0.22\textwidth]{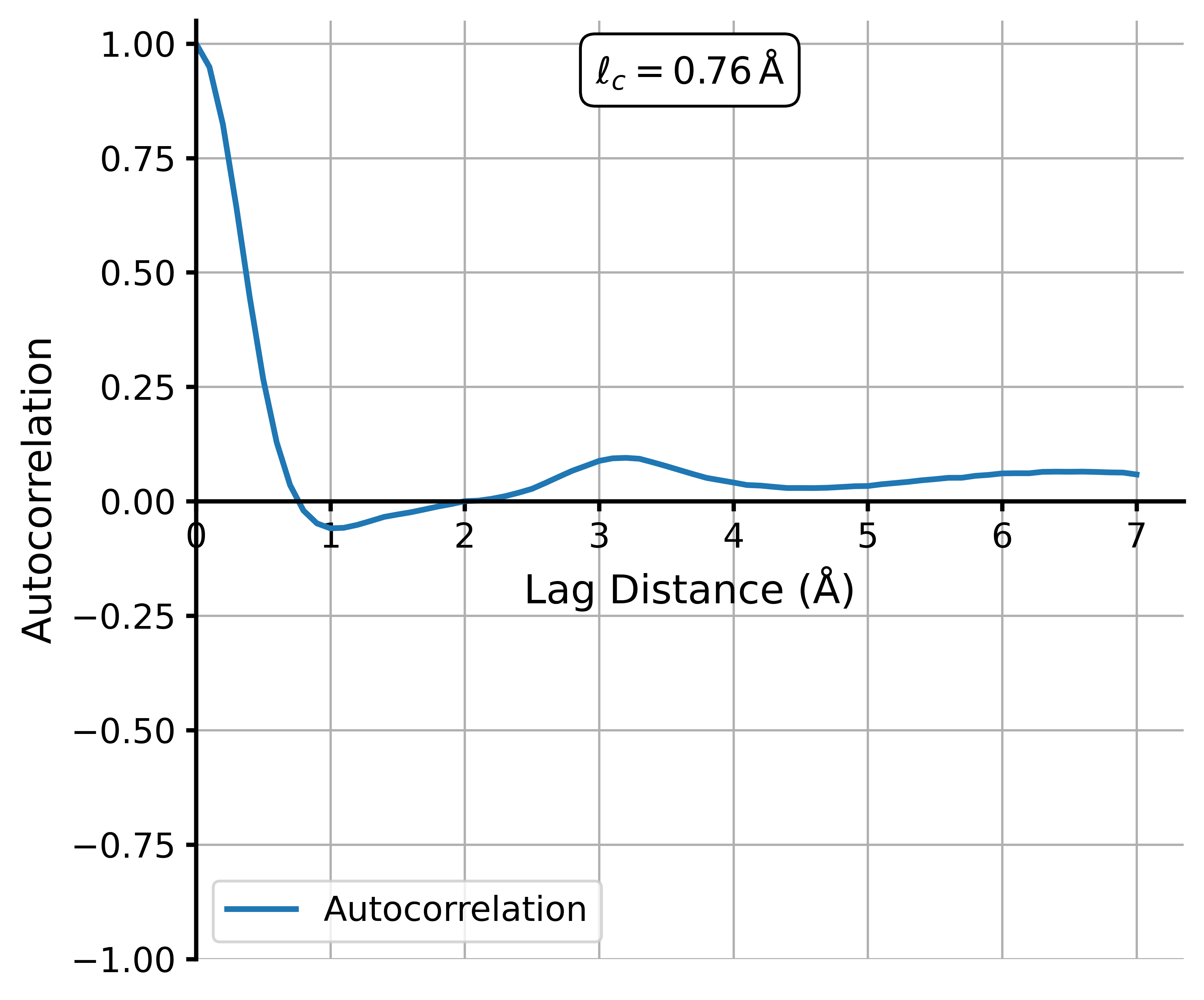} &
\includegraphics[width=0.22\textwidth]{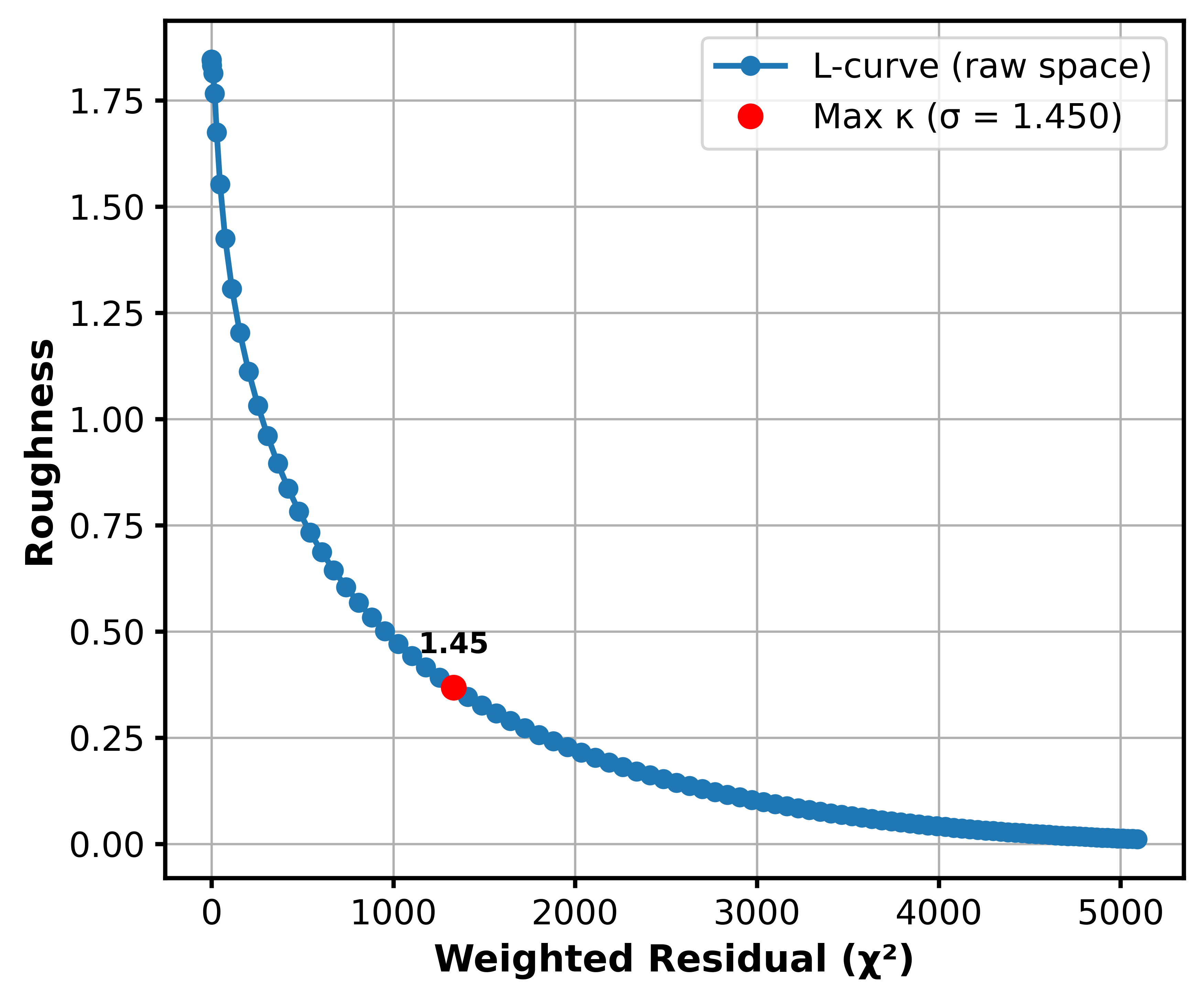} \\

\includegraphics[width=0.22\textwidth]{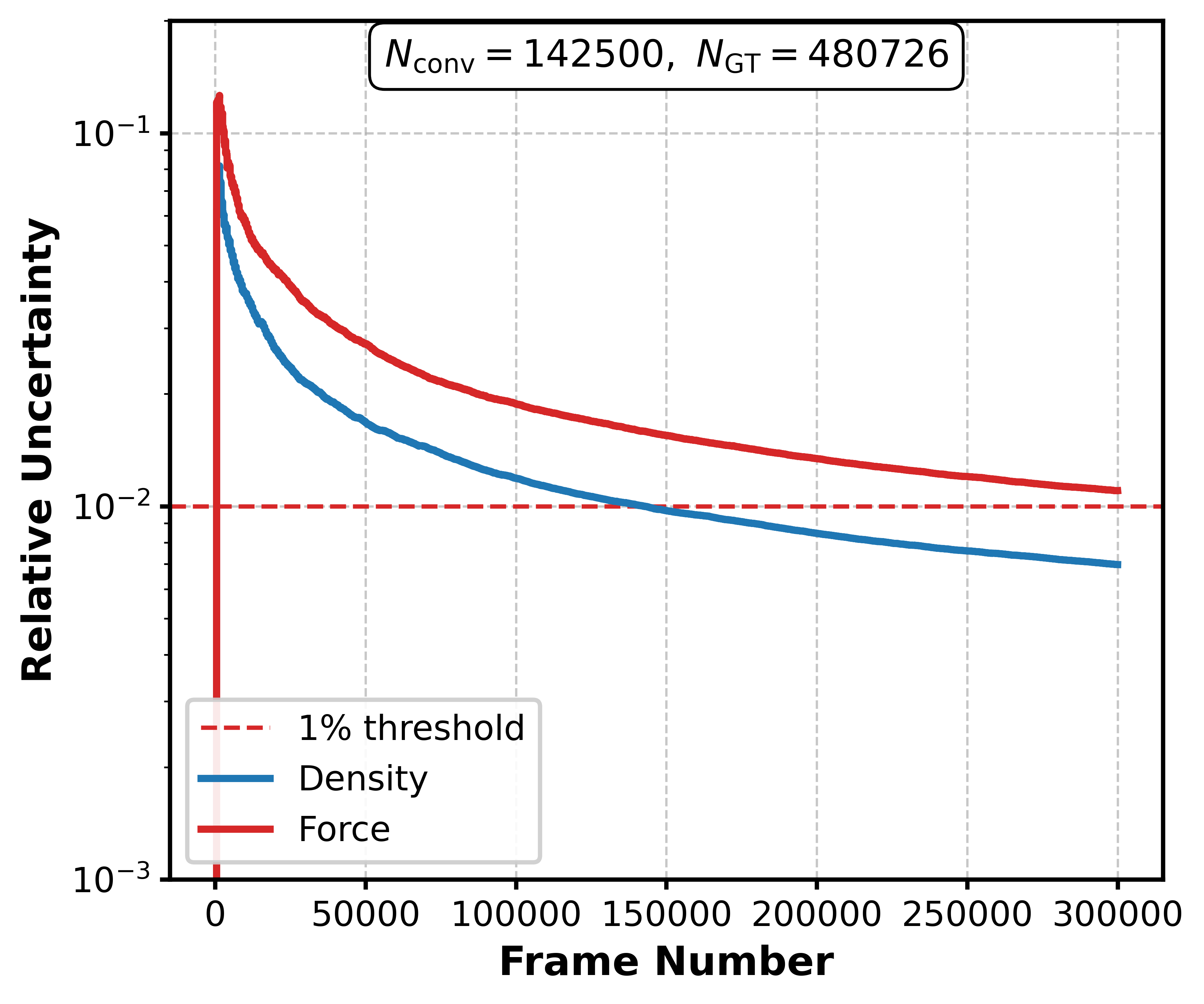} &
\includegraphics[width=0.22\textwidth]{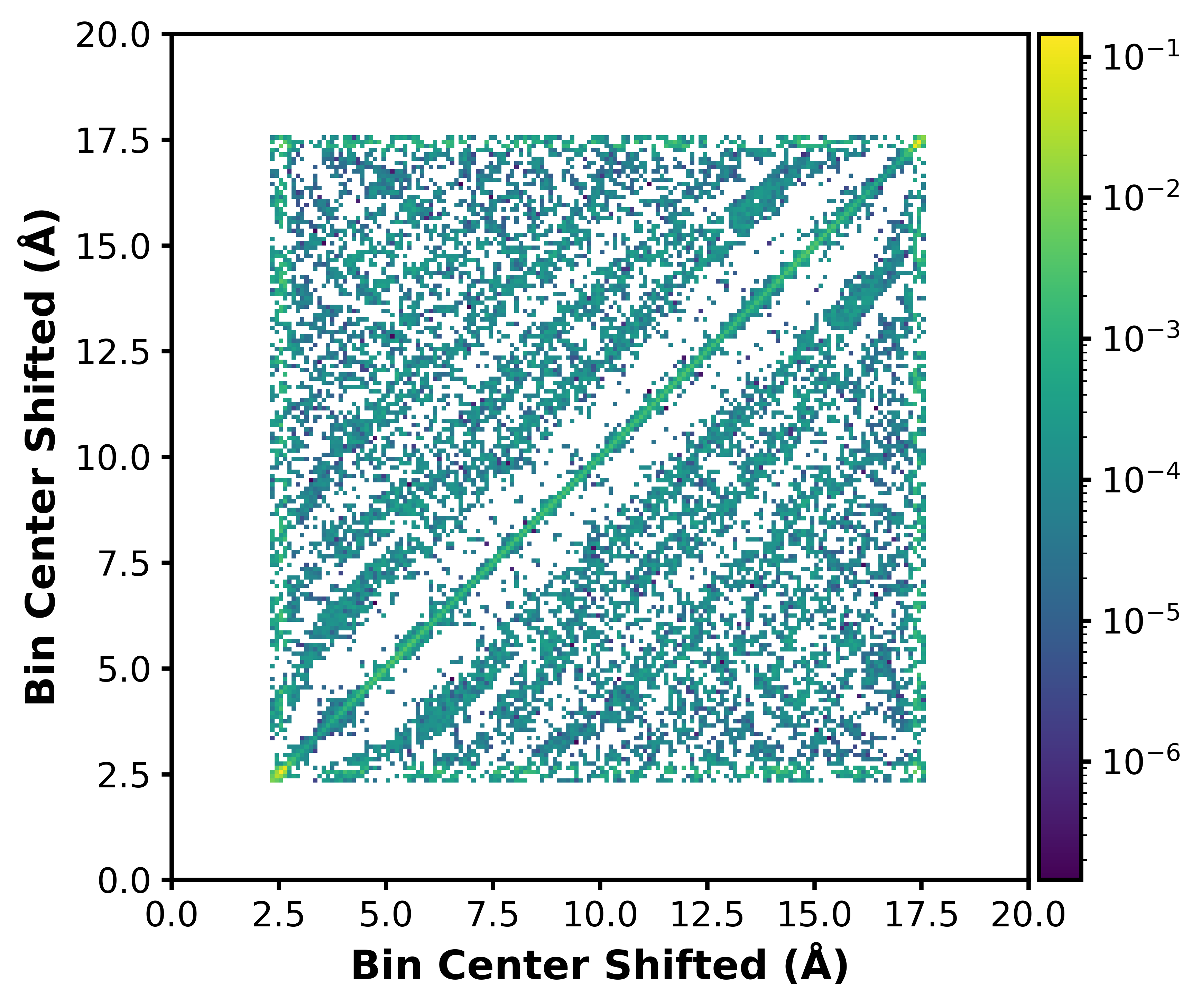} &
\includegraphics[width=0.22\textwidth]{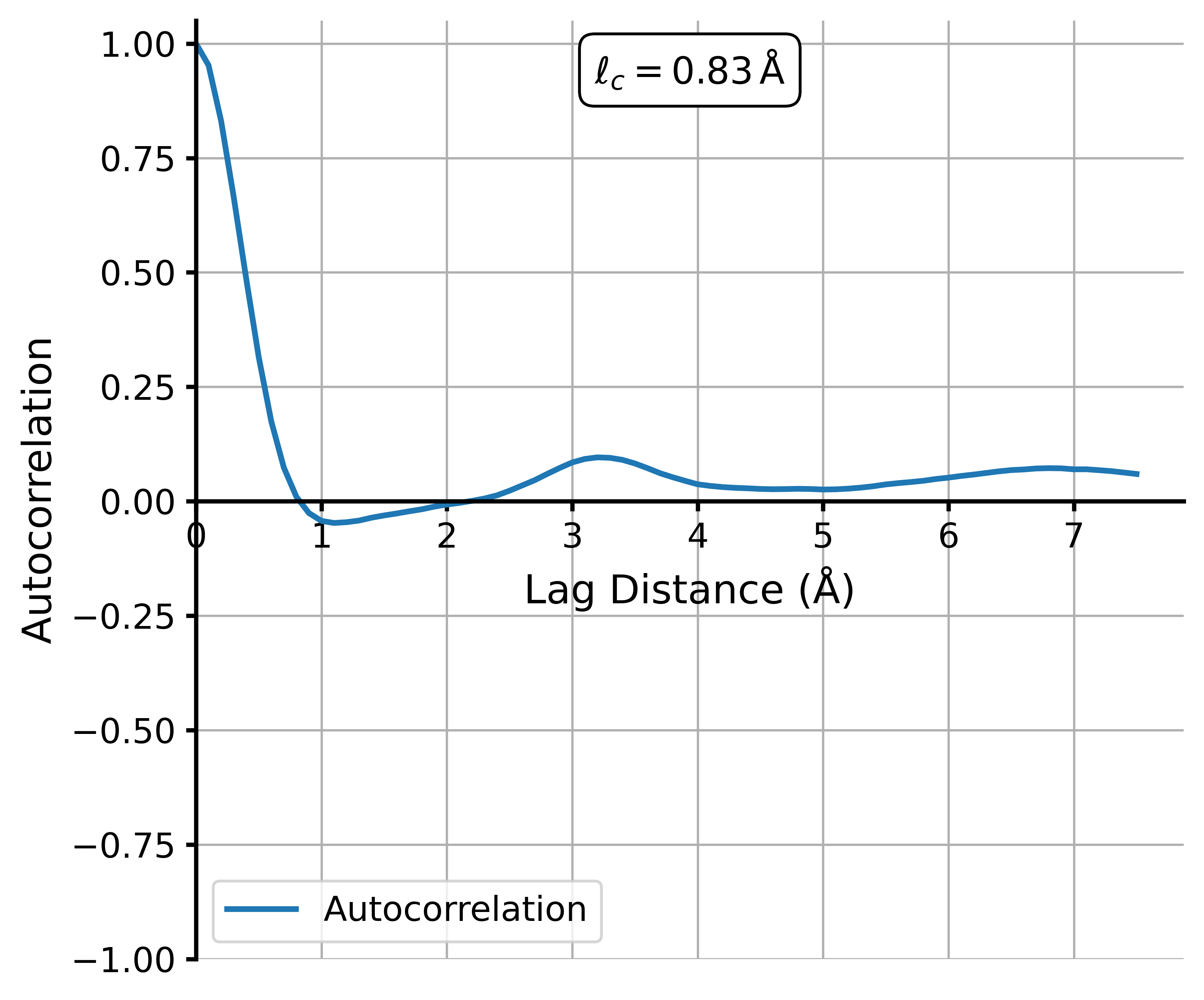} &
\includegraphics[width=0.22\textwidth]{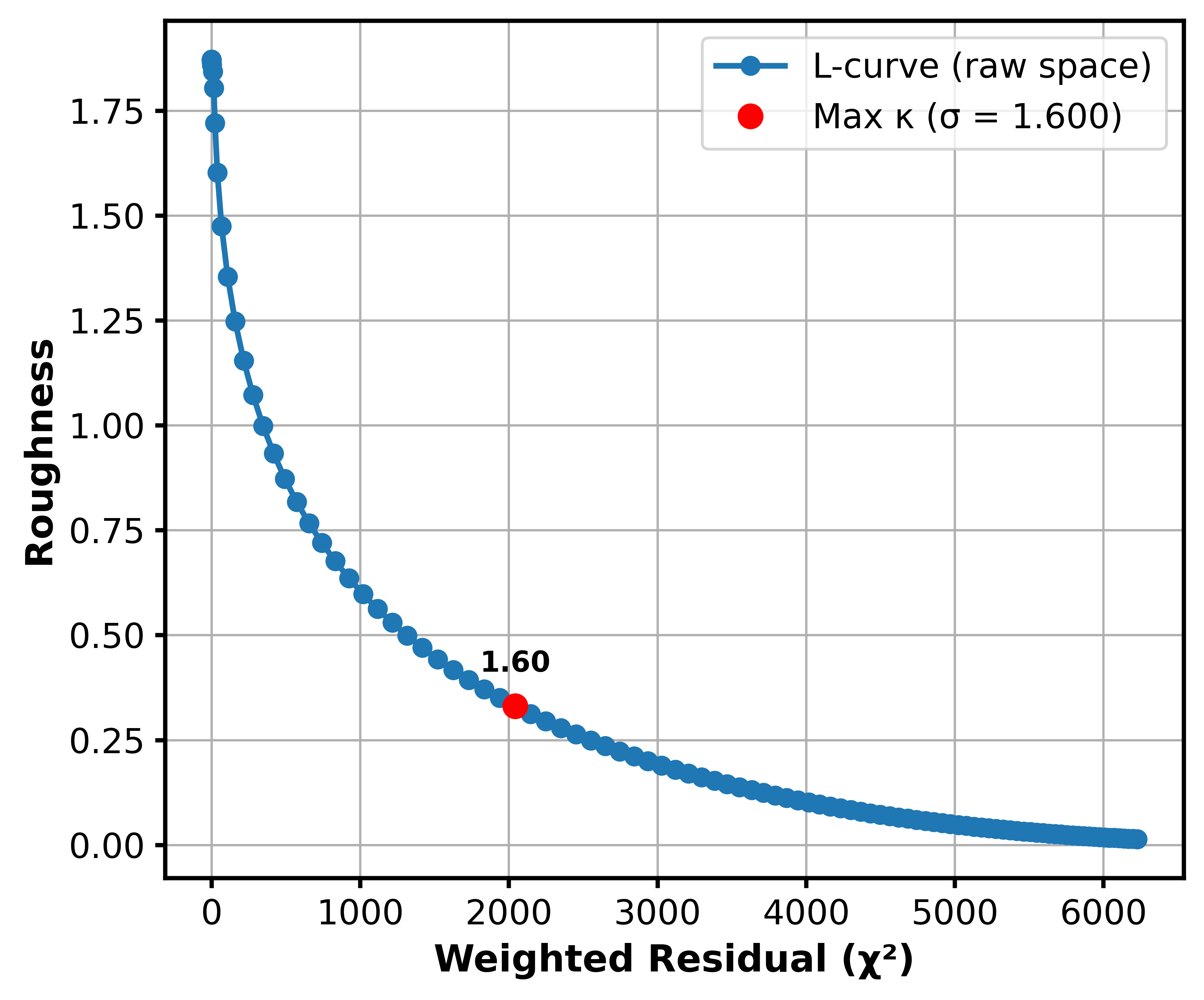} \\

\includegraphics[width=0.22\textwidth]{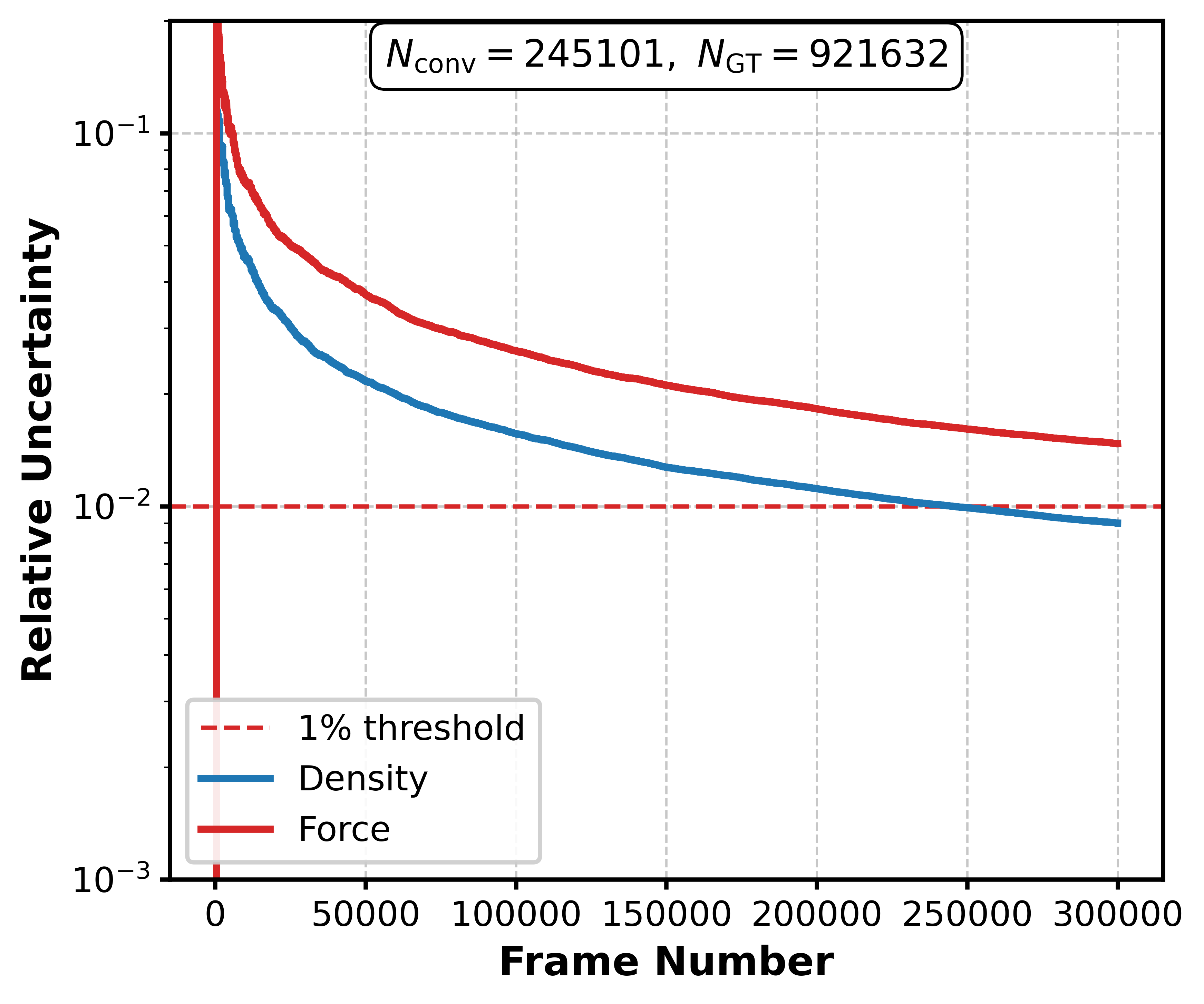} &
\includegraphics[width=0.22\textwidth]{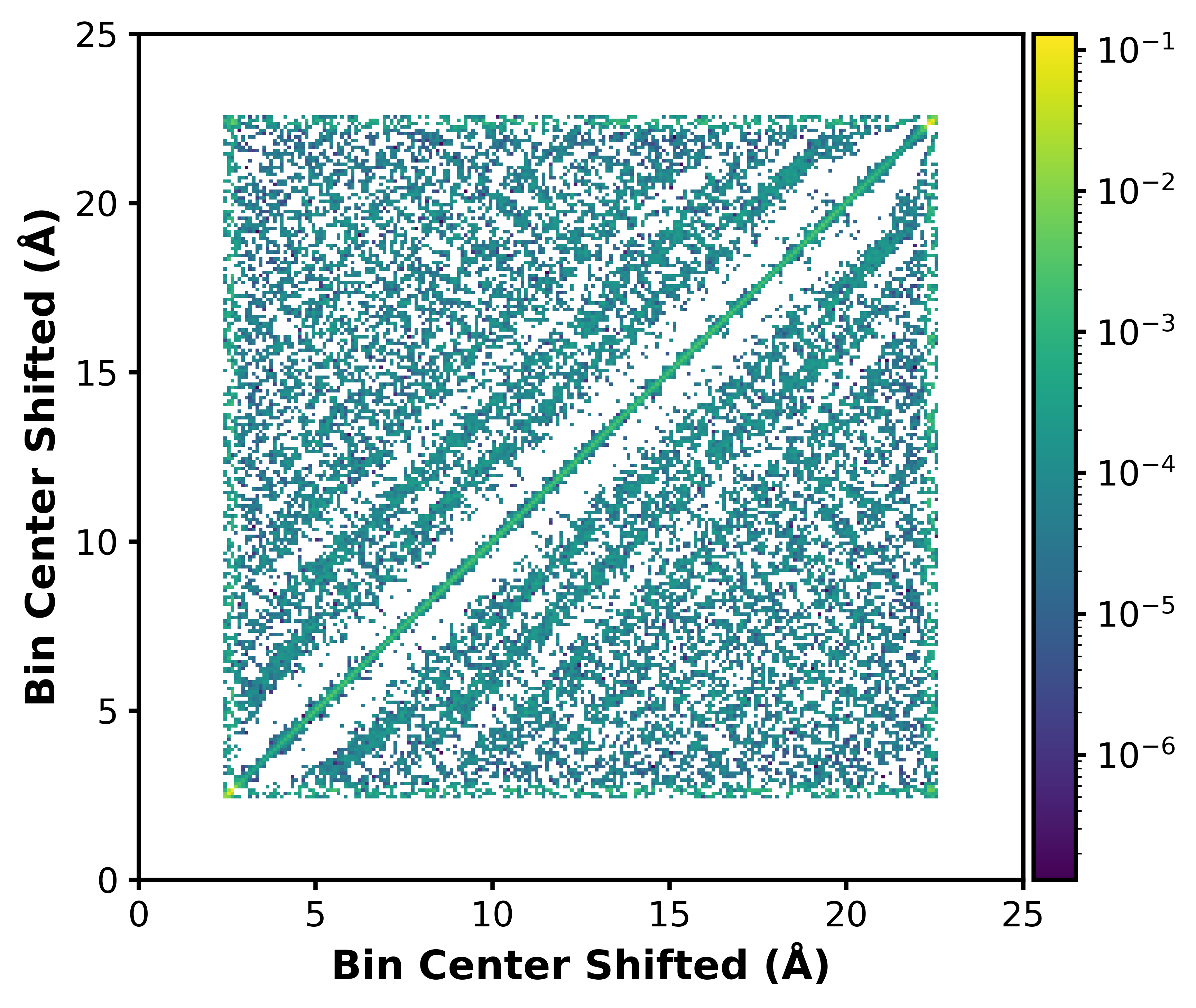} &
\includegraphics[width=0.22\textwidth]{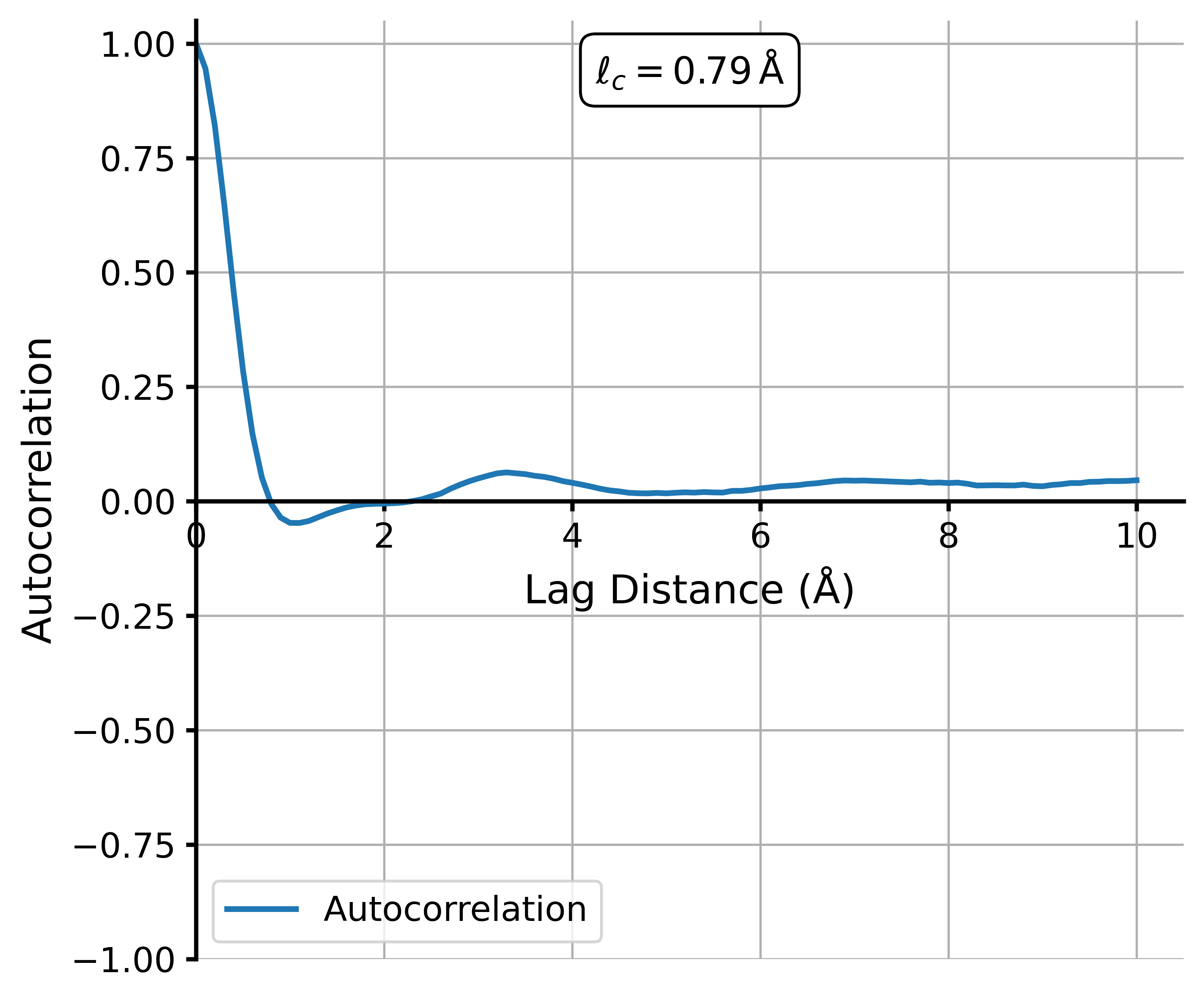} &
\includegraphics[width=0.22\textwidth]{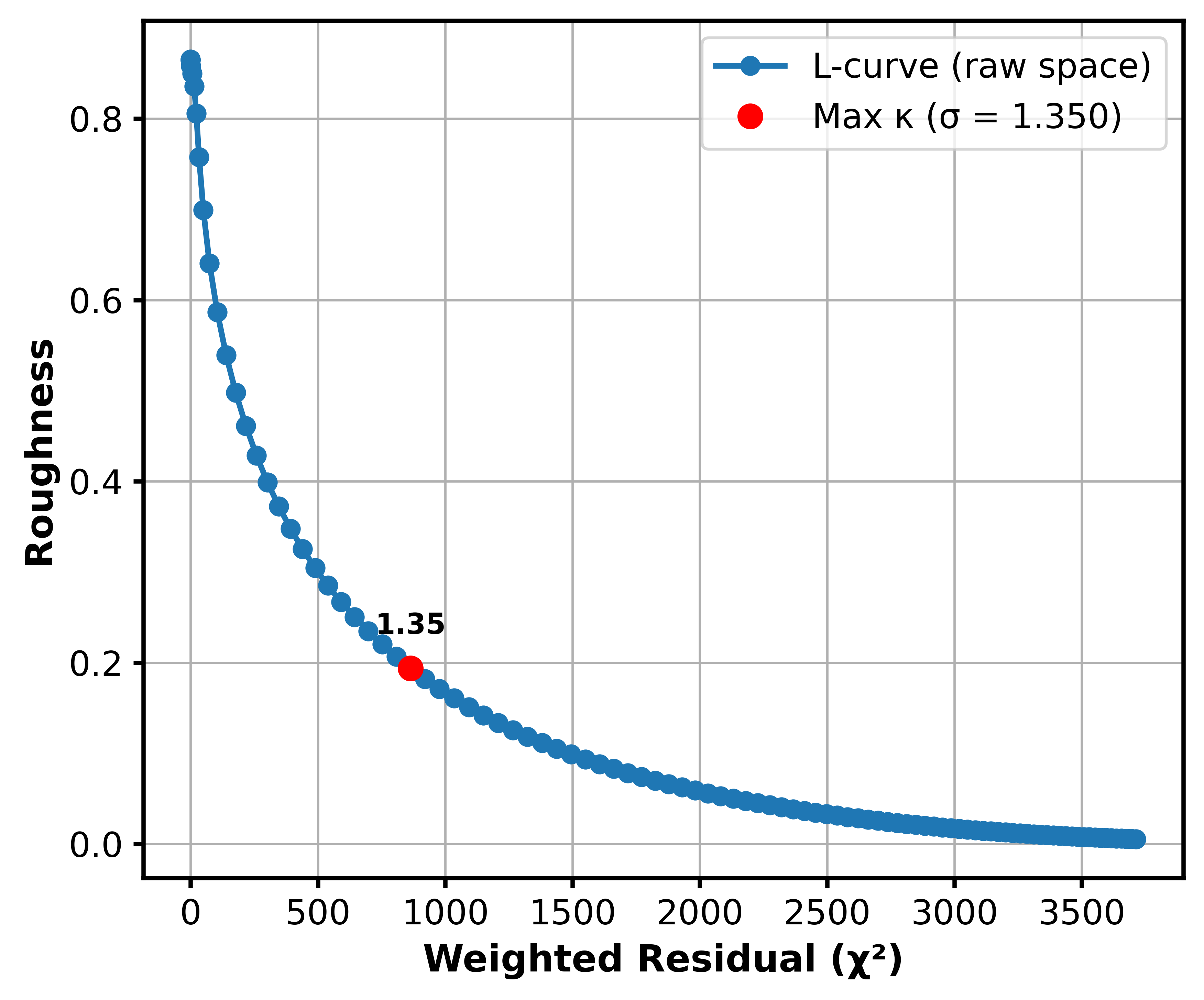} \\

\includegraphics[width=0.22\textwidth]{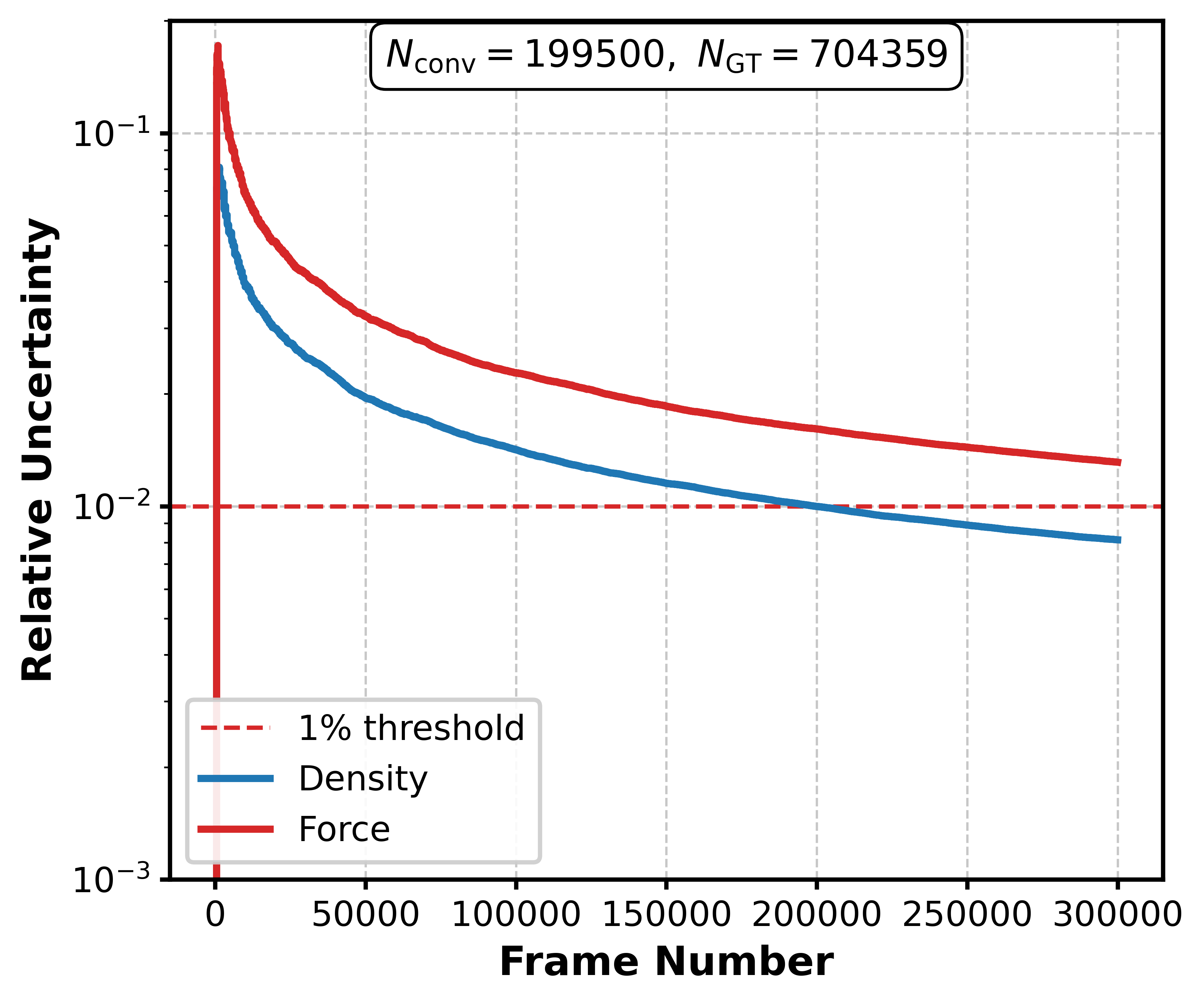} &
\includegraphics[width=0.22\textwidth]{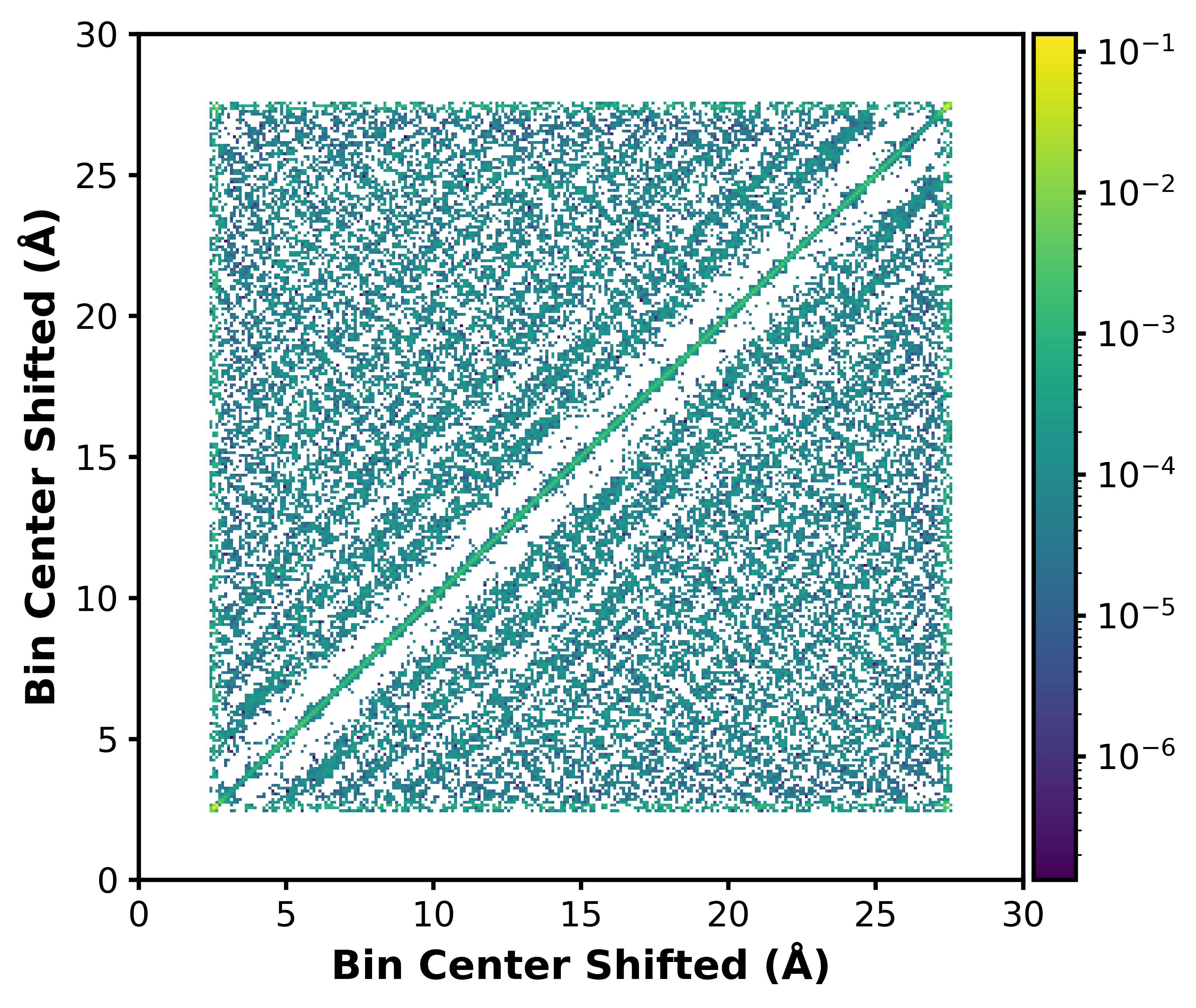} &
\includegraphics[width=0.22\textwidth]{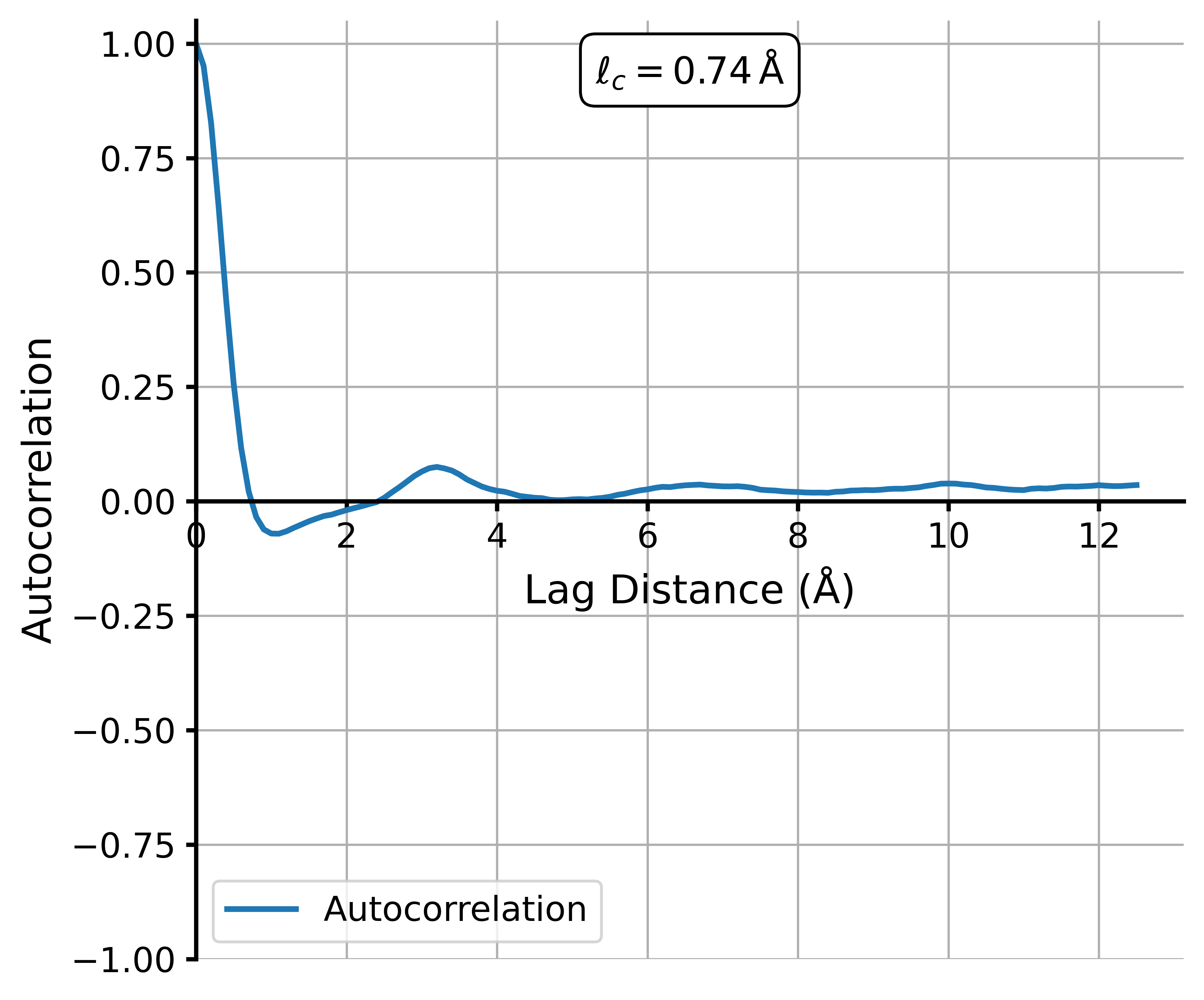} &
\includegraphics[width=0.22\textwidth]{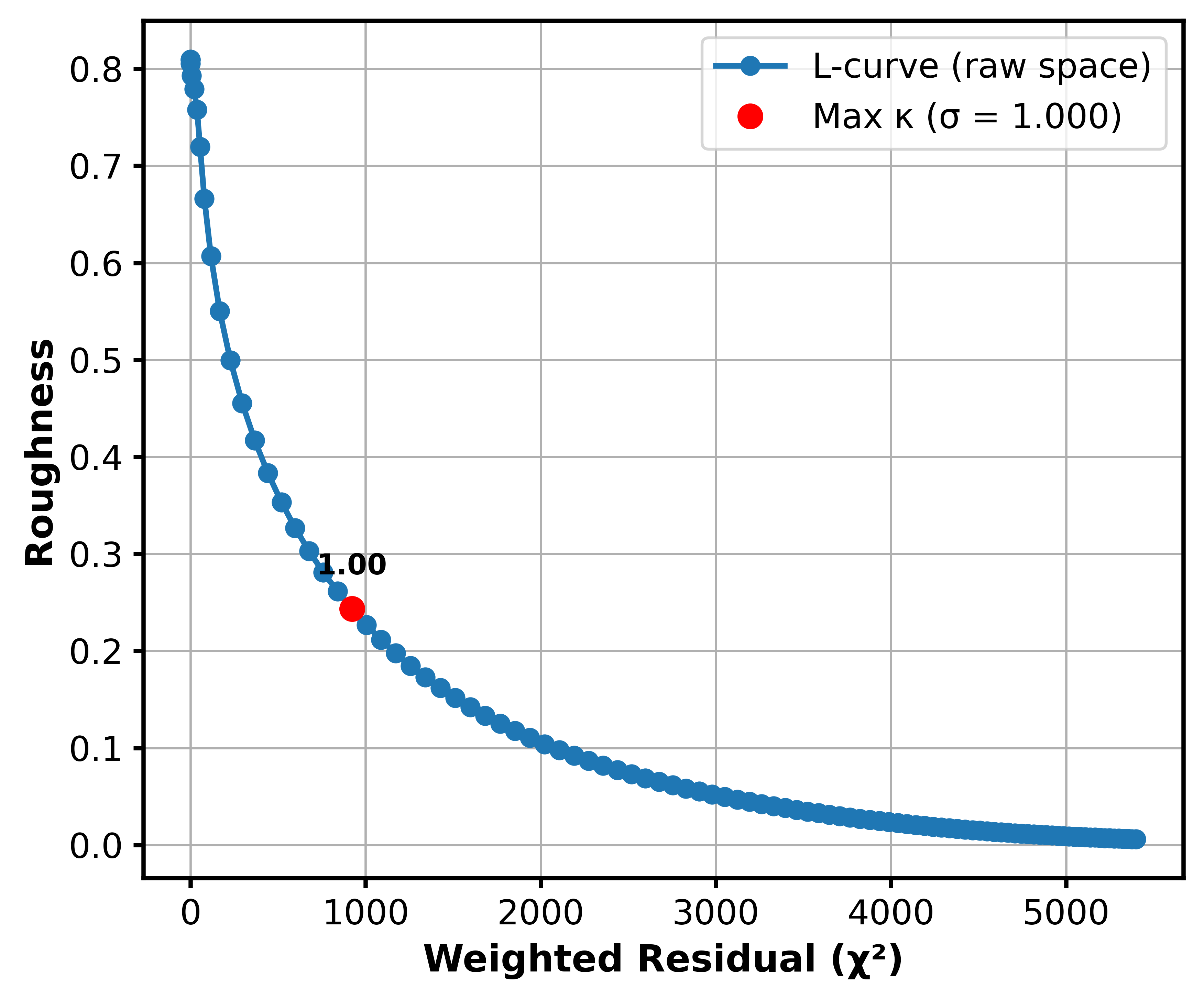} \\

\includegraphics[width=0.22\textwidth]{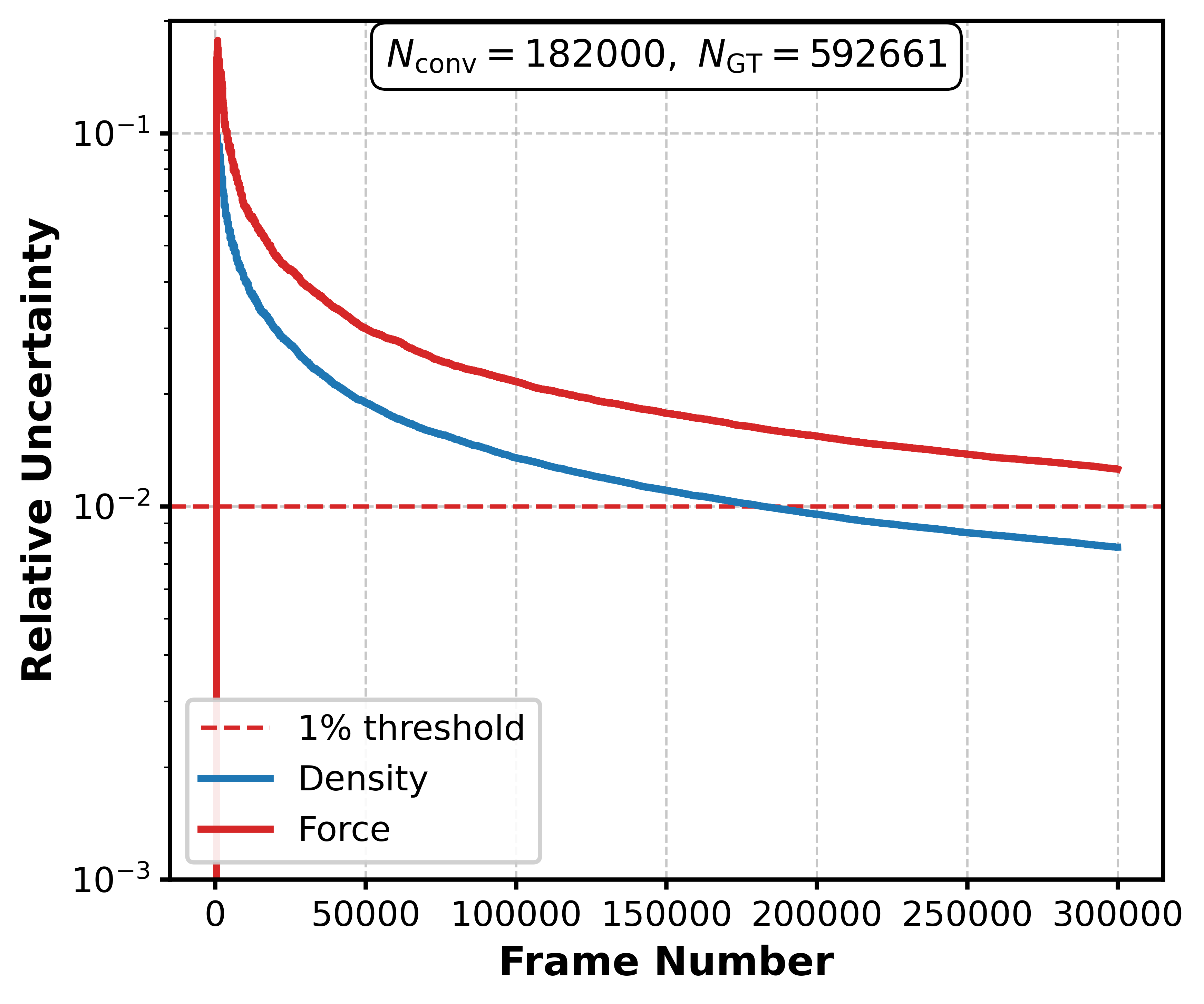} &
\includegraphics[width=0.22\textwidth]{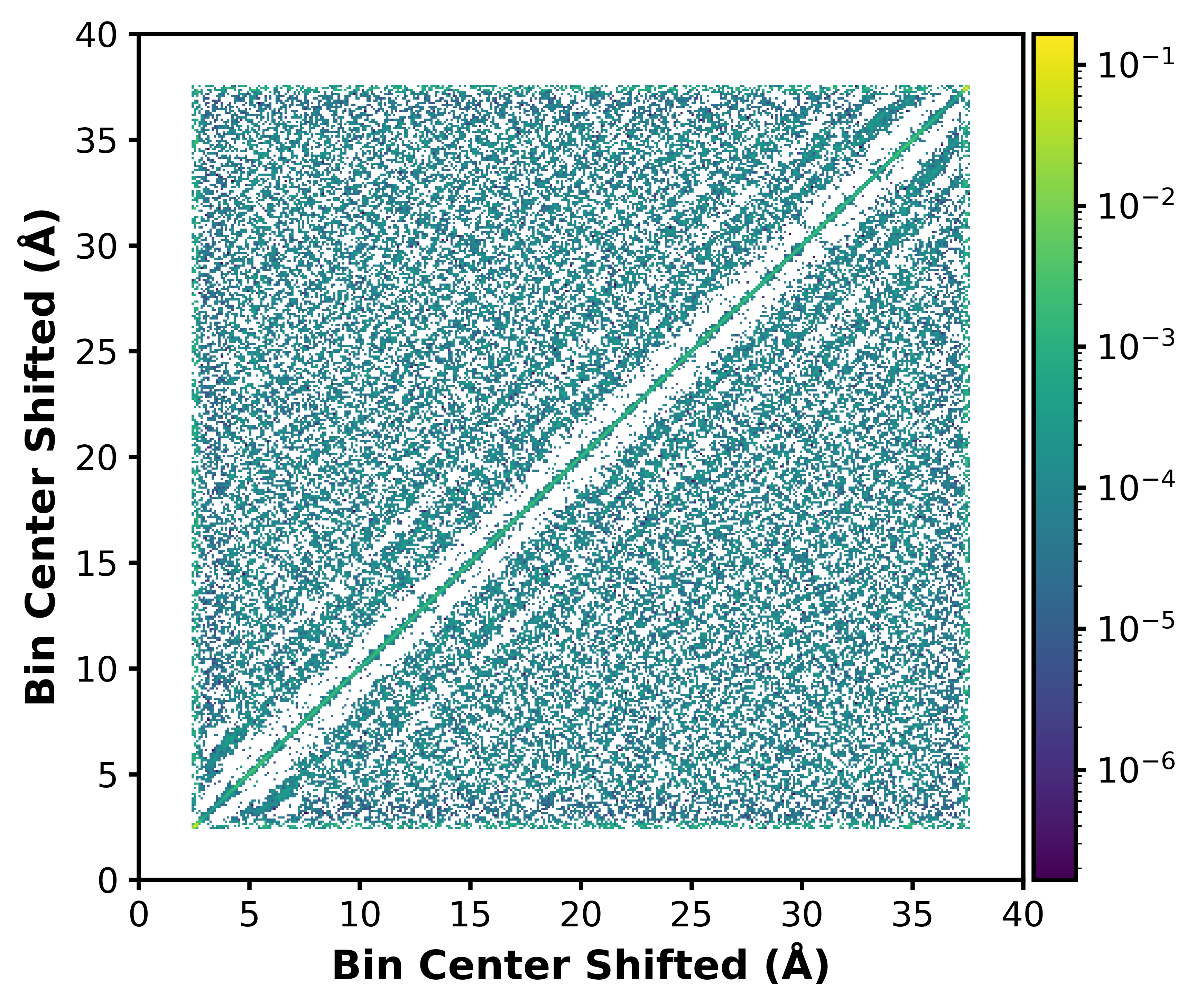} &
\includegraphics[width=0.22\textwidth]{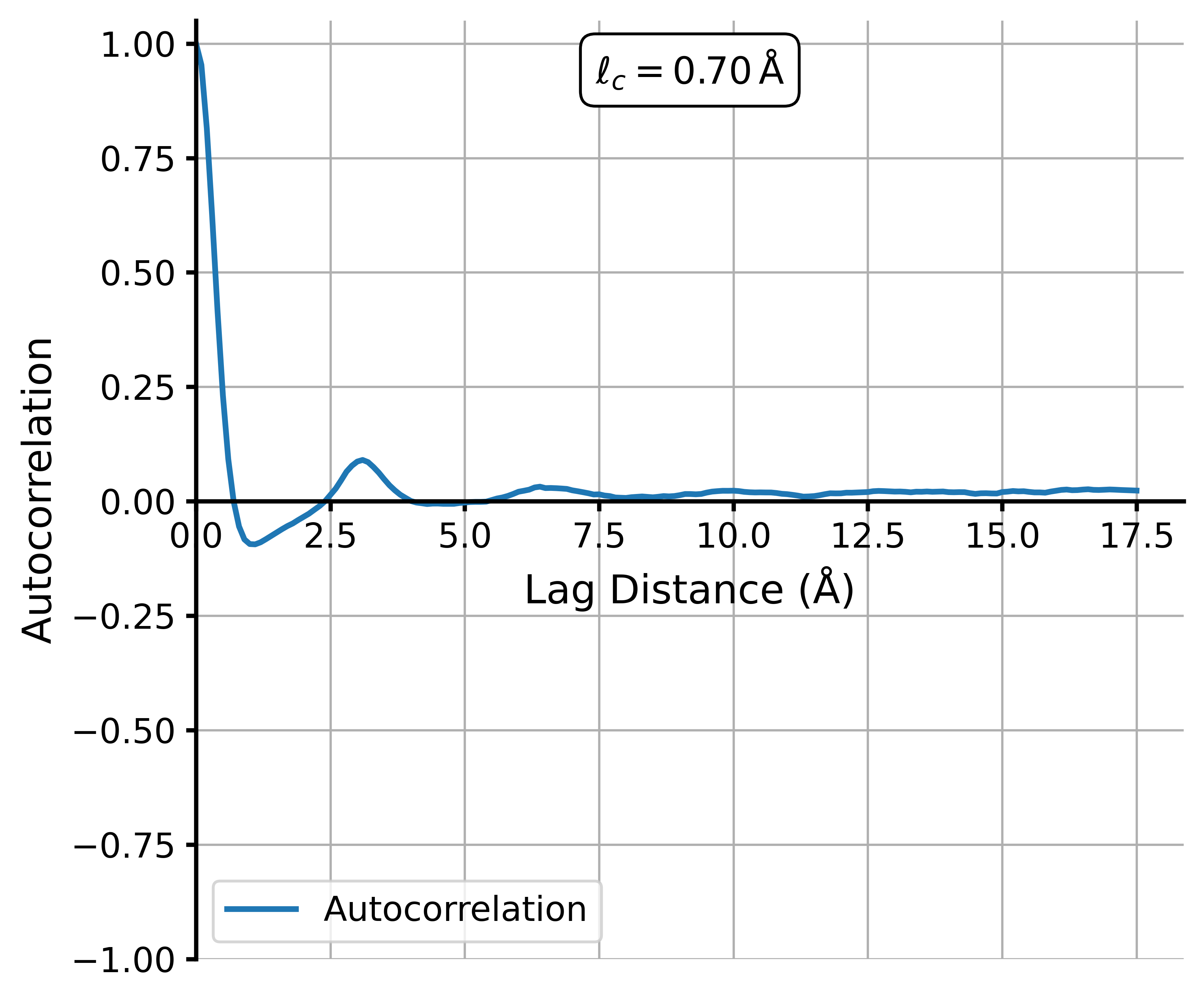} &
\includegraphics[width=0.22\textwidth]{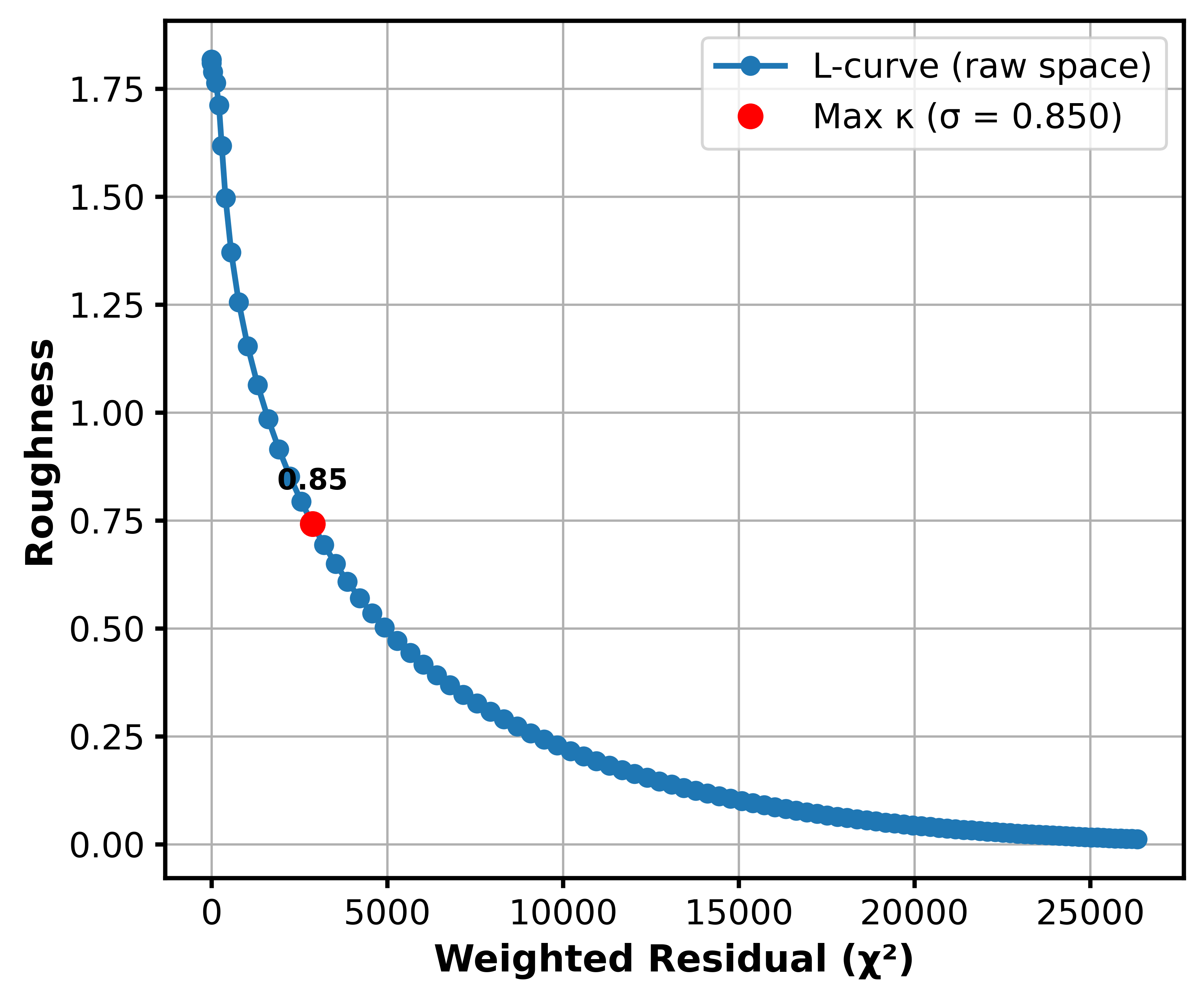} \\
\end{tabular}
\caption{Convergence, covariance, autocorrelation, and L-curve plots for channels 1.8\,nm (top) to 4\,nm. (bottom)}
\label{fig:lfa_convergence2}
\end{figure}

\FloatBarrier
\section{Diffusion Model Training Details}
\label{sec:AppendixD}
\begin{figure}[H]
    \centering
    \includegraphics[width=\columnwidth]{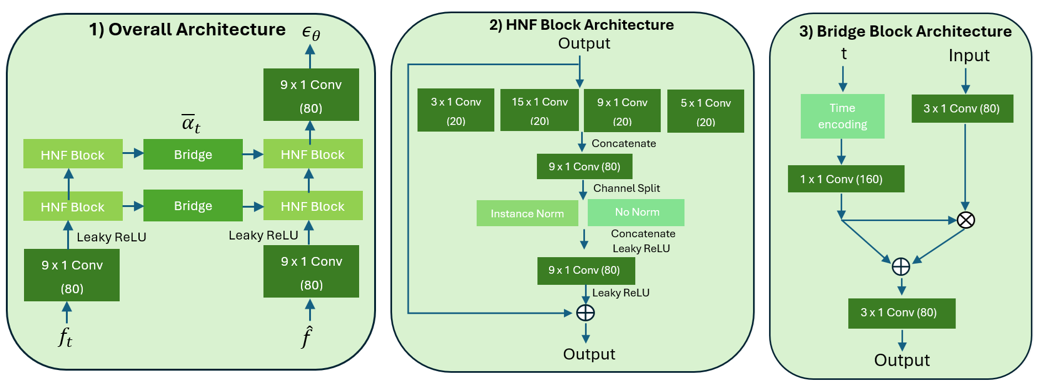}
    \caption{Schematic of the diffusion model architecture.
    (1) The denoising network $\epsilon_{\theta}$, originally introduced by Li \emph{et al.}~\cite{liDeScoDECGDeepScoreBased2024}, takes as input the triplet $(f_t, \hat{f}, t)$, where $f_t$ is the noisy signal at diffusion step $t$ and $\hat{f}$ is the original noisy observation.
    The architecture consists of two primary components:
    (i) a denoising backbone composed of specialized Half-Normalized-Filter (HNF) blocks, and
    (ii) Bridge blocks that facilitate information flow between HNF modules, enabling effective reconstruction of signal structure across diffusion steps.
    More architectural details can be found in Li \emph{et al.}~\cite{liDeScoDECGDeepScoreBased2024}.}
    \label{fig:diffusion_architecture}
\end{figure}

\noindent
The forward noise injection process was governed by a sigmoid noise schedule. The schedule ranged from an initial noise level of $\beta_0 = 1\times10^{-5}$ to a final value of $\beta_T = 1\times10^{-2}$. Model training was performed using the Adam optimizer~\cite{kingmaAdamMethodStochastic2014} with a learning rate of $1\times10^{-3}$ and a batch size of 128. To improve training stability and generalization, an exponential moving average (EMA) with a decay rate of 0.9 was maintained over the network parameters.

\clearpage

\section{Computational Details}
\label{sec:AppendixE}
\paragraph{AIMD simulation.}
The AIMD simulations are carried out using the Vienna \textit{Ab initio} Simulation Package (VASP)~\cite{kresseEfficiencyAbinitioTotal1996,kresseEfficientIterativeSchemes1996}. We employ the Perdew--Burke--Ernzerhof (PBE) exchange-correlation functional within the generalized gradient approximation (GGA), along with the DFT-D3 dispersion correction scheme to account for van der Waals interactions. The simulations use a time step of 0.5~fs and apply periodic boundary conditions in all three directions. A vacuum space of 1.6~nm is inserted along the $x$-direction to ensure that the periodic copies of the system do not interact across the channel width. A plane-wave energy cutoff of 450~eV is chosen, with a $\Gamma$-centered $1 \times 2 \times 2$ k-point mesh. The system temperature is maintained at 400~K using a Nos\'e--Hoover thermostat, and carbon atoms forming the graphene walls are held fixed.

\paragraph{MLMD simulation.} Initial configurations from the AIMD simulations are used for the corresponding nanochannels. All simulations are performed in the NVT ensemble at 400 K with a timestep of 0.5 fs. Periodic boundary conditions identical to those in the AIMD simulations are applied.

\end{document}